\documentclass[aps,letterpaper,superscriptaddress,twocolumn,longbibliography,nofootinbib]{revtex4-1}
\pdfoutput=1

\usepackage{bmpsize}

\usepackage{hyperref}
\usepackage[x11names]{xcolor}
\usepackage{graphicx}	%
\usepackage[utf8]{inputenc} 
\usepackage{bm}
\usepackage{amsmath}
\usepackage{amsfonts}

\usepackage{slashed}

\def\be{\begin{equation}}
\def\ee{\end{equation}}
\def\ba{\begin{eqnarray}}
\def\ea{\end{eqnarray}}

\usepackage{placeins}

\usepackage{soul}

\begin{document}

\title{Upper Bound on the Speed of Sound in Nuclear Matter from Transport}

\author{Mauricio Hippert}
\affiliation{Illinois Center for Advanced Studies of the Universe, University of Illinois Urbana-Champaign, Urbana, IL 61801, USA}
\affiliation{Department of Physics, University of Illinois Urbana-Champaign, Urbana, IL 61801, USA}

\author{Jorge Noronha}
\affiliation{Illinois Center for Advanced Studies of the Universe, University of Illinois Urbana-Champaign, Urbana, IL 61801, USA}
\affiliation{Department of Physics, University of Illinois Urbana-Champaign, Urbana, IL 61801, USA}

\author{Paul Romatschke}
\affiliation{Department of Physics, University of Colorado, Boulder, Colorado 80309, USA}
\affiliation{Center for Theory of Quantum Matter, University of Colorado, Boulder, Colorado 80309, USA}

\begin{abstract}
We point out that there is an upper bound on the speed of sound squared given by $c_s^2 \leq  0.781$ valid for all known systems described by relativistic transient hydrodynamics where calculations of certain ratios of hydrodynamic transport coefficients can be performed from first principles. Assuming this bound is valid for ultradense matter implies that the maximum mass of isolated (non-rotating) neutron stars cannot be larger than 2.7 solar masses. 
\end{abstract}

\maketitle
\section{Introduction}
The measurements of neutron stars with masses $\gtrsim 2\,M_\odot$ (where $M_\odot$ is the mass of the sun) \cite{Demorest:2010bx,Antoniadis:2013pzd, Cromartie:2019kug} suggest that the cold neutron star equation of state (EoS) is stiff enough such that its speed of sound, $c_s$, surpasses its conformal limit, $c_s\leq 1/\sqrt{3}$, at a given density \cite{Bedaque:2014sqa}. This is also supported by the binary neutron-star merger gravitational-wave event GW170817 \cite{TheLIGOScientific:2017qsa,Abbott:2018exr,Abbott:2018wiz}, which further constrained the masses and effective tidal deformability of the inspiraling neutron stars \cite{Tews:2018kmu,Tews:2019cap,Reed:2019ezm,Capano:2019eae,Annala:2019puf,Kanakis-Pegios:2020jnf,Huth:2020ozf,Han:2021kjx,Tan:2021ahl}. Further support to a supraconformal speed of sound may come from the observation of gravitational waves from the merger of a black hole with a compact object of mass $2.6 \,M_\odot$ \cite{Abbott:2020khf}, if the latter is a neutron star \cite{Tan:2020ics,Tsokaros:2020hli,Tews:2020ylw,Godzieba:2020tjn,Huang:2020cab,Biswas:2020xna}.

Due to asymptotic freedom \cite{Gross:1973id,Politzer:1973fx}, the speed of sound in QCD is expected to approach the conformal limit from below at sufficiently large densities \cite{Kurkela:2009gj,Graf:2015tda}. This fact, combined with the knowledge about the nuclear physics equation of state at low densities \cite{Epelbaum:2008ga,Drischler:2017wtt,Lu:2018bat,Tews:2018kmu} and the recent observation of large neutron star masses, implies that $c_s$ in QCD is not monotonic as this quantity must display at least one peak as a function of density (when taking into account the region of asymptotically large density). Assuming there is only a single peak, essential questions are: (i) How large is $c_s$ at the peak? (ii) At what density does $c_s$ peak? (iii) What is the correct effective theory of ultradense matter that explains the answers to (i) and (ii)? Ultimately, future observations \cite{Maggiore:2019uih,Ballmer:2022uxx,Bogdanov:2022faf,Patricelli:2022hhr} will constrain the answer to (i) and (ii) \cite{Landry:2018prl,Landry:2020vaw,Annala:2021gom,Huth:2021bsp,Legred:2021hdx,MUSES:2023hyz,Mroczek:2023zxo} and provide useful guidance towards answering (iii).   

Relativistic causality and covariant stability \cite{Gavassino:2023myj} impose that the speed of sound cannot surpass the speed of light, i.e., $c_s \leq 1$.
At high temperatures and zero baryon chemical potential, lattice QCD calculations \cite{Borsanyi:2013bia,Bazavov:2017dsy} and holographic models \cite{Gubser:2008yx,Cherman:2009tw,Finazzo:2014cna,Grefa:2021qvt} find that the conformal limit for the speed of sound is respected, $c_s \leq 1/\sqrt{3}$. At nonzero baryon density, however, discussions concerning the (subluminal) upper bound to the speed of sound go back several decades \cite{HARTLE1978201,Zeldovich:1962emp,Bludman:1968zz,zel2014stars}. It is now known that the speed of sound can surpass the conformal result in a variety of systems such as QCD at large isospin density \cite{Carignano:2016lxe}, two-color QCD \cite{Hands:2006ve}, holographic models \cite{Hoyos:2016cob,Ecker:2017fyh,Anabalon:2017eri,Chesler:2019osn,Ishii:2019gta}, resummed perturbation theory \cite{Gorda:2014vga,Fujimoto:2020tjc}, quarkyonic matter \cite{McLerran:2018hbz,Jeong:2019lhv,Zhao:2020dvu,Margueron:2021dtx,Duarte:2021tsx}, and other models at high density \cite{Blaschke:2007ri,Kojo:2014rca,Leonhardt:2019fua,Baym:2019iky,Roupas:2020nua,Malfatti:2020onm,Ayriyan:2021prr,Fadafa:2019euu,Xia:2019xax,Stone:2019abq,Li:2020dst,Pisarski:2021aoz,Pal:2021qav,Motta:2021xwo,Ma:2021zev,Hippert:2021gfs}. However, despite recent progress \cite{Hoyos:2020hmq,Hoyos:2021njg}, very little is known about the transport properties of ultradense matter when the speed of sound nears the speed of light.

\begin{figure}[t]
    \includegraphics[width=\linewidth]{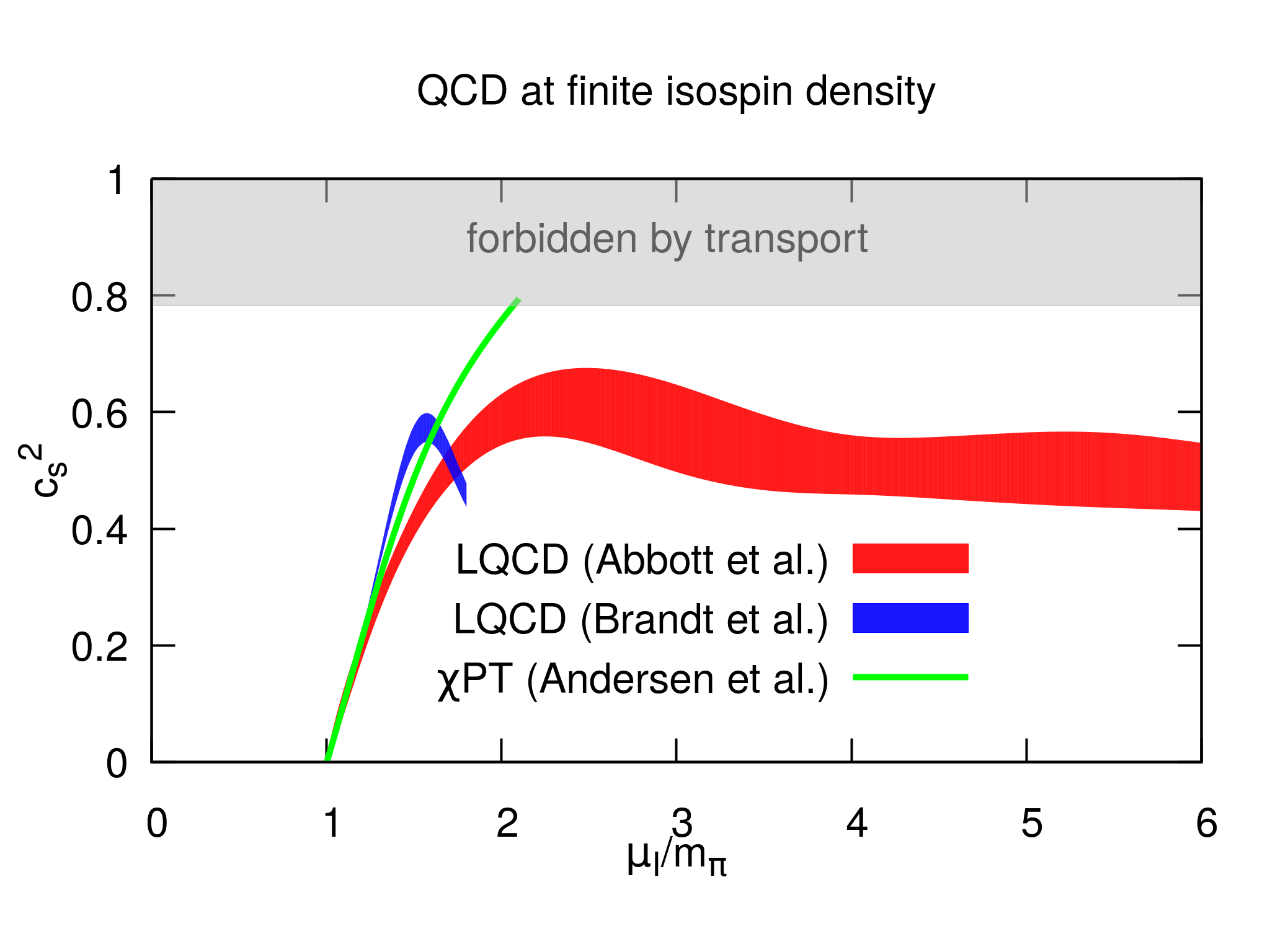}  
    \caption{\label{fig1}
    Speed of sound squared at $T=0$ in QCD as a function of isospin chemical potential $\mu_I$. Shown are results from lattice QCD (LQCD) for $N_f=2+1$ from Refs.~\cite{Detmold:2012wc,Abbott:2023coj} with $m_\pi\simeq 170$ MeV; LQCD for $N_f=2+1$ from Refs.~\cite{Brandt:2018bwq,Brandt:2022hwy} with $m_\pi\simeq 135$ MeV; and chiral perturbation theory ($\chi$PT) results for $N_f=2$ to NLO with $m_\pi\simeq 135$ MeV \cite{Andersen:2023ivj}.  The transport bound (\ref{bound}) is indicated by the grey region marked as ``forbidden by transport".}
  \end{figure}

In this work, we point out that ultradense matter with nearly luminal speed of sound must have very unusual transport properties. We discuss a scenario where fundamental properties of thermodynamics and transport in relativity can constrain the equation of state of ultradense matter. Assuming the validity of relativistic transient hydrodynamics  in which dissipative stresses obey additional relaxation equations \cite{Israel:1979wp}, we argue that causality and stability of matter imply that there is an upper bound on the speed of sound squared given by $c_s^2
    \leq 0.781$ in all known systems\footnote{We only consider systems where the equilibrium state is unique and the correlation length is finite. 
    } where calculations of certain ratios of
first-order hydrodynamic transport coefficients and their corresponding relaxation times can be performed from first principles. In this case, a violation of the bound implies that such systems either exhibit very unusual transport coefficients or  cannot be consistently described by transient fluid dynamics \footnote{This is not connected to the non-analytic behavior induced by the backreaction of sound waves \cite{Kovtun:2011np}.}, signaling the presence of exotic transport behavior that simple relaxation dynamics cannot describe.\footnote{While unusual transport coefficients can be expected from superfluid phases, the onset of superfluidity leads to multiple sound modes \cite{Schmitt:2014eka}, which makes causality and stability analyses considerably more convoluted \cite{Gavassino:2022zzz}. 
Therefore, we leave the exploration of superfluid states to future work. 
However, we note that our bound holds even in the case of a pion condensate, where 
a global $U(1)$ symmetry is spontaneously broken \cite{Son:2000xc,Kogut:2002zg,Carignano:2016lxe} (see Fig.~\ref{fig1}).} 
Given current astrophysical constraints, we show that imposing this bound on $c_s$ implies that the maximum mass of an isolated neutron star is bound by $M< 2.7 \, M_\odot$, assuming chiral effective theory is valid up to two times saturation density. Therefore, the new transport bound provides further support to the claim that the 2.6 $M_\odot$ compact object observed to merge with a 23 $M_\odot$ black hole causing the GW190814 event \cite{Abbott:2020khf} can be a neutron star.

\section{Speed of sound bound from transport}

Fluid dynamics describes the long-time, long-wavelength dynamics of conserved quantities in many-body systems \cite{LandauLifshitzFluids}. Ideal fluid dynamics describes the evolution of the energy density $\epsilon$, number density $n$, and flow velocity in the absence of dissipation, with the information about the system's microscopic properties encoded in the equation of state, e.g., $P=P(\epsilon,n)$, where $P$ is the pressure computed in thermodynamic equilibrium. In this limit, sound waves propagate with frequency $\omega = \pm c_s k$, where  $c_s = \sqrt{dP/d\epsilon}$ denotes the isentropic speed of sound in the fluid, and $k=|\mathbf{k}|$ is the wavenumber. 

Small dissipative corrections can be taken into account by expressing the dissipative fluxes in a systematic derivative expansion of the hydrodynamic fields \cite{LandauLifshitzFluids}, which introduces the well-known transport coefficients: the shear viscosity $\eta$, bulk viscosity $\zeta$, and charge conductivity $\sigma$. 

Constraining any transport coefficients in nuclear matter is challenging since it involves extracting near-equilibrium information at strong coupling from QCD. However, imposing causality can provide useful information about transport coefficients when the speed of sound nears the speed of light.

In causal and stable first-order theories \cite{Bemfica:2017wps,Kovtun:2019hdm,Bemfica:2019knx,Hoult:2020eho,Bemfica:2020zjp}, the long-wavelength behavior of the sound wave dispersion relation is given by\footnote{The $\mathcal{O}(k^3)$ terms are determined by the UV regulators parameterizing the choice of hydrodynamic frame.} $\omega = \pm c_s k - i \Gamma k^2/2 + \mathcal{O}(k^3)$, where the damping coefficient is 
\begin{equation}
    \Gamma = \frac{\zeta + \frac{4}{3}\eta}{\epsilon+P}+\frac{\sigma}{(\epsilon+P)c_s^2}\left(\frac{\partial P}{\partial n}\right)^2_\epsilon.
\end{equation}
Causality \cite{Heller:2022ejw} and the stability of the equilibrium state \cite{Gavassino:2023myj} impose that dispersion relations must obey the fundamental inequality $\mathrm{Im} \,\omega \leq |\mathrm{Im}\, k|$, $\omega,k \in \mathbb{C}$. Since the second law of thermodynamics imposes that $\eta,\zeta,\sigma$ are non-negative, $\Gamma\geq 0$, one can show\footnote{Take $k=i\alpha$ (with $0<\alpha \ll 1$) and impose the inequality $\mathrm{Im} \,\omega \leq |\mathrm{Im}\, k|$.} that $\Gamma \to 0$ when $c_s\to 1$ \cite{Gavassino:2023myj}. This statement was generalized in Ref.\ \cite{Heller:2023jtd}, where it was shown that the contribution from all the other terms also vanishes, making luminal sound waves exactly non-dispersive and described by the trivial dispersion relation, $\omega=k$. This implies that at densities where the speed of sound of ultradense matter approaches the speed of light, sound waves propagate without damping or dispersion.

In relativistic transient fluid dynamics~\cite{Israel:1979wp}, transport is further characterized by relaxation time coefficients that determine when the dissipative fluxes approach their first-order values. For instance, the shear relaxation time coefficient $\tau_\pi$ has been calculated for many theories (see, e.g., Tab. 2.1 in Ref.\ \cite{Romatschke:2017ejr}). In all of these systems, it was found to be tightly constrained as
\be
\tau_\pi = C_\pi \times \frac{\eta}{\epsilon+P}\,,\quad C_\pi \in  [2 (2-\ln 2),6.1]\,,
\ee
with the upper limit of $C_\pi$ originating from theories evaluated at \textit{weak coupling} \cite{York:2008rr}. We, therefore, find the following bound
\be
\label{a}
\frac{4 \eta}{3\tau_\pi (\epsilon+P)}=\frac{4}{3 C_\pi}\geq \frac{4}{3\times 6.1}\,,
\ee
from relativistic transport in theories where calculations can be done from first principles.

By contrast, while the bulk relaxation time $\tau_\Pi$ has also been calculated in many theories, the corresponding bound on the ratio of $\frac{\zeta}{\tau_\Pi (\epsilon+P)}$, where $\zeta$ is the bulk viscosity, is much less constrained,
\be
\label{b}
\frac{\zeta}{\tau_\Pi(\epsilon+P)}=\frac{1}{C_\Pi}>0\,,
\ee
because there are examples where $\zeta=0,\tau_\Pi\neq 0$ \cite{Jaiswal:2014isa}.  
Finally, not much is known from first principles about similar ratios involving the charge conductivity and the relaxation times, and thus, we shall not consider it further in this work.

Further general statements about transport can also be made in the context of transient fluid dynamics. By requiring that the entropy measured by all inertial observers is maximized in equilibrium in the presence of constraints (such as energy and/or particle number), one can derive a number of thermodynamic inequalities that ensure that the stability of the equilibrium state holds in all Lorentz reference frames \cite{Gavassino:2021cli}. These inequalities also imply causality in the linear regime around equilibrium \cite{Gavassino:2021kjm}. In transient fluid dynamics, one can obtain such thermodynamic inequalities by computing the so-called information current $E^\mu$ \cite{Gavassino:2021kjm}, which must be a time-like future-directed 4-vector with non-positive divergence, $\nabla_\mu E^\mu \leq 0$, to satisfy the covariant version of the second law of thermodynamics. In the case of Israel-Stewart theory with shear and bulk viscosity, this implies the following thermodynamic inequality \cite{Hiscock:1983zz,Olson:1990rzl}   
\be
c_s^2 \leq 1-\frac{4}{3 C_\pi}-\frac{1}{C_\Pi}\,,
\label{bound0}
\ee
which constitutes an upper bound on the speed of sound from relativistic thermodynamic principles\footnote{We note that \eqref{bound0} also follows from applying the inequality $\mathrm{Im} \,\omega \leq |\mathrm{Im}\, k|$, or by simply computing the maximum propagation speeds in linearized Israel-Stewart-like theories \cite[Eq.~(49)]{Romatschke:2009im}.}
Furthermore, we see that matter in a state where the equilibrium speed of sound $c_s\to 1$ must have unusual transport behavior as not only $\eta$ and $\zeta$ but also their corresponding ratios involving the relaxation times must vanish. 

It should be stressed that the bound (\ref{bound0}) relies on the assumption that transient fluid dynamics, where additional degrees of freedom (such as the shear and bulk stresses) are included as independent variables characterizing an extended quasi-equilibrium thermodynamic state \cite{landau5}, faithfully represent the thermodynamic properties of nuclear matter. This nontrivial assumption implies that the near-equilibrium properties of ultradense matter display not only typical fluid behavior but also visco-elastic properties, which are commonly found in liquids \cite{Frenkel:106808,trachenko2015collective,Baggioli:2019jcm}.

Using the additional information (\ref{a},\ref{b}) on $C_\pi,C_\Pi$, one can make the bound \eqref{bound0} sharp:
\be
\label{bound}
c_s^2\leq 0.781\,.
\ee
If the bound above is violated, this indicates either that the shear viscous coefficients evade Eq.~\eqref{a} somehow, or that transport cannot be studied using the framework of transient hydrodynamics without violating fundamental thermodynamic principles. This, per se, indicates the presence of nontrivial transport dynamics in ultradense matter that cannot be captured within well-known relativistic hydrodynamic formulations \cite{Rocha:2023ilf} where simple relaxation equations govern the evolution of the dissipative fluxes.

In the next section, we compare this bound to results in theories where $c_s^2$ can be calculated (or estimated) from first principles.

  \section{Speed of sound in nuclear matter}
\label{sec3}

\begin{figure}[t]    
   \includegraphics[width=\linewidth]{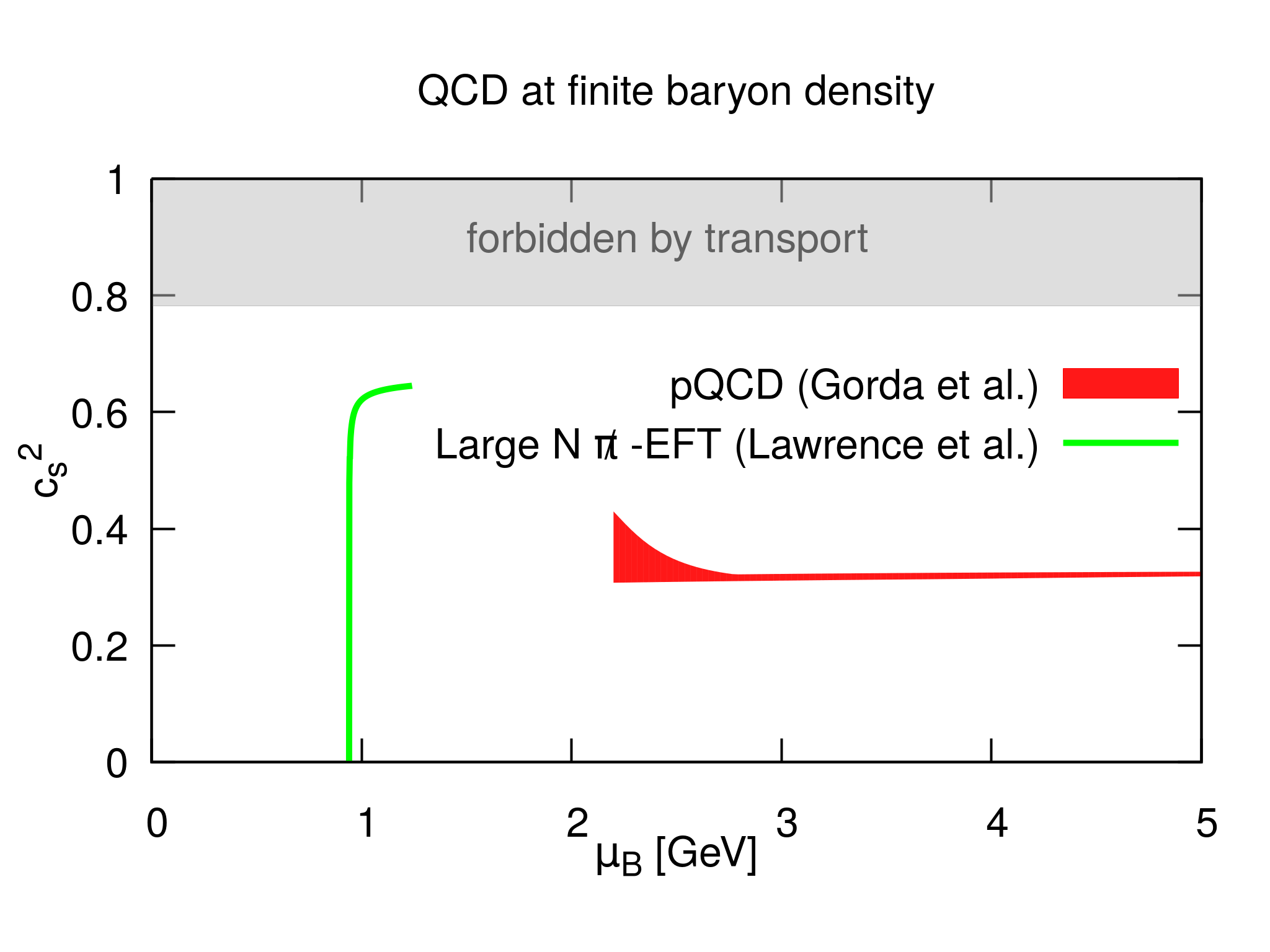}
    \caption{
    \label{fig2}
    Speed of sound squared at $T=0$ in QCD as a function of baryon chemical potential $\mu_B$. Shown are results from perturbative QCD (pQCD) for $N_f=3$ to partial N3LO Refs.~\cite{Gorda:2021kme,Gorda:2021znl,Gorda:2023usm} and large $N_f$ $\slashed{\pi}$ EFT results to LO from Ref.~\cite{Lawrence:2022vwa}.   The transport bound (\ref{bound}) is indicated by the grey region marked as ``forbidden by transport".}
  \end{figure}

\begin{figure}[t]
  \includegraphics[width=\linewidth]{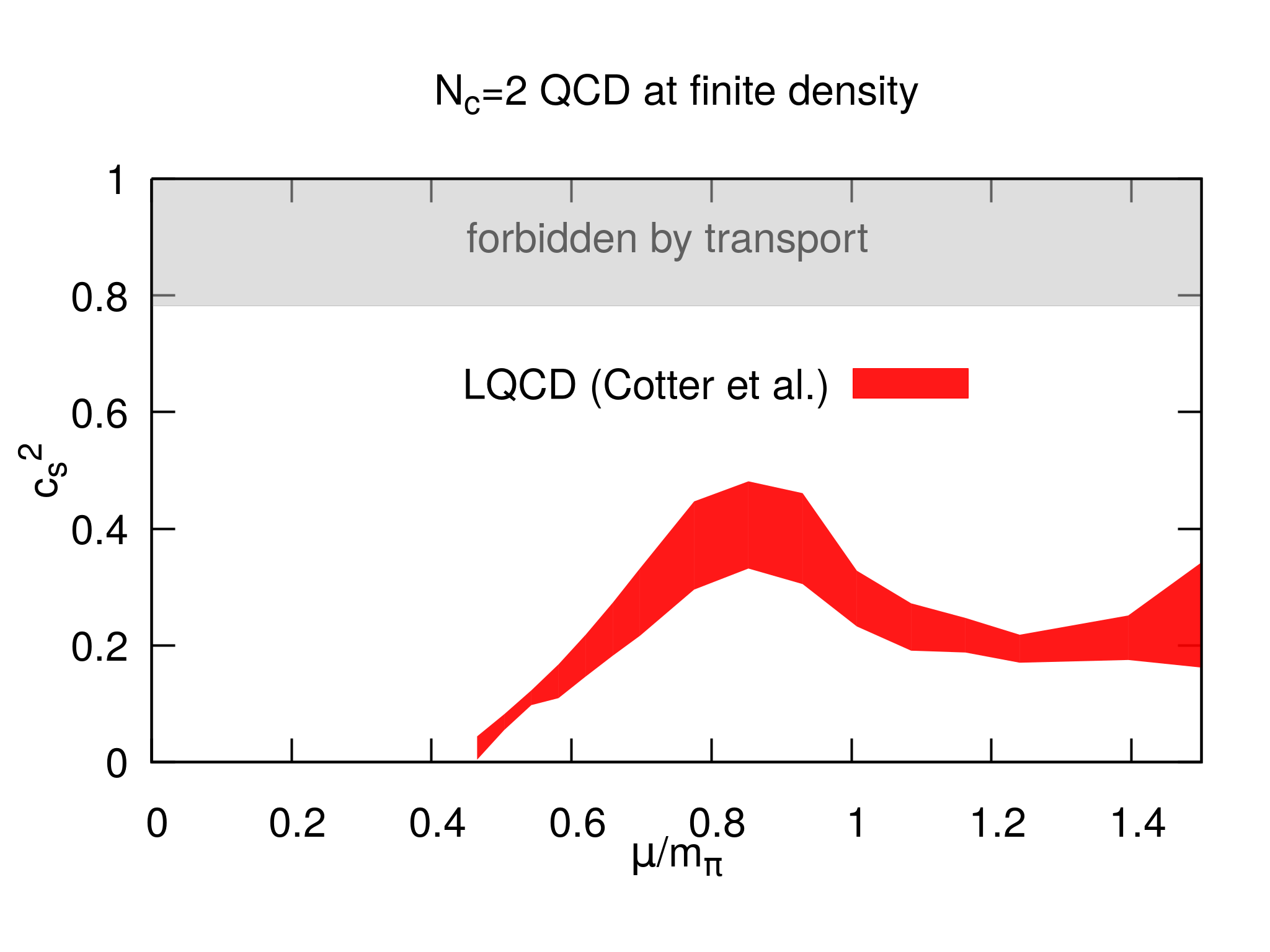}
    \caption{\label{fig3}
    Speed of sound squared at $T\simeq 50$ MeV in two-color QCD as a function of chemical potential $\mu$. Shown are results for $c_s^2$ extracted from data provided in Ref.~\cite{Cotter:2012mb} with $m_\pi\simeq 700$ MeV. The transport bound (\ref{bound}) is indicated by the grey region marked as ``forbidden by transport".}
  \end{figure}

Results for the speed of sound squared for cold nuclear matter are available from first-principles QCD calculations such as lattice QCD, effective field theory, and perturbative QCD. In particular, results for $c_s^2$ are shown in Figs.~\ref{fig1},\ref{fig2},\ref{fig3} for QCD at finite isospin chemical potential, QCD at finite baryon chemical potential, and two-color QCD at finite chemical potential, respectively. In addition to the results shown, results on thermodynamics at finite-density for QCD in the strong coupling limit from Ref.~\cite{Kim:2023dnq} and various model calculation results for $c_s^2$ are available \cite{Chiba:2023ftg,Ayala:2023mms}.

Comparing the results for $c_s^2$ from first principles calculations shown in Figs.\ \ref{fig1}, \ref{fig2}, and \ref{fig3} to the transport bound (\ref{bound}), one finds that the transport bound on the speed of sound is obeyed in all cases.

It is worth noticing that the bound in Eq.~\eqref{bound} is respected even for low temperature where superfluidity occurs, \cite{Son:2000xc,Kogut:2002zg,Carignano:2016lxe,Lawrence:2022vwa}, as can be seen in Figs.~\ref{fig1},\ref{fig2}.

\section{Implications for the neutron star mass -- radius relation}

   \begin{figure}[t]
    \includegraphics[width=\linewidth]{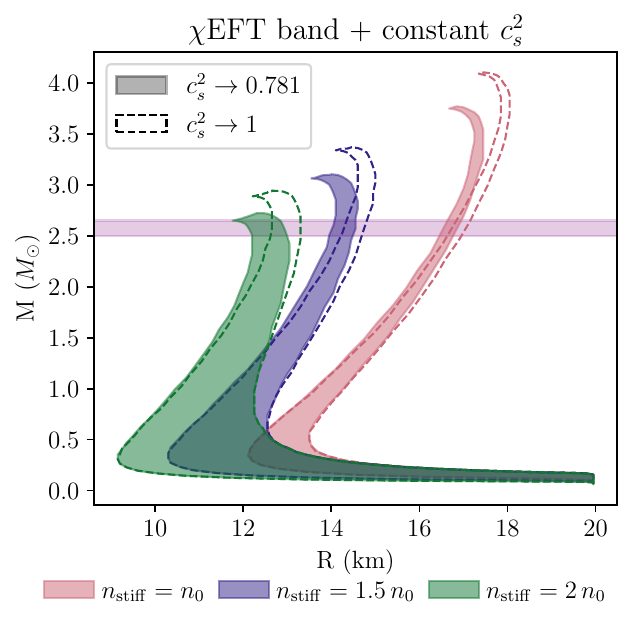}  
    \caption{\label{fig4} Mass--radius relations for EoSs that are in agreement with $\chi$EFT results at 
 baryon densities $n<n_{\textrm{stiff}}$ and saturate the new bound  $c_s^2 \leq 0.781$ at densities $n>n_{\textrm{stiff}}$. Different shaded bands represent results for switching densities $n_{\textrm{stiff}}$ of 1 (pink), 1.5 (purple), and 2 (green) times the baryon density at nuclear saturation, with the band thickness reflecting uncertainties in $\chi$EFT results. 
 Dashed lines represent results in the case where $c_s^2=1$ saturates the causal bound for $n>n_{\textrm{stiff}}$, with the same color scheme. 
 The horizontal magenta band indicates the uncertainty in the mass of the compact object of unknown nature observed in GW190814~\cite{Abbott:2020khf}. }
  \end{figure}
  
The speed of sound of ultradense matter plays a crucial role in the structure of neutron stars. A larger $c_s^2$ allows for larger values of the stellar mass --- as larger pressure gradients can counterbalance a larger gravitational force --- but also  gives rise to larger stars --- as matter becomes harder to compress. 
Observations of ``small'' neutron star radii of $R \sim 13$ km \cite{Miller:2019cac,Riley:2019yda,Miller:2021qha,Riley:2021pdl}, as well as tidal deformability constraints from GW170817 \cite{TheLIGOScientific:2017qsa,Abbott:2018exr,Abbott:2018wiz}, require speeds of sound well below the conformal value of $c_s^2 \leq 1/3$ at low densities, while very massive stars of $M \sim 2\,M_\odot$ \cite{Demorest:2010bx,Antoniadis:2013pzd,Cromartie:2019kug} require much larger values of  $c_s^2$ at high density, most likely above $c_s^2=1/3$ \cite{Bedaque:2014sqa,Alsing:2017bbc,Tews:2018kmu,Ma:2018qkg,Zhang:2019udy,Alford:2015dpa}.

We thus investigate the implications of the transport bound $c_s^2 \leq 0.781$ for the neutron star mass--radius relation. To find the most massive configurations allowed by this bound, we employ EoSs that saturate this bound above a threshold $n > n_{\textrm{stiff}}$ for the baryon density. 
At densities $n < n_{\textrm{stiff}}$ below this threshold, we generate $\sim 200$ low-density EoSs consistent with uncertainty bands from chiral effective theory ($\chi$EFT) results \cite{Tews:2018kmu}, combined with a GPPVA(TM1e) crust EoS \cite{Grill:2014aea,Shen:2020sec,Boukari:2020iut,Pearson:2018tkr,Oertel:2016bki} (we refer the reader to Appendix\ \ref{app:EoS} for details). 
We match each of these EoSs $\epsilon_{\textrm{soft}}^{(i)}(P)$, where superscript $i$ specifies which EoS, to a high-density EoS $\epsilon_{\textrm{stiff}}^{(i)}(P)$ with constant speed of sound $c_s^2 = {c_s^2}_{(\textrm{bound})} = 0.781$, while keeping $\epsilon$ a continuous function of $P$:
\be
\label{eq:constcs2eos}
\epsilon_{\textrm{stiff}}^{(i)}(P) = \epsilon_{\textrm{soft}}^{(i)}(P_{\textrm{stiff}}) +  \frac{P - P_{\textrm{stiff}}}{{c_s^2}_{(\textrm{bound})}},
\ee
where $P_{\textrm{stiff}}$ is the value of the pressure as $n \to n_{\textrm{stiff}}^-$ from the left. %

We then employ the Tolman-Oppenheimer-Volkoff (TOV) equation to obtain mass--radius relations 
for each of these piecewise equations of state. 
Similar procedures have been extensively employed in the literature to investigate the evidence in favor of supraconformal speeds of sound \cite{Bedaque:2014sqa,Kanakis-Pegios:2020jnf,Tews:2018kmu,Reed:2019ezm,Moustakidis:2016sab}. 
The region in the mass--radius diagram obtained via this process is shown in Fig.\ \ref{fig4} as shaded bands. 
We show results for different values of $n_{\textrm{stiff}}$, of $1$, $1.5$ and $2$ times the baryon density at nuclear saturation $n_0\approx 0.16$ fm$^{-3}$. 

To gauge the significance of the new $c_s^2$ bound, we compare the maximum allowed mass for ${c_s^2}_{(\textrm{bound})} = 0.781$ to the maximum mass when the less restrictive bound $c_s^2\leq1$ --- from relativistic causality \cite{HARTLE1978201} --- is saturated, which can be found by replacing ${c_s^2}_{(\textrm{bound})} = 1$ in  Eq.~\eqref{eq:constcs2eos}. 
Contours for the obtained regions in the mass--radius diagram, for each different value of $n_{\textrm{stiff}}$, are represented as dashed lines in Fig.~\ref{fig4}.

We find that  the transport bound $c_s^2\leq0.781$ decreases the maximum mass by $\sim 8 \%$ with respect to the maximum value allowed by the causal bound. 
This is the case for all the different soft EoSs, regardless of the density $n_{\textrm{stiff}}$ at which the EoS becomes stiff, with an uncertainty of $0.2\%$. 
The new bound also tends to decrease the radius of the maximally massive configuration by $3.6 \pm 0.3 \%$. 
Effects of the same bound become small for lighter stellar masses, below $\sim 2\, M_\odot$.

 \section{Discussion}

Relativistic causality and covariant stability impose strong constraints on transport quantities and the speed of sound in relativistic fluids. Combining these constraints with bounds on transport coefficients observed in a wide range of systems, we have found that, for all of these systems, assuming the validity of transient relativistic fluid dynamics implies that the speed of sound squared is bounded from above by $c_s^2 \leq 0.781$. 
 We investigated the applicability of this bound in first-principles calculations of the speed of sound in QCD and two-color QCD.  
 All of the known first-principles calculations obey the new bound. This bound can be sharpened by further constraining the ratios $C_\pi,C_\Pi$ from shear and bulk viscosity (and performing a similar analysis including effects from charge conductivity), especially in systems at finite baryon density. 

Not much is known about transport properties beyond first order at low temperature and finite density, the exception being calculations of so-called thermodynamic transport coefficients \cite{Kovtun:2018dvd,Grieninger:2021rxd,Lawrence:2022vwa,Weiner:2023wew}. 
Nonetheless, in Sec.~\ref{sec3}, we have checked that our new bound on the speed of sound is respected in first-principles results both for QCD at zero temperature and for two-color QCD at a temperature around $50$ MeV. 
The results discussed in this work can lead to a renewed interest in calculating transport coefficients, particularly the shear and bulk relaxation times, in order to assess to what extent the bounds in Eq.~\eqref{a} and \eqref{b} apply in the cold and dense regime. 
 
It would be interesting to extend our analysis to the case where superfluids may be present, which we leave for future work. While our analysis is not directly applicable in this case, we have seen that our bound on $c_s^2$ is not violated in first-principles results for QCD at high isospin density, where a superfluid pion condensate is expected \cite{Son:2000xc,Kogut:2002zg,Carignano:2016lxe}, and non-relativistic chiral QCD where the analogues of Cooper pairs form \cite{Lawrence:2022vwa}.\footnote{A superfluid state may also be present in two-color QCD \cite{Ratti:2004ra}, where the bound $c_s^2 \leq 0.781$ also seems to hold.}

We also investigated the consequences of this bound for the neutron star mass--radius relation. 
Our results show that, if $\chi$EFT is taken to be a good description of nuclear matter up to two times saturation density ($n_{\textrm{stiff}}=2\,n_0$), then a single neutron star mass above $2.7\, M_\odot$ (and below $\sim 2.9\, M_\odot$) would provide strong evidence for the breaking of the  $c_s^2\leq0.781$ bound, indicating that ultradense matter must exhibit exotic transport properties. 
On the other hand, 
our new bound is fully compatible with the possibility that the $\sim 2.6\, M_\odot$ compact object observed via gravitational wave event GW190814 could be a neutron star.

 Future Bayesian analyses of neutron star and gravitational wave observations could be employed to more systematically assess the implications of our new bound on $c_s^2$ for the cold and dense nuclear matter EoS and for the neutron star mass--radius relation. Such analyses could also estimate the probability that this new bound is violated inside neutron stars and, therefore, help establish how likely it is that the matter in the core of massive neutron stars displays unusual transport properties.

  \section{Acknowledgments}
We want to thank Ryan Abbot, Bastian Brandt, Tyler Gorda, Jon-Ivar Skullerud, Ingo Tews, and Qing Yu for providing tabulated data for the equations of state in Refs.~\cite{Abbott:2023coj,Brandt:2022hwy,Andersen:2023ivj,Gorda:2023usm,Cotter:2012mb}.
M.H. and J.N. thank J.~Noronha-Hostler, D.~Mroczek, and G.~S.~Denicol for helpful discussions. J.N especially thanks L.~Gavassino for a number of insightful and clarifying comments about thermodynamic stability and causality in relativistic fluids.  
M.H. and J.N. were supported in part by the National Science Foundation (NSF) within the framework of the MUSES Collaboration, under grant number OAC-2103680. J.N. was partially supported by the U.S.
Department of Energy, Office of Science, Office for Nuclear Physics under Award No. DE-SC0023861. 
  P.R. was supported by the Department of Energy, DOE award No DE-SC0017905.

\appendix
\renewcommand{\thefigure}{A\arabic{figure}}

\setcounter{figure}{0}
\section{Neutron-star equation of state from chiral effective theory}
\label{app:EoS}

\begin{figure}[h!]
    \centering
    \includegraphics[width=\linewidth]{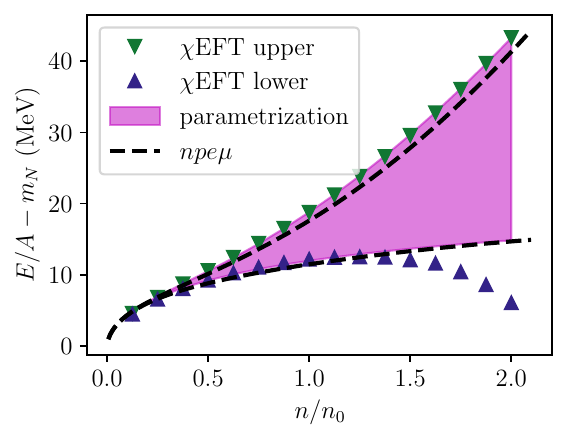}
    \caption{Energy per nucleon as a function of baryon density for the upper (green triangles down) and lower (blue triangles up) limits of the $\chi$EFT band for pure neutron matter. The band of fits to the functional form in Eq.~\ref{eq:parametrization} with $x=0$ is shown as the magenta band, while the same band after including leptons under weak equilibrium is delimited by the black dashed lines.}
    \label{figapp-1}
\end{figure}

For our low-density, soft EoS, we start with $\chi$EFT results for the energy per nucleon $E/A$ of pure neutron matter \cite{Tews:2018kmu}. 
We consider the entire band of EoSs from Ref.~\cite{Tews:2018kmu}, which spans the uncertainty bands for three different nuclear Hamiltonians.  
The upper and lower limits of this band are shown ad the green downward and blue upward triangles in Fig.\ \ref{figapp-1}, respectively.

As a straightforward way to cover this $E/A$ band, we linearly interpolate between the upper and lower limits of the $\chi$EFT band, $(E/A)_{\textrm{up}}$ and $(E/A)_{\textrm{low}}$:
\be
(E/A)_{\sigma} = (1-\sigma)\,(E/A)_{\textrm{up}} + \sigma\,(E/A)_{\textrm{low}},
\ee
where we consider 2001 values of $\sigma \in [0,1]$. 
To avoid instability against contraction, we discard, for each $\sigma$, any points where the energy per nucleon decreases with baryon density. 

To interpolate between points and to extrapolate them towards symmetric nuclear matter, 
we employ the parametrization from \cite{Hebeler:2010jx,Bedaque:2014sqa,Tews:2018kmu}:
\begin{multline}
\label{eq:parametrization}
    \frac{E}{A}(n,x) - m_N = T_0\,\bigg[ 
    \frac{3}{5}\left( x^{3/5} + (1-x)^{3/5} \right)\left(\frac{2 n}{n_0}\right)^{2/3}\\
    - [(2\,\alpha - 4\,\alpha_L)\,x\,(1-x) + \alpha_L]\frac{n}{n_0}\\
    + [(2\,\eta - 4\,\eta_L)\,x\,(1-x) + \alpha_L]\left(\frac{2 n}{n_0}\right)^\gamma
    \bigg]    ,
\end{multline}
where $m_N$ is the nucleon mass, $n$ is the baryon density and $x$ is the proton fraction. 
For each $\sigma$ value, $\sigma_i$, we fit $(E/A)_{\sigma_i}$ with this parametrization, taking $x=0$. This yields parameters $\alpha_L$, $\eta_L$ and $\gamma$.\footnote{Fitting residues are within $5\%$.} 
By further imposing that symmetric nuclear matter at saturation ($n=n_0$, $x=0.5$) has a binding energy of $-16.3$ MeV  and zero pressure, we obtain parameters $\alpha$ and $\eta$ in Eq.~\eqref{eq:parametrization}. 
The $E/A$ band obtained by fitting each $(E/A)_{\sigma_i}$ with Eq.~\eqref{eq:parametrization} for $x=0$ is shown as the magenta band in Fig.\ \ref{figapp-1}. Visible differences with respect to the lower $\chi$EFT limit are due to the discarding of points with negative derivatives. 

Using the parametrized form for each $E/A$ curve within the band, we can add electrons and muons and impose beta equilibrium to obtain a set of EoSs for $npe\mu$ matter.\footnote{Before doing so, we discard a small number of parametrizations with spurious phase transitions.} The limits of the obtained band are shown in Fig.~\ref{figapp-1} as black dashed lines.

\begin{figure}
    \centering
    \includegraphics[width=\linewidth]{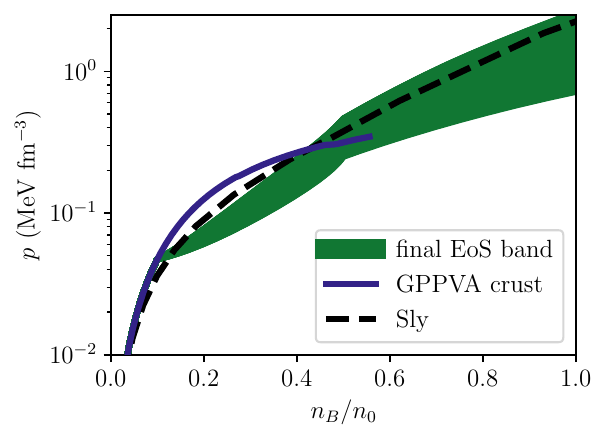}
    \caption{Final band of low-density EoSs used in this work (green band), after merging with a GPPVA crust \cite{Grill:2014aea,Shen:2020sec,Boukari:2020iut,Pearson:2018tkr} (blue solid line) at densities $n\leq 0.1 \, n_0$. For reference, the Sly EoS \cite{Douchin:2001sv} is shown as the black dashed line.}
    \label{figapp-2}
\end{figure}

Finally, we interpolate between these $npe\mu$ EoSs, compatible with the $\chi$EFT band of Ref.~\cite{Tews:2018kmu}, above $0.5\,n_0$, and a crust EoS below $0.1\,n_0$. 
For that purpose, we use the GPPVA(TM1e) crust EoS \cite{Grill:2014aea,Shen:2020sec,Boukari:2020iut,Pearson:2018tkr} from the \texttt{CompOSE} website \cite{Oertel:2016bki,compose}. 
For densities of $0.1\,n_0$ to $0.5\,n_0$, we interpolate between the two EoSs with a cubic for $P(\mu)$,  with coefficients chosen so as to guarantee the continuity of both $P(\mu)$ and $n(\mu)$, where $\mu$ is the baryon chemical potential. The resulting final EoS band is shown in Fig.~\ref{figapp-2} as the green band.

\FloatBarrier

\bibliography{cs2,speedofsound}

\begin{thebibliography}{145}%
\makeatletter
\providecommand \@ifxundefined [1]{%
 \@ifx{#1\undefined}
}%
\providecommand \@ifnum [1]{%
 \ifnum #1\expandafter \@firstoftwo
 \else \expandafter \@secondoftwo
 \fi
}%
\providecommand \@ifx [1]{%
 \ifx #1\expandafter \@firstoftwo
 \else \expandafter \@secondoftwo
 \fi
}%
\providecommand \natexlab [1]{#1}%
\providecommand \enquote  [1]{``#1''}%
\providecommand \bibnamefont  [1]{#1}%
\providecommand \bibfnamefont [1]{#1}%
\providecommand \citenamefont [1]{#1}%
\providecommand \href@noop [0]{\@secondoftwo}%
\providecommand \href [0]{\begingroup \@sanitize@url \@href}%
\providecommand \@href[1]{\@@startlink{#1}\@@href}%
\providecommand \@@href[1]{\endgroup#1\@@endlink}%
\providecommand \@sanitize@url [0]{\catcode `\\12\catcode `\$12\catcode `\&12\catcode `\#12\catcode `\^12\catcode `\_12\catcode `\%12\relax}%
\providecommand \@@startlink[1]{}%
\providecommand \@@endlink[0]{}%
\providecommand \url  [0]{\begingroup\@sanitize@url \@url }%
\providecommand \@url [1]{\endgroup\@href {#1}{\urlprefix }}%
\providecommand \urlprefix  [0]{URL }%
\providecommand \Eprint [0]{\href }%
\providecommand \doibase [0]{http://dx.doi.org/}%
\providecommand \selectlanguage [0]{\@gobble}%
\providecommand \bibinfo  [0]{\@secondoftwo}%
\providecommand \bibfield  [0]{\@secondoftwo}%
\providecommand \translation [1]{[#1]}%
\providecommand \BibitemOpen [0]{}%
\providecommand \bibitemStop [0]{}%
\providecommand \bibitemNoStop [0]{.\EOS\space}%
\providecommand \EOS [0]{\spacefactor3000\relax}%
\providecommand \BibitemShut  [1]{\csname bibitem#1\endcsname}%
\let\auto@bib@innerbib\@empty
\bibitem [{\citenamefont {Demorest}\ \emph {et~al.}(2010)\citenamefont {Demorest}, \citenamefont {Pennucci}, \citenamefont {Ransom}, \citenamefont {Roberts},\ and\ \citenamefont {Hessels}}]{Demorest:2010bx}%
  \BibitemOpen
  \bibfield  {author} {\bibinfo {author} {\bibfnamefont {Paul}\ \bibnamefont {Demorest}}, \bibinfo {author} {\bibfnamefont {Tim}\ \bibnamefont {Pennucci}}, \bibinfo {author} {\bibfnamefont {Scott}\ \bibnamefont {Ransom}}, \bibinfo {author} {\bibfnamefont {Mallory}\ \bibnamefont {Roberts}}, \ and\ \bibinfo {author} {\bibfnamefont {Jason}\ \bibnamefont {Hessels}},\ }\bibfield  {title} {\enquote {\bibinfo {title} {{Shapiro Delay Measurement of A Two Solar Mass Neutron Star}},}\ }\href {\doibase 10.1038/nature09466} {\bibfield  {journal} {\bibinfo  {journal} {Nature}\ }\textbf {\bibinfo {volume} {467}},\ \bibinfo {pages} {1081--1083} (\bibinfo {year} {2010})},\ \Eprint {http://arxiv.org/abs/1010.5788} {arXiv:1010.5788 [astro-ph.HE]} \BibitemShut {NoStop}%
\bibitem [{\citenamefont {Antoniadis}\ \emph {et~al.}(2013)\citenamefont {Antoniadis} \emph {et~al.}}]{Antoniadis:2013pzd}%
  \BibitemOpen
  \bibfield  {author} {\bibinfo {author} {\bibfnamefont {John}\ \bibnamefont {Antoniadis}} \emph {et~al.},\ }\bibfield  {title} {\enquote {\bibinfo {title} {{A Massive Pulsar in a Compact Relativistic Binary}},}\ }\href {\doibase 10.1126/science.1233232} {\bibfield  {journal} {\bibinfo  {journal} {Science}\ }\textbf {\bibinfo {volume} {340}},\ \bibinfo {pages} {6131} (\bibinfo {year} {2013})},\ \Eprint {http://arxiv.org/abs/1304.6875} {arXiv:1304.6875 [astro-ph.HE]} \BibitemShut {NoStop}%
\bibitem [{\citenamefont {Cromartie}\ \emph {et~al.}(2019)\citenamefont {Cromartie} \emph {et~al.}}]{Cromartie:2019kug}%
  \BibitemOpen
  \bibfield  {author} {\bibinfo {author} {\bibfnamefont {H.~T.}\ \bibnamefont {Cromartie}} \emph {et~al.} (\bibinfo {collaboration} {NANOGrav}),\ }\bibfield  {title} {\enquote {\bibinfo {title} {{Relativistic Shapiro delay measurements of an extremely massive millisecond pulsar}},}\ }\href {\doibase 10.1038/s41550-019-0880-2} {\bibfield  {journal} {\bibinfo  {journal} {Nature Astron.}\ }\textbf {\bibinfo {volume} {4}},\ \bibinfo {pages} {72--76} (\bibinfo {year} {2019})},\ \Eprint {http://arxiv.org/abs/1904.06759} {arXiv:1904.06759 [astro-ph.HE]} \BibitemShut {NoStop}%
\bibitem [{\citenamefont {Bedaque}\ and\ \citenamefont {Steiner}(2015)}]{Bedaque:2014sqa}%
  \BibitemOpen
  \bibfield  {author} {\bibinfo {author} {\bibfnamefont {Paulo}\ \bibnamefont {Bedaque}}\ and\ \bibinfo {author} {\bibfnamefont {Andrew~W.}\ \bibnamefont {Steiner}},\ }\bibfield  {title} {\enquote {\bibinfo {title} {{Sound velocity bound and neutron stars}},}\ }\href {\doibase 10.1103/PhysRevLett.114.031103} {\bibfield  {journal} {\bibinfo  {journal} {Phys. Rev. Lett.}\ }\textbf {\bibinfo {volume} {114}},\ \bibinfo {pages} {031103} (\bibinfo {year} {2015})},\ \Eprint {http://arxiv.org/abs/1408.5116} {arXiv:1408.5116 [nucl-th]} \BibitemShut {NoStop}%
\bibitem [{\citenamefont {Abbott}\ \emph {et~al.}(2017)\citenamefont {Abbott} \emph {et~al.}}]{TheLIGOScientific:2017qsa}%
  \BibitemOpen
  \bibfield  {author} {\bibinfo {author} {\bibfnamefont {B.~P.}\ \bibnamefont {Abbott}} \emph {et~al.} (\bibinfo {collaboration} {LIGO Scientific, Virgo}),\ }\bibfield  {title} {\enquote {\bibinfo {title} {{GW170817: Observation of Gravitational Waves from a Binary Neutron Star Inspiral}},}\ }\href {\doibase 10.1103/PhysRevLett.119.161101} {\bibfield  {journal} {\bibinfo  {journal} {Phys. Rev. Lett.}\ }\textbf {\bibinfo {volume} {119}},\ \bibinfo {pages} {161101} (\bibinfo {year} {2017})},\ \Eprint {http://arxiv.org/abs/1710.05832} {arXiv:1710.05832 [gr-qc]} \BibitemShut {NoStop}%
\bibitem [{\citenamefont {Abbott}\ \emph {et~al.}(2018)\citenamefont {Abbott} \emph {et~al.}}]{Abbott:2018exr}%
  \BibitemOpen
  \bibfield  {author} {\bibinfo {author} {\bibfnamefont {B.~P.}\ \bibnamefont {Abbott}} \emph {et~al.} (\bibinfo {collaboration} {LIGO Scientific, Virgo}),\ }\bibfield  {title} {\enquote {\bibinfo {title} {{GW170817: Measurements of neutron star radii and equation of state}},}\ }\href {\doibase 10.1103/PhysRevLett.121.161101} {\bibfield  {journal} {\bibinfo  {journal} {Phys. Rev. Lett.}\ }\textbf {\bibinfo {volume} {121}},\ \bibinfo {pages} {161101} (\bibinfo {year} {2018})},\ \Eprint {http://arxiv.org/abs/1805.11581} {arXiv:1805.11581 [gr-qc]} \BibitemShut {NoStop}%
\bibitem [{\citenamefont {Abbott}\ \emph {et~al.}(2019)\citenamefont {Abbott} \emph {et~al.}}]{Abbott:2018wiz}%
  \BibitemOpen
  \bibfield  {author} {\bibinfo {author} {\bibfnamefont {B.~P.}\ \bibnamefont {Abbott}} \emph {et~al.} (\bibinfo {collaboration} {LIGO Scientific, Virgo}),\ }\bibfield  {title} {\enquote {\bibinfo {title} {{Properties of the binary neutron star merger GW170817}},}\ }\href {\doibase 10.1103/PhysRevX.9.011001} {\bibfield  {journal} {\bibinfo  {journal} {Phys. Rev. X}\ }\textbf {\bibinfo {volume} {9}},\ \bibinfo {pages} {011001} (\bibinfo {year} {2019})},\ \Eprint {http://arxiv.org/abs/1805.11579} {arXiv:1805.11579 [gr-qc]} \BibitemShut {NoStop}%
\bibitem [{\citenamefont {Tews}\ \emph {et~al.}(2018)\citenamefont {Tews}, \citenamefont {Carlson}, \citenamefont {Gandolfi},\ and\ \citenamefont {Reddy}}]{Tews:2018kmu}%
  \BibitemOpen
  \bibfield  {author} {\bibinfo {author} {\bibfnamefont {Ingo}\ \bibnamefont {Tews}}, \bibinfo {author} {\bibfnamefont {Joseph}\ \bibnamefont {Carlson}}, \bibinfo {author} {\bibfnamefont {Stefano}\ \bibnamefont {Gandolfi}}, \ and\ \bibinfo {author} {\bibfnamefont {Sanjay}\ \bibnamefont {Reddy}},\ }\bibfield  {title} {\enquote {\bibinfo {title} {{Constraining the speed of sound inside neutron stars with chiral effective field theory interactions and observations}},}\ }\href {\doibase 10.3847/1538-4357/aac267} {\bibfield  {journal} {\bibinfo  {journal} {Astrophys. J.}\ }\textbf {\bibinfo {volume} {860}},\ \bibinfo {pages} {149} (\bibinfo {year} {2018})},\ \Eprint {http://arxiv.org/abs/1801.01923} {arXiv:1801.01923 [nucl-th]} \BibitemShut {NoStop}%
\bibitem [{\citenamefont {Tews}\ \emph {et~al.}(2019)\citenamefont {Tews}, \citenamefont {Margueron},\ and\ \citenamefont {Reddy}}]{Tews:2019cap}%
  \BibitemOpen
  \bibfield  {author} {\bibinfo {author} {\bibfnamefont {I.}~\bibnamefont {Tews}}, \bibinfo {author} {\bibfnamefont {J.}~\bibnamefont {Margueron}}, \ and\ \bibinfo {author} {\bibfnamefont {S.}~\bibnamefont {Reddy}},\ }\bibfield  {title} {\enquote {\bibinfo {title} {{Confronting gravitational-wave observations with modern nuclear physics constraints}},}\ }\href {\doibase 10.1140/epja/i2019-12774-6} {\bibfield  {journal} {\bibinfo  {journal} {Eur. Phys. J. A}\ }\textbf {\bibinfo {volume} {55}},\ \bibinfo {pages} {97} (\bibinfo {year} {2019})},\ \Eprint {http://arxiv.org/abs/1901.09874} {arXiv:1901.09874 [nucl-th]} \BibitemShut {NoStop}%
\bibitem [{\citenamefont {Reed}\ and\ \citenamefont {Horowitz}(2020)}]{Reed:2019ezm}%
  \BibitemOpen
  \bibfield  {author} {\bibinfo {author} {\bibfnamefont {Brendan}\ \bibnamefont {Reed}}\ and\ \bibinfo {author} {\bibfnamefont {C.~J.}\ \bibnamefont {Horowitz}},\ }\bibfield  {title} {\enquote {\bibinfo {title} {{Large sound speed in dense matter and the deformability of neutron stars}},}\ }\href {\doibase 10.1103/PhysRevC.101.045803} {\bibfield  {journal} {\bibinfo  {journal} {Phys. Rev. C}\ }\textbf {\bibinfo {volume} {101}},\ \bibinfo {pages} {045803} (\bibinfo {year} {2020})},\ \Eprint {http://arxiv.org/abs/1910.05463} {arXiv:1910.05463 [astro-ph.HE]} \BibitemShut {NoStop}%
\bibitem [{\citenamefont {Capano}\ \emph {et~al.}(2020)\citenamefont {Capano}, \citenamefont {Tews}, \citenamefont {Brown}, \citenamefont {Margalit}, \citenamefont {De}, \citenamefont {Kumar}, \citenamefont {Brown}, \citenamefont {Krishnan},\ and\ \citenamefont {Reddy}}]{Capano:2019eae}%
  \BibitemOpen
  \bibfield  {author} {\bibinfo {author} {\bibfnamefont {Collin~D.}\ \bibnamefont {Capano}}, \bibinfo {author} {\bibfnamefont {Ingo}\ \bibnamefont {Tews}}, \bibinfo {author} {\bibfnamefont {Stephanie~M.}\ \bibnamefont {Brown}}, \bibinfo {author} {\bibfnamefont {Ben}\ \bibnamefont {Margalit}}, \bibinfo {author} {\bibfnamefont {Soumi}\ \bibnamefont {De}}, \bibinfo {author} {\bibfnamefont {Sumit}\ \bibnamefont {Kumar}}, \bibinfo {author} {\bibfnamefont {Duncan~A.}\ \bibnamefont {Brown}}, \bibinfo {author} {\bibfnamefont {Badri}\ \bibnamefont {Krishnan}}, \ and\ \bibinfo {author} {\bibfnamefont {Sanjay}\ \bibnamefont {Reddy}},\ }\bibfield  {title} {\enquote {\bibinfo {title} {{Stringent constraints on neutron-star radii from multimessenger observations and nuclear theory}},}\ }\href {\doibase 10.1038/s41550-020-1014-6} {\bibfield  {journal} {\bibinfo  {journal} {Nature Astron.}\ }\textbf {\bibinfo {volume} {4}},\ \bibinfo {pages} {625--632} (\bibinfo {year} {2020})},\ \Eprint {http://arxiv.org/abs/1908.10352}
  {arXiv:1908.10352 [astro-ph.HE]} \BibitemShut {NoStop}%
\bibitem [{\citenamefont {Annala}\ \emph {et~al.}(2020)\citenamefont {Annala}, \citenamefont {Gorda}, \citenamefont {Kurkela}, \citenamefont {N\"attil\"a},\ and\ \citenamefont {Vuorinen}}]{Annala:2019puf}%
  \BibitemOpen
  \bibfield  {author} {\bibinfo {author} {\bibfnamefont {Eemeli}\ \bibnamefont {Annala}}, \bibinfo {author} {\bibfnamefont {Tyler}\ \bibnamefont {Gorda}}, \bibinfo {author} {\bibfnamefont {Aleksi}\ \bibnamefont {Kurkela}}, \bibinfo {author} {\bibfnamefont {Joonas}\ \bibnamefont {N\"attil\"a}}, \ and\ \bibinfo {author} {\bibfnamefont {Aleksi}\ \bibnamefont {Vuorinen}},\ }\bibfield  {title} {\enquote {\bibinfo {title} {{Evidence for quark-matter cores in massive neutron stars}},}\ }\href {\doibase 10.1038/s41567-020-0914-9} {\bibfield  {journal} {\bibinfo  {journal} {Nature Phys.}\ }\textbf {\bibinfo {volume} {16}},\ \bibinfo {pages} {907--910} (\bibinfo {year} {2020})},\ \Eprint {http://arxiv.org/abs/1903.09121} {arXiv:1903.09121 [astro-ph.HE]} \BibitemShut {NoStop}%
\bibitem [{\citenamefont {Kanakis-Pegios}\ \emph {et~al.}(2020)\citenamefont {Kanakis-Pegios}, \citenamefont {Koliogiannis},\ and\ \citenamefont {Moustakidis}}]{Kanakis-Pegios:2020jnf}%
  \BibitemOpen
  \bibfield  {author} {\bibinfo {author} {\bibfnamefont {A.}~\bibnamefont {Kanakis-Pegios}}, \bibinfo {author} {\bibfnamefont {P.~S.}\ \bibnamefont {Koliogiannis}}, \ and\ \bibinfo {author} {\bibfnamefont {Ch.~C.}\ \bibnamefont {Moustakidis}},\ }\bibfield  {title} {\enquote {\bibinfo {title} {{Speed of sound constraints from tidal deformability of neutron stars}},}\ }\href {\doibase 10.1103/PhysRevC.102.055801} {\bibfield  {journal} {\bibinfo  {journal} {Phys. Rev. C}\ }\textbf {\bibinfo {volume} {102}},\ \bibinfo {pages} {055801} (\bibinfo {year} {2020})},\ \Eprint {http://arxiv.org/abs/2007.13399} {arXiv:2007.13399 [nucl-th]} \BibitemShut {NoStop}%
\bibitem [{\citenamefont {Huth}\ \emph {et~al.}(2021)\citenamefont {Huth}, \citenamefont {Wellenhofer},\ and\ \citenamefont {Schwenk}}]{Huth:2020ozf}%
  \BibitemOpen
  \bibfield  {author} {\bibinfo {author} {\bibfnamefont {S.}~\bibnamefont {Huth}}, \bibinfo {author} {\bibfnamefont {C.}~\bibnamefont {Wellenhofer}}, \ and\ \bibinfo {author} {\bibfnamefont {A.}~\bibnamefont {Schwenk}},\ }\bibfield  {title} {\enquote {\bibinfo {title} {{New equations of state constrained by nuclear physics, observations, and QCD calculations of high-density nuclear matter}},}\ }\href {\doibase 10.1103/PhysRevC.103.025803} {\bibfield  {journal} {\bibinfo  {journal} {Phys. Rev. C}\ }\textbf {\bibinfo {volume} {103}},\ \bibinfo {pages} {025803} (\bibinfo {year} {2021})},\ \Eprint {http://arxiv.org/abs/2009.08885} {arXiv:2009.08885 [nucl-th]} \BibitemShut {NoStop}%
\bibitem [{\citenamefont {Han}\ \emph {et~al.}(2021)\citenamefont {Han}, \citenamefont {Jiang}, \citenamefont {Tang},\ and\ \citenamefont {Fan}}]{Han:2021kjx}%
  \BibitemOpen
  \bibfield  {author} {\bibinfo {author} {\bibfnamefont {Ming-Zhe}\ \bibnamefont {Han}}, \bibinfo {author} {\bibfnamefont {Jin-Liang}\ \bibnamefont {Jiang}}, \bibinfo {author} {\bibfnamefont {Shao-Peng}\ \bibnamefont {Tang}}, \ and\ \bibinfo {author} {\bibfnamefont {Yi-Zhong}\ \bibnamefont {Fan}},\ }\bibfield  {title} {\enquote {\bibinfo {title} {{Bayesian Nonparametric Inference of the Neutron Star Equation of State via a Neural Network}},}\ }\href {\doibase 10.3847/1538-4357/ac11f8} {\bibfield  {journal} {\bibinfo  {journal} {Astrophys. J.}\ }\textbf {\bibinfo {volume} {919}},\ \bibinfo {pages} {11} (\bibinfo {year} {2021})},\ \Eprint {http://arxiv.org/abs/2103.05408} {arXiv:2103.05408 [hep-ph]} \BibitemShut {NoStop}%
\bibitem [{\citenamefont {Tan}\ \emph {et~al.}(2022)\citenamefont {Tan}, \citenamefont {Dore}, \citenamefont {Dexheimer}, \citenamefont {Noronha-Hostler},\ and\ \citenamefont {Yunes}}]{Tan:2021ahl}%
  \BibitemOpen
  \bibfield  {author} {\bibinfo {author} {\bibfnamefont {Hung}\ \bibnamefont {Tan}}, \bibinfo {author} {\bibfnamefont {Travis}\ \bibnamefont {Dore}}, \bibinfo {author} {\bibfnamefont {Veronica}\ \bibnamefont {Dexheimer}}, \bibinfo {author} {\bibfnamefont {Jacquelyn}\ \bibnamefont {Noronha-Hostler}}, \ and\ \bibinfo {author} {\bibfnamefont {Nicol\'as}\ \bibnamefont {Yunes}},\ }\bibfield  {title} {\enquote {\bibinfo {title} {{Extreme matter meets extreme gravity: Ultraheavy neutron stars with phase transitions}},}\ }\href {\doibase 10.1103/PhysRevD.105.023018} {\bibfield  {journal} {\bibinfo  {journal} {Phys. Rev. D}\ }\textbf {\bibinfo {volume} {105}},\ \bibinfo {pages} {023018} (\bibinfo {year} {2022})},\ \Eprint {http://arxiv.org/abs/2106.03890} {arXiv:2106.03890 [astro-ph.HE]} \BibitemShut {NoStop}%
\bibitem [{\citenamefont {Abbott}\ \emph {et~al.}(2020)\citenamefont {Abbott} \emph {et~al.}}]{Abbott:2020khf}%
  \BibitemOpen
  \bibfield  {author} {\bibinfo {author} {\bibfnamefont {R.}~\bibnamefont {Abbott}} \emph {et~al.} (\bibinfo {collaboration} {LIGO Scientific, Virgo}),\ }\bibfield  {title} {\enquote {\bibinfo {title} {{GW190814: Gravitational Waves from the Coalescence of a 23 Solar Mass Black Hole with a 2.6 Solar Mass Compact Object}},}\ }\href {\doibase 10.3847/2041-8213/ab960f} {\bibfield  {journal} {\bibinfo  {journal} {Astrophys. J. Lett.}\ }\textbf {\bibinfo {volume} {896}},\ \bibinfo {pages} {L44} (\bibinfo {year} {2020})},\ \Eprint {http://arxiv.org/abs/2006.12611} {arXiv:2006.12611 [astro-ph.HE]} \BibitemShut {NoStop}%
\bibitem [{\citenamefont {Tan}\ \emph {et~al.}(2020)\citenamefont {Tan}, \citenamefont {Noronha-Hostler},\ and\ \citenamefont {Yunes}}]{Tan:2020ics}%
  \BibitemOpen
  \bibfield  {author} {\bibinfo {author} {\bibfnamefont {Hung}\ \bibnamefont {Tan}}, \bibinfo {author} {\bibfnamefont {Jacquelyn}\ \bibnamefont {Noronha-Hostler}}, \ and\ \bibinfo {author} {\bibfnamefont {Nico}\ \bibnamefont {Yunes}},\ }\bibfield  {title} {\enquote {\bibinfo {title} {{Neutron Star Equation of State in light of GW190814}},}\ }\href {\doibase 10.1103/PhysRevLett.125.261104} {\bibfield  {journal} {\bibinfo  {journal} {Phys. Rev. Lett.}\ }\textbf {\bibinfo {volume} {125}},\ \bibinfo {pages} {261104} (\bibinfo {year} {2020})},\ \Eprint {http://arxiv.org/abs/2006.16296} {arXiv:2006.16296 [astro-ph.HE]} \BibitemShut {NoStop}%
\bibitem [{\citenamefont {Tsokaros}\ \emph {et~al.}(2020)\citenamefont {Tsokaros}, \citenamefont {Ruiz},\ and\ \citenamefont {Shapiro}}]{Tsokaros:2020hli}%
  \BibitemOpen
  \bibfield  {author} {\bibinfo {author} {\bibfnamefont {Antonios}\ \bibnamefont {Tsokaros}}, \bibinfo {author} {\bibfnamefont {Milton}\ \bibnamefont {Ruiz}}, \ and\ \bibinfo {author} {\bibfnamefont {Stuart~L.}\ \bibnamefont {Shapiro}},\ }\bibfield  {title} {\enquote {\bibinfo {title} {{GW190814: Spin and equation of state of a neutron star companion}},}\ }\href {\doibase 10.3847/1538-4357/abc421} {\bibfield  {journal} {\bibinfo  {journal} {Astrophys. J.}\ }\textbf {\bibinfo {volume} {905}},\ \bibinfo {pages} {48} (\bibinfo {year} {2020})},\ \Eprint {http://arxiv.org/abs/2007.05526} {arXiv:2007.05526 [astro-ph.HE]} \BibitemShut {NoStop}%
\bibitem [{\citenamefont {Tews}\ \emph {et~al.}(2021)\citenamefont {Tews}, \citenamefont {Pang}, \citenamefont {Dietrich}, \citenamefont {Coughlin}, \citenamefont {Antier}, \citenamefont {Bulla}, \citenamefont {Heinzel},\ and\ \citenamefont {Issa}}]{Tews:2020ylw}%
  \BibitemOpen
  \bibfield  {author} {\bibinfo {author} {\bibfnamefont {Ingo}\ \bibnamefont {Tews}}, \bibinfo {author} {\bibfnamefont {Peter T.~H.}\ \bibnamefont {Pang}}, \bibinfo {author} {\bibfnamefont {Tim}\ \bibnamefont {Dietrich}}, \bibinfo {author} {\bibfnamefont {Michael~W.}\ \bibnamefont {Coughlin}}, \bibinfo {author} {\bibfnamefont {Sarah}\ \bibnamefont {Antier}}, \bibinfo {author} {\bibfnamefont {Mattia}\ \bibnamefont {Bulla}}, \bibinfo {author} {\bibfnamefont {Jack}\ \bibnamefont {Heinzel}}, \ and\ \bibinfo {author} {\bibfnamefont {Lina}\ \bibnamefont {Issa}},\ }\bibfield  {title} {\enquote {\bibinfo {title} {{On the Nature of GW190814 and Its Impact on the Understanding of Supranuclear Matter}},}\ }\href {\doibase 10.3847/2041-8213/abdaae} {\bibfield  {journal} {\bibinfo  {journal} {Astrophys. J. Lett.}\ }\textbf {\bibinfo {volume} {908}},\ \bibinfo {pages} {L1} (\bibinfo {year} {2021})},\ \Eprint {http://arxiv.org/abs/2007.06057} {arXiv:2007.06057 [astro-ph.HE]} \BibitemShut {NoStop}%
\bibitem [{\citenamefont {Godzieba}\ \emph {et~al.}(2021)\citenamefont {Godzieba}, \citenamefont {Radice},\ and\ \citenamefont {Bernuzzi}}]{Godzieba:2020tjn}%
  \BibitemOpen
  \bibfield  {author} {\bibinfo {author} {\bibfnamefont {Daniel~A.}\ \bibnamefont {Godzieba}}, \bibinfo {author} {\bibfnamefont {David}\ \bibnamefont {Radice}}, \ and\ \bibinfo {author} {\bibfnamefont {Sebastiano}\ \bibnamefont {Bernuzzi}},\ }\bibfield  {title} {\enquote {\bibinfo {title} {{On the maximum mass of neutron stars and GW190814}},}\ }\href {\doibase 10.3847/1538-4357/abd4dd} {\bibfield  {journal} {\bibinfo  {journal} {Astrophys. J.}\ }\textbf {\bibinfo {volume} {908}},\ \bibinfo {pages} {122} (\bibinfo {year} {2021})},\ \Eprint {http://arxiv.org/abs/2007.10999} {arXiv:2007.10999 [astro-ph.HE]} \BibitemShut {NoStop}%
\bibitem [{\citenamefont {Huang}\ \emph {et~al.}(2020)\citenamefont {Huang}, \citenamefont {Hu}, \citenamefont {Zhang},\ and\ \citenamefont {Shen}}]{Huang:2020cab}%
  \BibitemOpen
  \bibfield  {author} {\bibinfo {author} {\bibfnamefont {Kaixuan}\ \bibnamefont {Huang}}, \bibinfo {author} {\bibfnamefont {Jinniu}\ \bibnamefont {Hu}}, \bibinfo {author} {\bibfnamefont {Ying}\ \bibnamefont {Zhang}}, \ and\ \bibinfo {author} {\bibfnamefont {Hong}\ \bibnamefont {Shen}},\ }\bibfield  {title} {\enquote {\bibinfo {title} {{The possibility of the secondary object in GW190814 as a neutron star}},}\ }\href {\doibase 10.3847/1538-4357/abbb37} {\bibfield  {journal} {\bibinfo  {journal} {Astrophys. J.}\ }\textbf {\bibinfo {volume} {904}},\ \bibinfo {pages} {39} (\bibinfo {year} {2020})},\ \Eprint {http://arxiv.org/abs/2008.04491} {arXiv:2008.04491 [nucl-th]} \BibitemShut {NoStop}%
\bibitem [{\citenamefont {Biswas}\ \emph {et~al.}(2021)\citenamefont {Biswas}, \citenamefont {Nandi}, \citenamefont {Char}, \citenamefont {Bose},\ and\ \citenamefont {Stergioulas}}]{Biswas:2020xna}%
  \BibitemOpen
  \bibfield  {author} {\bibinfo {author} {\bibfnamefont {Bhaskar}\ \bibnamefont {Biswas}}, \bibinfo {author} {\bibfnamefont {Rana}\ \bibnamefont {Nandi}}, \bibinfo {author} {\bibfnamefont {Prasanta}\ \bibnamefont {Char}}, \bibinfo {author} {\bibfnamefont {Sukanta}\ \bibnamefont {Bose}}, \ and\ \bibinfo {author} {\bibfnamefont {Nikolaos}\ \bibnamefont {Stergioulas}},\ }\bibfield  {title} {\enquote {\bibinfo {title} {{GW190814: on the properties of the secondary component of the binary}},}\ }\href {\doibase 10.1093/mnras/stab1383} {\bibfield  {journal} {\bibinfo  {journal} {Mon. Not. Roy. Astron. Soc.}\ }\textbf {\bibinfo {volume} {505}},\ \bibinfo {pages} {1600--1606} (\bibinfo {year} {2021})},\ \Eprint {http://arxiv.org/abs/2010.02090} {arXiv:2010.02090 [astro-ph.HE]} \BibitemShut {NoStop}%
\bibitem [{\citenamefont {Gross}\ and\ \citenamefont {Wilczek}(1973)}]{Gross:1973id}%
  \BibitemOpen
  \bibfield  {author} {\bibinfo {author} {\bibfnamefont {David~J.}\ \bibnamefont {Gross}}\ and\ \bibinfo {author} {\bibfnamefont {Frank}\ \bibnamefont {Wilczek}},\ }\bibfield  {title} {\enquote {\bibinfo {title} {{Ultraviolet Behavior of Nonabelian Gauge Theories}},}\ }\href {\doibase 10.1103/PhysRevLett.30.1343} {\bibfield  {journal} {\bibinfo  {journal} {Phys. Rev. Lett.}\ }\textbf {\bibinfo {volume} {30}},\ \bibinfo {pages} {1343--1346} (\bibinfo {year} {1973})}\BibitemShut {NoStop}%
\bibitem [{\citenamefont {Politzer}(1973)}]{Politzer:1973fx}%
  \BibitemOpen
  \bibfield  {author} {\bibinfo {author} {\bibfnamefont {H.~David}\ \bibnamefont {Politzer}},\ }\bibfield  {title} {\enquote {\bibinfo {title} {{Reliable Perturbative Results for Strong Interactions?}}}\ }\href {\doibase 10.1103/PhysRevLett.30.1346} {\bibfield  {journal} {\bibinfo  {journal} {Phys. Rev. Lett.}\ }\textbf {\bibinfo {volume} {30}},\ \bibinfo {pages} {1346--1349} (\bibinfo {year} {1973})}\BibitemShut {NoStop}%
\bibitem [{\citenamefont {Kurkela}\ \emph {et~al.}(2010)\citenamefont {Kurkela}, \citenamefont {Romatschke},\ and\ \citenamefont {Vuorinen}}]{Kurkela:2009gj}%
  \BibitemOpen
  \bibfield  {author} {\bibinfo {author} {\bibfnamefont {Aleksi}\ \bibnamefont {Kurkela}}, \bibinfo {author} {\bibfnamefont {Paul}\ \bibnamefont {Romatschke}}, \ and\ \bibinfo {author} {\bibfnamefont {Aleksi}\ \bibnamefont {Vuorinen}},\ }\bibfield  {title} {\enquote {\bibinfo {title} {{Cold Quark Matter}},}\ }\href {\doibase 10.1103/PhysRevD.81.105021} {\bibfield  {journal} {\bibinfo  {journal} {Phys. Rev. D}\ }\textbf {\bibinfo {volume} {81}},\ \bibinfo {pages} {105021} (\bibinfo {year} {2010})},\ \Eprint {http://arxiv.org/abs/0912.1856} {arXiv:0912.1856 [hep-ph]} \BibitemShut {NoStop}%
\bibitem [{\citenamefont {Graf}\ \emph {et~al.}(2016)\citenamefont {Graf}, \citenamefont {Schaffner-Bielich},\ and\ \citenamefont {Fraga}}]{Graf:2015tda}%
  \BibitemOpen
  \bibfield  {author} {\bibinfo {author} {\bibfnamefont {Thorben}\ \bibnamefont {Graf}}, \bibinfo {author} {\bibfnamefont {Juergen}\ \bibnamefont {Schaffner-Bielich}}, \ and\ \bibinfo {author} {\bibfnamefont {Eduardo~S.}\ \bibnamefont {Fraga}},\ }\bibfield  {title} {\enquote {\bibinfo {title} {{The impact of quark masses on pQCD thermodynamics}},}\ }\href {\doibase 10.1140/epja/i2016-16208-9} {\bibfield  {journal} {\bibinfo  {journal} {Eur. Phys. J. A}\ }\textbf {\bibinfo {volume} {52}},\ \bibinfo {pages} {208} (\bibinfo {year} {2016})},\ \Eprint {http://arxiv.org/abs/1507.08941} {arXiv:1507.08941 [hep-ph]} \BibitemShut {NoStop}%
\bibitem [{\citenamefont {Epelbaum}\ \emph {et~al.}(2009)\citenamefont {Epelbaum}, \citenamefont {Hammer},\ and\ \citenamefont {Meissner}}]{Epelbaum:2008ga}%
  \BibitemOpen
  \bibfield  {author} {\bibinfo {author} {\bibfnamefont {Evgeny}\ \bibnamefont {Epelbaum}}, \bibinfo {author} {\bibfnamefont {Hans-Werner}\ \bibnamefont {Hammer}}, \ and\ \bibinfo {author} {\bibfnamefont {Ulf-G.}\ \bibnamefont {Meissner}},\ }\bibfield  {title} {\enquote {\bibinfo {title} {{Modern Theory of Nuclear Forces}},}\ }\href {\doibase 10.1103/RevModPhys.81.1773} {\bibfield  {journal} {\bibinfo  {journal} {Rev. Mod. Phys.}\ }\textbf {\bibinfo {volume} {81}},\ \bibinfo {pages} {1773--1825} (\bibinfo {year} {2009})},\ \Eprint {http://arxiv.org/abs/0811.1338} {arXiv:0811.1338 [nucl-th]} \BibitemShut {NoStop}%
\bibitem [{\citenamefont {Drischler}\ \emph {et~al.}(2019)\citenamefont {Drischler}, \citenamefont {Hebeler},\ and\ \citenamefont {Schwenk}}]{Drischler:2017wtt}%
  \BibitemOpen
  \bibfield  {author} {\bibinfo {author} {\bibfnamefont {C.}~\bibnamefont {Drischler}}, \bibinfo {author} {\bibfnamefont {K.}~\bibnamefont {Hebeler}}, \ and\ \bibinfo {author} {\bibfnamefont {A.}~\bibnamefont {Schwenk}},\ }\bibfield  {title} {\enquote {\bibinfo {title} {{Chiral interactions up to next-to-next-to-next-to-leading order and nuclear saturation}},}\ }\href {\doibase 10.1103/PhysRevLett.122.042501} {\bibfield  {journal} {\bibinfo  {journal} {Phys. Rev. Lett.}\ }\textbf {\bibinfo {volume} {122}},\ \bibinfo {pages} {042501} (\bibinfo {year} {2019})},\ \Eprint {http://arxiv.org/abs/1710.08220} {arXiv:1710.08220 [nucl-th]} \BibitemShut {NoStop}%
\bibitem [{\citenamefont {Lu}\ \emph {et~al.}(2019)\citenamefont {Lu}, \citenamefont {Li}, \citenamefont {Elhatisari}, \citenamefont {Lee}, \citenamefont {Epelbaum},\ and\ \citenamefont {Mei\ss{}ner}}]{Lu:2018bat}%
  \BibitemOpen
  \bibfield  {author} {\bibinfo {author} {\bibfnamefont {Bing-Nan}\ \bibnamefont {Lu}}, \bibinfo {author} {\bibfnamefont {Ning}\ \bibnamefont {Li}}, \bibinfo {author} {\bibfnamefont {Serdar}\ \bibnamefont {Elhatisari}}, \bibinfo {author} {\bibfnamefont {Dean}\ \bibnamefont {Lee}}, \bibinfo {author} {\bibfnamefont {Evgeny}\ \bibnamefont {Epelbaum}}, \ and\ \bibinfo {author} {\bibfnamefont {Ulf-G.}\ \bibnamefont {Mei\ss{}ner}},\ }\bibfield  {title} {\enquote {\bibinfo {title} {{Essential elements for nuclear binding}},}\ }\href {\doibase 10.1016/j.physletb.2019.134863} {\bibfield  {journal} {\bibinfo  {journal} {Phys. Lett. B}\ }\textbf {\bibinfo {volume} {797}},\ \bibinfo {pages} {134863} (\bibinfo {year} {2019})},\ \Eprint {http://arxiv.org/abs/1812.10928} {arXiv:1812.10928 [nucl-th]} \BibitemShut {NoStop}%
\bibitem [{\citenamefont {Maggiore}\ \emph {et~al.}(2020)\citenamefont {Maggiore} \emph {et~al.}}]{Maggiore:2019uih}%
  \BibitemOpen
  \bibfield  {author} {\bibinfo {author} {\bibfnamefont {Michele}\ \bibnamefont {Maggiore}} \emph {et~al.},\ }\bibfield  {title} {\enquote {\bibinfo {title} {{Science Case for the Einstein Telescope}},}\ }\href {\doibase 10.1088/1475-7516/2020/03/050} {\bibfield  {journal} {\bibinfo  {journal} {JCAP}\ }\textbf {\bibinfo {volume} {03}},\ \bibinfo {pages} {050} (\bibinfo {year} {2020})},\ \Eprint {http://arxiv.org/abs/1912.02622} {arXiv:1912.02622 [astro-ph.CO]} \BibitemShut {NoStop}%
\bibitem [{\citenamefont {Ballmer}\ \emph {et~al.}(2022)\citenamefont {Ballmer} \emph {et~al.}}]{Ballmer:2022uxx}%
  \BibitemOpen
  \bibfield  {author} {\bibinfo {author} {\bibfnamefont {Stefan~W.}\ \bibnamefont {Ballmer}} \emph {et~al.},\ }\bibfield  {title} {\enquote {\bibinfo {title} {{Snowmass2021 Cosmic Frontier White Paper: Future Gravitational-Wave Detector Facilities}},}\ }in\ \href@noop {} {\emph {\bibinfo {booktitle} {{Snowmass 2021}}}}\ (\bibinfo {year} {2022})\ \Eprint {http://arxiv.org/abs/2203.08228} {arXiv:2203.08228 [gr-qc]} \BibitemShut {NoStop}%
\bibitem [{\citenamefont {Bogdanov}\ \emph {et~al.}(2022)\citenamefont {Bogdanov} \emph {et~al.}}]{Bogdanov:2022faf}%
  \BibitemOpen
  \bibfield  {author} {\bibinfo {author} {\bibfnamefont {Slavko}\ \bibnamefont {Bogdanov}} \emph {et~al.},\ }\bibfield  {title} {\enquote {\bibinfo {title} {{Snowmass 2021 Cosmic Frontier White Paper: The Dense Matter Equation of State and QCD Phase Transitions}},}\ }in\ \href@noop {} {\emph {\bibinfo {booktitle} {{Snowmass 2021}}}}\ (\bibinfo {year} {2022})\ \Eprint {http://arxiv.org/abs/2209.07412} {arXiv:2209.07412 [astro-ph.HE]} \BibitemShut {NoStop}%
\bibitem [{\citenamefont {Patricelli}\ \emph {et~al.}(2022)\citenamefont {Patricelli}, \citenamefont {Bernardini}, \citenamefont {Mapelli}, \citenamefont {D'Avanzo}, \citenamefont {Santoliquido}, \citenamefont {Cella}, \citenamefont {Razzano},\ and\ \citenamefont {Cuoco}}]{Patricelli:2022hhr}%
  \BibitemOpen
  \bibfield  {author} {\bibinfo {author} {\bibfnamefont {Barbara}\ \bibnamefont {Patricelli}}, \bibinfo {author} {\bibfnamefont {Maria~Grazia}\ \bibnamefont {Bernardini}}, \bibinfo {author} {\bibfnamefont {Michela}\ \bibnamefont {Mapelli}}, \bibinfo {author} {\bibfnamefont {Paolo}\ \bibnamefont {D'Avanzo}}, \bibinfo {author} {\bibfnamefont {Filippo}\ \bibnamefont {Santoliquido}}, \bibinfo {author} {\bibfnamefont {Giancarlo}\ \bibnamefont {Cella}}, \bibinfo {author} {\bibfnamefont {Massimiliano}\ \bibnamefont {Razzano}}, \ and\ \bibinfo {author} {\bibfnamefont {Elena}\ \bibnamefont {Cuoco}},\ }\bibfield  {title} {\enquote {\bibinfo {title} {{Prospects for multimessenger detection of binary neutron star mergers in the fourth LIGO\textendash{}Virgo\textendash{}KAGRA observing run}},}\ }\href {\doibase 10.1093/mnras/stac1167} {\bibfield  {journal} {\bibinfo  {journal} {Mon. Not. Roy. Astron. Soc.}\ }\textbf {\bibinfo {volume} {513}},\ \bibinfo {pages} {4159--4168} (\bibinfo {year} {2022})},\ \bibinfo {note}
  {[Erratum: Mon.Not.Roy.Astron.Soc. 514, 3395 (2022)]},\ \Eprint {http://arxiv.org/abs/2204.12504} {arXiv:2204.12504 [astro-ph.HE]} \BibitemShut {NoStop}%
\bibitem [{\citenamefont {Landry}\ and\ \citenamefont {Essick}(2019)}]{Landry:2018prl}%
  \BibitemOpen
  \bibfield  {author} {\bibinfo {author} {\bibfnamefont {Philippe}\ \bibnamefont {Landry}}\ and\ \bibinfo {author} {\bibfnamefont {Reed}\ \bibnamefont {Essick}},\ }\bibfield  {title} {\enquote {\bibinfo {title} {{Nonparametric inference of the neutron star equation of state from gravitational wave observations}},}\ }\href {\doibase 10.1103/PhysRevD.99.084049} {\bibfield  {journal} {\bibinfo  {journal} {Phys. Rev. D}\ }\textbf {\bibinfo {volume} {99}},\ \bibinfo {pages} {084049} (\bibinfo {year} {2019})},\ \Eprint {http://arxiv.org/abs/1811.12529} {arXiv:1811.12529 [gr-qc]} \BibitemShut {NoStop}%
\bibitem [{\citenamefont {Landry}\ \emph {et~al.}(2020)\citenamefont {Landry}, \citenamefont {Essick},\ and\ \citenamefont {Chatziioannou}}]{Landry:2020vaw}%
  \BibitemOpen
  \bibfield  {author} {\bibinfo {author} {\bibfnamefont {Philippe}\ \bibnamefont {Landry}}, \bibinfo {author} {\bibfnamefont {Reed}\ \bibnamefont {Essick}}, \ and\ \bibinfo {author} {\bibfnamefont {Katerina}\ \bibnamefont {Chatziioannou}},\ }\bibfield  {title} {\enquote {\bibinfo {title} {{Nonparametric constraints on neutron star matter with existing and upcoming gravitational wave and pulsar observations}},}\ }\href {\doibase 10.1103/PhysRevD.101.123007} {\bibfield  {journal} {\bibinfo  {journal} {Phys. Rev. D}\ }\textbf {\bibinfo {volume} {101}},\ \bibinfo {pages} {123007} (\bibinfo {year} {2020})},\ \Eprint {http://arxiv.org/abs/2003.04880} {arXiv:2003.04880 [astro-ph.HE]} \BibitemShut {NoStop}%
\bibitem [{\citenamefont {Annala}\ \emph {et~al.}(2022)\citenamefont {Annala}, \citenamefont {Gorda}, \citenamefont {Katerini}, \citenamefont {Kurkela}, \citenamefont {N\"attil\"a}, \citenamefont {Paschalidis},\ and\ \citenamefont {Vuorinen}}]{Annala:2021gom}%
  \BibitemOpen
  \bibfield  {author} {\bibinfo {author} {\bibfnamefont {Eemeli}\ \bibnamefont {Annala}}, \bibinfo {author} {\bibfnamefont {Tyler}\ \bibnamefont {Gorda}}, \bibinfo {author} {\bibfnamefont {Evangelia}\ \bibnamefont {Katerini}}, \bibinfo {author} {\bibfnamefont {Aleksi}\ \bibnamefont {Kurkela}}, \bibinfo {author} {\bibfnamefont {Joonas}\ \bibnamefont {N\"attil\"a}}, \bibinfo {author} {\bibfnamefont {Vasileios}\ \bibnamefont {Paschalidis}}, \ and\ \bibinfo {author} {\bibfnamefont {Aleksi}\ \bibnamefont {Vuorinen}},\ }\bibfield  {title} {\enquote {\bibinfo {title} {{Multimessenger Constraints for Ultradense Matter}},}\ }\href {\doibase 10.1103/PhysRevX.12.011058} {\bibfield  {journal} {\bibinfo  {journal} {Phys. Rev. X}\ }\textbf {\bibinfo {volume} {12}},\ \bibinfo {pages} {011058} (\bibinfo {year} {2022})},\ \Eprint {http://arxiv.org/abs/2105.05132} {arXiv:2105.05132 [astro-ph.HE]} \BibitemShut {NoStop}%
\bibitem [{\citenamefont {Huth}\ \emph {et~al.}(2022)\citenamefont {Huth} \emph {et~al.}}]{Huth:2021bsp}%
  \BibitemOpen
  \bibfield  {author} {\bibinfo {author} {\bibfnamefont {S.}~\bibnamefont {Huth}} \emph {et~al.},\ }\bibfield  {title} {\enquote {\bibinfo {title} {{Constraining Neutron-Star Matter with Microscopic and Macroscopic Collisions}},}\ }\href {\doibase 10.1038/s41586-022-04750-w} {\bibfield  {journal} {\bibinfo  {journal} {Nature}\ }\textbf {\bibinfo {volume} {606}},\ \bibinfo {pages} {276--280} (\bibinfo {year} {2022})},\ \Eprint {http://arxiv.org/abs/2107.06229} {arXiv:2107.06229 [nucl-th]} \BibitemShut {NoStop}%
\bibitem [{\citenamefont {Legred}\ \emph {et~al.}(2021)\citenamefont {Legred}, \citenamefont {Chatziioannou}, \citenamefont {Essick}, \citenamefont {Han},\ and\ \citenamefont {Landry}}]{Legred:2021hdx}%
  \BibitemOpen
  \bibfield  {author} {\bibinfo {author} {\bibfnamefont {Isaac}\ \bibnamefont {Legred}}, \bibinfo {author} {\bibfnamefont {Katerina}\ \bibnamefont {Chatziioannou}}, \bibinfo {author} {\bibfnamefont {Reed}\ \bibnamefont {Essick}}, \bibinfo {author} {\bibfnamefont {Sophia}\ \bibnamefont {Han}}, \ and\ \bibinfo {author} {\bibfnamefont {Philippe}\ \bibnamefont {Landry}},\ }\bibfield  {title} {\enquote {\bibinfo {title} {{Impact of the PSR J0740+6620 radius constraint on the properties of high-density matter}},}\ }\href {\doibase 10.1103/PhysRevD.104.063003} {\bibfield  {journal} {\bibinfo  {journal} {Phys. Rev. D}\ }\textbf {\bibinfo {volume} {104}},\ \bibinfo {pages} {063003} (\bibinfo {year} {2021})},\ \Eprint {http://arxiv.org/abs/2106.05313} {arXiv:2106.05313 [astro-ph.HE]} \BibitemShut {NoStop}%
\bibitem [{\citenamefont {Kumar}\ \emph {et~al.}(2023)\citenamefont {Kumar} \emph {et~al.}}]{MUSES:2023hyz}%
  \BibitemOpen
  \bibfield  {author} {\bibinfo {author} {\bibfnamefont {Rajesh}\ \bibnamefont {Kumar}} \emph {et~al.} (\bibinfo {collaboration} {MUSES}),\ }\bibfield  {title} {\enquote {\bibinfo {title} {{Theoretical and Experimental Constraints for the Equation of State of Dense and Hot Matter}},}\ }\href@noop {} {\  (\bibinfo {year} {2023})},\ \Eprint {http://arxiv.org/abs/2303.17021} {arXiv:2303.17021 [nucl-th]} \BibitemShut {NoStop}%
\bibitem [{\citenamefont {Mroczek}\ \emph {et~al.}(2023)\citenamefont {Mroczek}, \citenamefont {Miller}, \citenamefont {Noronha-Hostler},\ and\ \citenamefont {Yunes}}]{Mroczek:2023zxo}%
  \BibitemOpen
  \bibfield  {author} {\bibinfo {author} {\bibfnamefont {Debora}\ \bibnamefont {Mroczek}}, \bibinfo {author} {\bibfnamefont {M.~Coleman}\ \bibnamefont {Miller}}, \bibinfo {author} {\bibfnamefont {Jacquelyn}\ \bibnamefont {Noronha-Hostler}}, \ and\ \bibinfo {author} {\bibfnamefont {Nicolas}\ \bibnamefont {Yunes}},\ }\bibfield  {title} {\enquote {\bibinfo {title} {{Nontrivial features in the speed of sound inside neutron stars}},}\ }\href@noop {} {\  (\bibinfo {year} {2023})},\ \Eprint {http://arxiv.org/abs/2309.02345} {arXiv:2309.02345 [astro-ph.HE]} \BibitemShut {NoStop}%
\bibitem [{\citenamefont {Gavassino}(2023)}]{Gavassino:2023myj}%
  \BibitemOpen
  \bibfield  {author} {\bibinfo {author} {\bibfnamefont {L.}~\bibnamefont {Gavassino}},\ }\bibfield  {title} {\enquote {\bibinfo {title} {{Bounds on transport from hydrodynamic stability}},}\ }\href {\doibase 10.1016/j.physletb.2023.137854} {\bibfield  {journal} {\bibinfo  {journal} {Phys. Lett. B}\ }\textbf {\bibinfo {volume} {840}},\ \bibinfo {pages} {137854} (\bibinfo {year} {2023})},\ \Eprint {http://arxiv.org/abs/2301.06651} {arXiv:2301.06651 [hep-th]} \BibitemShut {NoStop}%
\bibitem [{\citenamefont {Borsanyi}\ \emph {et~al.}(2014)\citenamefont {Borsanyi}, \citenamefont {Fodor}, \citenamefont {Hoelbling}, \citenamefont {Katz}, \citenamefont {Krieg},\ and\ \citenamefont {Szabo}}]{Borsanyi:2013bia}%
  \BibitemOpen
  \bibfield  {author} {\bibinfo {author} {\bibfnamefont {Szabocls}\ \bibnamefont {Borsanyi}}, \bibinfo {author} {\bibfnamefont {Zoltan}\ \bibnamefont {Fodor}}, \bibinfo {author} {\bibfnamefont {Christian}\ \bibnamefont {Hoelbling}}, \bibinfo {author} {\bibfnamefont {Sandor~D.}\ \bibnamefont {Katz}}, \bibinfo {author} {\bibfnamefont {Stefan}\ \bibnamefont {Krieg}}, \ and\ \bibinfo {author} {\bibfnamefont {Kalman~K.}\ \bibnamefont {Szabo}},\ }\bibfield  {title} {\enquote {\bibinfo {title} {{Full result for the QCD equation of state with 2+1 flavors}},}\ }\href {\doibase 10.1016/j.physletb.2014.01.007} {\bibfield  {journal} {\bibinfo  {journal} {Phys. Lett. B}\ }\textbf {\bibinfo {volume} {730}},\ \bibinfo {pages} {99--104} (\bibinfo {year} {2014})},\ \Eprint {http://arxiv.org/abs/1309.5258} {arXiv:1309.5258 [hep-lat]} \BibitemShut {NoStop}%
\bibitem [{\citenamefont {Bazavov}\ \emph {et~al.}(2018)\citenamefont {Bazavov}, \citenamefont {Petreczky},\ and\ \citenamefont {Weber}}]{Bazavov:2017dsy}%
  \BibitemOpen
  \bibfield  {author} {\bibinfo {author} {\bibfnamefont {A.}~\bibnamefont {Bazavov}}, \bibinfo {author} {\bibfnamefont {P.}~\bibnamefont {Petreczky}}, \ and\ \bibinfo {author} {\bibfnamefont {J.~H.}\ \bibnamefont {Weber}},\ }\bibfield  {title} {\enquote {\bibinfo {title} {{Equation of State in 2+1 Flavor QCD at High Temperatures}},}\ }\href {\doibase 10.1103/PhysRevD.97.014510} {\bibfield  {journal} {\bibinfo  {journal} {Phys. Rev. D}\ }\textbf {\bibinfo {volume} {97}},\ \bibinfo {pages} {014510} (\bibinfo {year} {2018})},\ \Eprint {http://arxiv.org/abs/1710.05024} {arXiv:1710.05024 [hep-lat]} \BibitemShut {NoStop}%
\bibitem [{\citenamefont {Gubser}\ \emph {et~al.}(2008)\citenamefont {Gubser}, \citenamefont {Nellore}, \citenamefont {Pufu},\ and\ \citenamefont {Rocha}}]{Gubser:2008yx}%
  \BibitemOpen
  \bibfield  {author} {\bibinfo {author} {\bibfnamefont {Steven~S.}\ \bibnamefont {Gubser}}, \bibinfo {author} {\bibfnamefont {Abhinav}\ \bibnamefont {Nellore}}, \bibinfo {author} {\bibfnamefont {Silviu~S.}\ \bibnamefont {Pufu}}, \ and\ \bibinfo {author} {\bibfnamefont {Fabio~D.}\ \bibnamefont {Rocha}},\ }\bibfield  {title} {\enquote {\bibinfo {title} {{Thermodynamics and bulk viscosity of approximate black hole duals to finite temperature quantum chromodynamics}},}\ }\href {\doibase 10.1103/PhysRevLett.101.131601} {\bibfield  {journal} {\bibinfo  {journal} {Phys. Rev. Lett.}\ }\textbf {\bibinfo {volume} {101}},\ \bibinfo {pages} {131601} (\bibinfo {year} {2008})},\ \Eprint {http://arxiv.org/abs/0804.1950} {arXiv:0804.1950 [hep-th]} \BibitemShut {NoStop}%
\bibitem [{\citenamefont {Cherman}\ \emph {et~al.}(2009)\citenamefont {Cherman}, \citenamefont {Cohen},\ and\ \citenamefont {Nellore}}]{Cherman:2009tw}%
  \BibitemOpen
  \bibfield  {author} {\bibinfo {author} {\bibfnamefont {Aleksey}\ \bibnamefont {Cherman}}, \bibinfo {author} {\bibfnamefont {Thomas~D.}\ \bibnamefont {Cohen}}, \ and\ \bibinfo {author} {\bibfnamefont {Abhinav}\ \bibnamefont {Nellore}},\ }\bibfield  {title} {\enquote {\bibinfo {title} {{A Bound on the speed of sound from holography}},}\ }\href {\doibase 10.1103/PhysRevD.80.066003} {\bibfield  {journal} {\bibinfo  {journal} {Phys. Rev. D}\ }\textbf {\bibinfo {volume} {80}},\ \bibinfo {pages} {066003} (\bibinfo {year} {2009})},\ \Eprint {http://arxiv.org/abs/0905.0903} {arXiv:0905.0903 [hep-th]} \BibitemShut {NoStop}%
\bibitem [{\citenamefont {Finazzo}\ \emph {et~al.}(2015)\citenamefont {Finazzo}, \citenamefont {Rougemont}, \citenamefont {Marrochio},\ and\ \citenamefont {Noronha}}]{Finazzo:2014cna}%
  \BibitemOpen
  \bibfield  {author} {\bibinfo {author} {\bibfnamefont {Stefano~I.}\ \bibnamefont {Finazzo}}, \bibinfo {author} {\bibfnamefont {Romulo}\ \bibnamefont {Rougemont}}, \bibinfo {author} {\bibfnamefont {Hugo}\ \bibnamefont {Marrochio}}, \ and\ \bibinfo {author} {\bibfnamefont {Jorge}\ \bibnamefont {Noronha}},\ }\bibfield  {title} {\enquote {\bibinfo {title} {{Hydrodynamic transport coefficients for the non-conformal quark-gluon plasma from holography}},}\ }\href {\doibase 10.1007/JHEP02(2015)051} {\bibfield  {journal} {\bibinfo  {journal} {JHEP}\ }\textbf {\bibinfo {volume} {02}},\ \bibinfo {pages} {051} (\bibinfo {year} {2015})},\ \Eprint {http://arxiv.org/abs/1412.2968} {arXiv:1412.2968 [hep-ph]} \BibitemShut {NoStop}%
\bibitem [{\citenamefont {Grefa}\ \emph {et~al.}(2021)\citenamefont {Grefa}, \citenamefont {Noronha}, \citenamefont {Noronha-Hostler}, \citenamefont {Portillo}, \citenamefont {Ratti},\ and\ \citenamefont {Rougemont}}]{Grefa:2021qvt}%
  \BibitemOpen
  \bibfield  {author} {\bibinfo {author} {\bibfnamefont {Joaquin}\ \bibnamefont {Grefa}}, \bibinfo {author} {\bibfnamefont {Jorge}\ \bibnamefont {Noronha}}, \bibinfo {author} {\bibfnamefont {Jacquelyn}\ \bibnamefont {Noronha-Hostler}}, \bibinfo {author} {\bibfnamefont {Israel}\ \bibnamefont {Portillo}}, \bibinfo {author} {\bibfnamefont {Claudia}\ \bibnamefont {Ratti}}, \ and\ \bibinfo {author} {\bibfnamefont {Romulo}\ \bibnamefont {Rougemont}},\ }\bibfield  {title} {\enquote {\bibinfo {title} {{Hot and dense quark-gluon plasma thermodynamics from holographic black holes}},}\ }\href {\doibase 10.1103/PhysRevD.104.034002} {\bibfield  {journal} {\bibinfo  {journal} {Phys. Rev. D}\ }\textbf {\bibinfo {volume} {104}},\ \bibinfo {pages} {034002} (\bibinfo {year} {2021})},\ \Eprint {http://arxiv.org/abs/2102.12042} {arXiv:2102.12042 [nucl-th]} \BibitemShut {NoStop}%
\bibitem [{\citenamefont {Hartle}(1978)}]{HARTLE1978201}%
  \BibitemOpen
  \bibfield  {author} {\bibinfo {author} {\bibfnamefont {James~B.}\ \bibnamefont {Hartle}},\ }\bibfield  {title} {\enquote {\bibinfo {title} {Bounds on the mass and moment of inertia of non-rotating neutron stars},}\ }\href {\doibase https://doi.org/10.1016/0370-1573(78)90140-0} {\bibfield  {journal} {\bibinfo  {journal} {Physics Reports}\ }\textbf {\bibinfo {volume} {46}},\ \bibinfo {pages} {201--247} (\bibinfo {year} {1978})}\BibitemShut {NoStop}%
\bibitem [{\citenamefont {Zel'dovich}(1961)}]{Zeldovich:1962emp}%
  \BibitemOpen
  \bibfield  {author} {\bibinfo {author} {\bibfnamefont {Ya.~B.}\ \bibnamefont {Zel'dovich}},\ }\bibfield  {title} {\enquote {\bibinfo {title} {{The equation of state at ultrahigh densities and its relativistic limitations}},}\ }\href@noop {} {\bibfield  {journal} {\bibinfo  {journal} {Zh. Eksp. Teor. Fiz.}\ }\textbf {\bibinfo {volume} {41}},\ \bibinfo {pages} {1609--1615} (\bibinfo {year} {1961})}\BibitemShut {NoStop}%
\bibitem [{\citenamefont {Bludman}\ and\ \citenamefont {Ruderman}(1968)}]{Bludman:1968zz}%
  \BibitemOpen
  \bibfield  {author} {\bibinfo {author} {\bibfnamefont {Sidney~A.}\ \bibnamefont {Bludman}}\ and\ \bibinfo {author} {\bibfnamefont {M.~A.}\ \bibnamefont {Ruderman}},\ }\bibfield  {title} {\enquote {\bibinfo {title} {{Possibility of the Speed of Sound Exceeding the Speed of Light in Ultradense Matter}},}\ }\href {\doibase 10.1103/PhysRev.170.1176} {\bibfield  {journal} {\bibinfo  {journal} {Phys. Rev.}\ }\textbf {\bibinfo {volume} {170}},\ \bibinfo {pages} {1176--1184} (\bibinfo {year} {1968})}\BibitemShut {NoStop}%
\bibitem [{\citenamefont {Zel’dovich}\ and\ \citenamefont {Novikov}(2014)}]{zel2014stars}%
  \BibitemOpen
  \bibfield  {author} {\bibinfo {author} {\bibfnamefont {Y.B.}\ \bibnamefont {Zel’dovich}}\ and\ \bibinfo {author} {\bibfnamefont {I.D.}\ \bibnamefont {Novikov}},\ }\href@noop {} {\emph {\bibinfo {title} {Stars and Relativity}}},\ Dover Books on Physics\ (\bibinfo  {publisher} {Dover Publications},\ \bibinfo {year} {2014})\BibitemShut {NoStop}%
\bibitem [{\citenamefont {Carignano}\ \emph {et~al.}(2017)\citenamefont {Carignano}, \citenamefont {Lepori}, \citenamefont {Mammarella}, \citenamefont {Mannarelli},\ and\ \citenamefont {Pagliaroli}}]{Carignano:2016lxe}%
  \BibitemOpen
  \bibfield  {author} {\bibinfo {author} {\bibfnamefont {Stefano}\ \bibnamefont {Carignano}}, \bibinfo {author} {\bibfnamefont {Luca}\ \bibnamefont {Lepori}}, \bibinfo {author} {\bibfnamefont {Andrea}\ \bibnamefont {Mammarella}}, \bibinfo {author} {\bibfnamefont {Massimo}\ \bibnamefont {Mannarelli}}, \ and\ \bibinfo {author} {\bibfnamefont {Giulia}\ \bibnamefont {Pagliaroli}},\ }\bibfield  {title} {\enquote {\bibinfo {title} {{Scrutinizing the pion condensed phase}},}\ }\href {\doibase 10.1140/epja/i2017-12221-x} {\bibfield  {journal} {\bibinfo  {journal} {Eur. Phys. J. A}\ }\textbf {\bibinfo {volume} {53}},\ \bibinfo {pages} {35} (\bibinfo {year} {2017})},\ \Eprint {http://arxiv.org/abs/1610.06097} {arXiv:1610.06097 [hep-ph]} \BibitemShut {NoStop}%
\bibitem [{\citenamefont {Hands}\ \emph {et~al.}(2006)\citenamefont {Hands}, \citenamefont {Kim},\ and\ \citenamefont {Skullerud}}]{Hands:2006ve}%
  \BibitemOpen
  \bibfield  {author} {\bibinfo {author} {\bibfnamefont {Simon}\ \bibnamefont {Hands}}, \bibinfo {author} {\bibfnamefont {Seyong}\ \bibnamefont {Kim}}, \ and\ \bibinfo {author} {\bibfnamefont {Jon-Ivar}\ \bibnamefont {Skullerud}},\ }\bibfield  {title} {\enquote {\bibinfo {title} {{Deconfinement in dense 2-color QCD}},}\ }\href {\doibase 10.1140/epjc/s2006-02621-8} {\bibfield  {journal} {\bibinfo  {journal} {Eur. Phys. J. C}\ }\textbf {\bibinfo {volume} {48}},\ \bibinfo {pages} {193} (\bibinfo {year} {2006})},\ \Eprint {http://arxiv.org/abs/hep-lat/0604004} {arXiv:hep-lat/0604004} \BibitemShut {NoStop}%
\bibitem [{\citenamefont {Hoyos}\ \emph {et~al.}(2016)\citenamefont {Hoyos}, \citenamefont {Jokela}, \citenamefont {Rodr\'\i{}guez~Fern\'andez},\ and\ \citenamefont {Vuorinen}}]{Hoyos:2016cob}%
  \BibitemOpen
  \bibfield  {author} {\bibinfo {author} {\bibfnamefont {Carlos}\ \bibnamefont {Hoyos}}, \bibinfo {author} {\bibfnamefont {Niko}\ \bibnamefont {Jokela}}, \bibinfo {author} {\bibfnamefont {David}\ \bibnamefont {Rodr\'\i{}guez~Fern\'andez}}, \ and\ \bibinfo {author} {\bibfnamefont {Aleksi}\ \bibnamefont {Vuorinen}},\ }\bibfield  {title} {\enquote {\bibinfo {title} {{Breaking the sound barrier in AdS/CFT}},}\ }\href {\doibase 10.1103/PhysRevD.94.106008} {\bibfield  {journal} {\bibinfo  {journal} {Phys. Rev. D}\ }\textbf {\bibinfo {volume} {94}},\ \bibinfo {pages} {106008} (\bibinfo {year} {2016})},\ \Eprint {http://arxiv.org/abs/1609.03480} {arXiv:1609.03480 [hep-th]} \BibitemShut {NoStop}%
\bibitem [{\citenamefont {Ecker}\ \emph {et~al.}(2017)\citenamefont {Ecker}, \citenamefont {Hoyos}, \citenamefont {Jokela}, \citenamefont {Rodr\'\i{}guez~Fern\'andez},\ and\ \citenamefont {Vuorinen}}]{Ecker:2017fyh}%
  \BibitemOpen
  \bibfield  {author} {\bibinfo {author} {\bibfnamefont {Christian}\ \bibnamefont {Ecker}}, \bibinfo {author} {\bibfnamefont {Carlos}\ \bibnamefont {Hoyos}}, \bibinfo {author} {\bibfnamefont {Niko}\ \bibnamefont {Jokela}}, \bibinfo {author} {\bibfnamefont {David}\ \bibnamefont {Rodr\'\i{}guez~Fern\'andez}}, \ and\ \bibinfo {author} {\bibfnamefont {Aleksi}\ \bibnamefont {Vuorinen}},\ }\bibfield  {title} {\enquote {\bibinfo {title} {{Stiff phases in strongly coupled gauge theories with holographic duals}},}\ }\href {\doibase 10.1007/JHEP11(2017)031} {\bibfield  {journal} {\bibinfo  {journal} {JHEP}\ }\textbf {\bibinfo {volume} {11}},\ \bibinfo {pages} {031} (\bibinfo {year} {2017})},\ \Eprint {http://arxiv.org/abs/1707.00521} {arXiv:1707.00521 [hep-th]} \BibitemShut {NoStop}%
\bibitem [{\citenamefont {Anabalon}\ \emph {et~al.}(2018)\citenamefont {Anabalon}, \citenamefont {Andrade}, \citenamefont {Astefanesei},\ and\ \citenamefont {Mann}}]{Anabalon:2017eri}%
  \BibitemOpen
  \bibfield  {author} {\bibinfo {author} {\bibfnamefont {Andres}\ \bibnamefont {Anabalon}}, \bibinfo {author} {\bibfnamefont {Tomas}\ \bibnamefont {Andrade}}, \bibinfo {author} {\bibfnamefont {Dumitru}\ \bibnamefont {Astefanesei}}, \ and\ \bibinfo {author} {\bibfnamefont {Robert}\ \bibnamefont {Mann}},\ }\bibfield  {title} {\enquote {\bibinfo {title} {{Universal Formula for the Holographic Speed of Sound}},}\ }\href {\doibase 10.1016/j.physletb.2018.04.028} {\bibfield  {journal} {\bibinfo  {journal} {Phys. Lett. B}\ }\textbf {\bibinfo {volume} {781}},\ \bibinfo {pages} {547--552} (\bibinfo {year} {2018})},\ \Eprint {http://arxiv.org/abs/1702.00017} {arXiv:1702.00017 [hep-th]} \BibitemShut {NoStop}%
\bibitem [{\citenamefont {Chesler}\ \emph {et~al.}(2019)\citenamefont {Chesler}, \citenamefont {Jokela}, \citenamefont {Loeb},\ and\ \citenamefont {Vuorinen}}]{Chesler:2019osn}%
  \BibitemOpen
  \bibfield  {author} {\bibinfo {author} {\bibfnamefont {Paul~M.}\ \bibnamefont {Chesler}}, \bibinfo {author} {\bibfnamefont {Niko}\ \bibnamefont {Jokela}}, \bibinfo {author} {\bibfnamefont {Abraham}\ \bibnamefont {Loeb}}, \ and\ \bibinfo {author} {\bibfnamefont {Aleksi}\ \bibnamefont {Vuorinen}},\ }\bibfield  {title} {\enquote {\bibinfo {title} {{Finite-temperature Equations of State for Neutron Star Mergers}},}\ }\href {\doibase 10.1103/PhysRevD.100.066027} {\bibfield  {journal} {\bibinfo  {journal} {Phys. Rev. D}\ }\textbf {\bibinfo {volume} {100}},\ \bibinfo {pages} {066027} (\bibinfo {year} {2019})},\ \Eprint {http://arxiv.org/abs/1906.08440} {arXiv:1906.08440 [astro-ph.HE]} \BibitemShut {NoStop}%
\bibitem [{\citenamefont {Ishii}\ \emph {et~al.}(2019)\citenamefont {Ishii}, \citenamefont {J\"arvinen},\ and\ \citenamefont {Nijs}}]{Ishii:2019gta}%
  \BibitemOpen
  \bibfield  {author} {\bibinfo {author} {\bibfnamefont {Takaaki}\ \bibnamefont {Ishii}}, \bibinfo {author} {\bibfnamefont {Matti}\ \bibnamefont {J\"arvinen}}, \ and\ \bibinfo {author} {\bibfnamefont {Govert}\ \bibnamefont {Nijs}},\ }\bibfield  {title} {\enquote {\bibinfo {title} {{Cool baryon and quark matter in holographic QCD}},}\ }\href {\doibase 10.1007/JHEP07(2019)003} {\bibfield  {journal} {\bibinfo  {journal} {JHEP}\ }\textbf {\bibinfo {volume} {07}},\ \bibinfo {pages} {003} (\bibinfo {year} {2019})},\ \Eprint {http://arxiv.org/abs/1903.06169} {arXiv:1903.06169 [hep-ph]} \BibitemShut {NoStop}%
\bibitem [{\citenamefont {Gorda}\ and\ \citenamefont {Romatschke}(2015)}]{Gorda:2014vga}%
  \BibitemOpen
  \bibfield  {author} {\bibinfo {author} {\bibfnamefont {Tyler}\ \bibnamefont {Gorda}}\ and\ \bibinfo {author} {\bibfnamefont {Paul}\ \bibnamefont {Romatschke}},\ }\bibfield  {title} {\enquote {\bibinfo {title} {{Equation of state in two-, three-, and four-color QCD at nonzero temperature and density}},}\ }\href {\doibase 10.1103/PhysRevD.92.014019} {\bibfield  {journal} {\bibinfo  {journal} {Phys. Rev. D}\ }\textbf {\bibinfo {volume} {92}},\ \bibinfo {pages} {014019} (\bibinfo {year} {2015})},\ \Eprint {http://arxiv.org/abs/1412.6712} {arXiv:1412.6712 [hep-ph]} \BibitemShut {NoStop}%
\bibitem [{\citenamefont {Fujimoto}\ and\ \citenamefont {Fukushima}(2022)}]{Fujimoto:2020tjc}%
  \BibitemOpen
  \bibfield  {author} {\bibinfo {author} {\bibfnamefont {Yuki}\ \bibnamefont {Fujimoto}}\ and\ \bibinfo {author} {\bibfnamefont {Kenji}\ \bibnamefont {Fukushima}},\ }\bibfield  {title} {\enquote {\bibinfo {title} {{Equation of state of cold and dense QCD matter in resummed perturbation theory}},}\ }\href {\doibase 10.1103/PhysRevD.105.014025} {\bibfield  {journal} {\bibinfo  {journal} {Phys. Rev. D}\ }\textbf {\bibinfo {volume} {105}},\ \bibinfo {pages} {014025} (\bibinfo {year} {2022})},\ \Eprint {http://arxiv.org/abs/2011.10891} {arXiv:2011.10891 [hep-ph]} \BibitemShut {NoStop}%
\bibitem [{\citenamefont {McLerran}\ and\ \citenamefont {Reddy}(2019)}]{McLerran:2018hbz}%
  \BibitemOpen
  \bibfield  {author} {\bibinfo {author} {\bibfnamefont {Larry}\ \bibnamefont {McLerran}}\ and\ \bibinfo {author} {\bibfnamefont {Sanjay}\ \bibnamefont {Reddy}},\ }\bibfield  {title} {\enquote {\bibinfo {title} {{Quarkyonic Matter and Neutron Stars}},}\ }\href {\doibase 10.1103/PhysRevLett.122.122701} {\bibfield  {journal} {\bibinfo  {journal} {Phys. Rev. Lett.}\ }\textbf {\bibinfo {volume} {122}},\ \bibinfo {pages} {122701} (\bibinfo {year} {2019})},\ \Eprint {http://arxiv.org/abs/1811.12503} {arXiv:1811.12503 [nucl-th]} \BibitemShut {NoStop}%
\bibitem [{\citenamefont {Jeong}\ \emph {et~al.}(2020)\citenamefont {Jeong}, \citenamefont {McLerran},\ and\ \citenamefont {Sen}}]{Jeong:2019lhv}%
  \BibitemOpen
  \bibfield  {author} {\bibinfo {author} {\bibfnamefont {Kie~Sang}\ \bibnamefont {Jeong}}, \bibinfo {author} {\bibfnamefont {Larry}\ \bibnamefont {McLerran}}, \ and\ \bibinfo {author} {\bibfnamefont {Srimoyee}\ \bibnamefont {Sen}},\ }\bibfield  {title} {\enquote {\bibinfo {title} {{Dynamically generated momentum space shell structure of quarkyonic matter via an excluded volume model}},}\ }\href {\doibase 10.1103/PhysRevC.101.035201} {\bibfield  {journal} {\bibinfo  {journal} {Phys. Rev. C}\ }\textbf {\bibinfo {volume} {101}},\ \bibinfo {pages} {035201} (\bibinfo {year} {2020})},\ \Eprint {http://arxiv.org/abs/1908.04799} {arXiv:1908.04799 [nucl-th]} \BibitemShut {NoStop}%
\bibitem [{\citenamefont {Zhao}\ and\ \citenamefont {Lattimer}(2020)}]{Zhao:2020dvu}%
  \BibitemOpen
  \bibfield  {author} {\bibinfo {author} {\bibfnamefont {Tianqi}\ \bibnamefont {Zhao}}\ and\ \bibinfo {author} {\bibfnamefont {James~M.}\ \bibnamefont {Lattimer}},\ }\bibfield  {title} {\enquote {\bibinfo {title} {{Quarkyonic Matter Equation of State in Beta-Equilibrium}},}\ }\href {\doibase 10.1103/PhysRevD.102.023021} {\bibfield  {journal} {\bibinfo  {journal} {Phys. Rev. D}\ }\textbf {\bibinfo {volume} {102}},\ \bibinfo {pages} {023021} (\bibinfo {year} {2020})},\ \Eprint {http://arxiv.org/abs/2004.08293} {arXiv:2004.08293 [astro-ph.HE]} \BibitemShut {NoStop}%
\bibitem [{\citenamefont {Margueron}\ \emph {et~al.}(2021)\citenamefont {Margueron}, \citenamefont {Hansen}, \citenamefont {Proust},\ and\ \citenamefont {Chanfray}}]{Margueron:2021dtx}%
  \BibitemOpen
  \bibfield  {author} {\bibinfo {author} {\bibfnamefont {J.}~\bibnamefont {Margueron}}, \bibinfo {author} {\bibfnamefont {H.}~\bibnamefont {Hansen}}, \bibinfo {author} {\bibfnamefont {P.}~\bibnamefont {Proust}}, \ and\ \bibinfo {author} {\bibfnamefont {G.}~\bibnamefont {Chanfray}},\ }\bibfield  {title} {\enquote {\bibinfo {title} {{Quarkyonic stars with isospin-flavor asymmetry}},}\ }\href {\doibase 10.1103/PhysRevC.104.055803} {\bibfield  {journal} {\bibinfo  {journal} {Phys. Rev. C}\ }\textbf {\bibinfo {volume} {104}},\ \bibinfo {pages} {055803} (\bibinfo {year} {2021})},\ \Eprint {http://arxiv.org/abs/2103.10209} {arXiv:2103.10209 [nucl-th]} \BibitemShut {NoStop}%
\bibitem [{\citenamefont {Duarte}\ \emph {et~al.}(2021)\citenamefont {Duarte}, \citenamefont {Hernandez-Ortiz}, \citenamefont {Jeong},\ and\ \citenamefont {McLerran}}]{Duarte:2021tsx}%
  \BibitemOpen
  \bibfield  {author} {\bibinfo {author} {\bibfnamefont {Dyana~C.}\ \bibnamefont {Duarte}}, \bibinfo {author} {\bibfnamefont {Saul}\ \bibnamefont {Hernandez-Ortiz}}, \bibinfo {author} {\bibfnamefont {Kie~Sang}\ \bibnamefont {Jeong}}, \ and\ \bibinfo {author} {\bibfnamefont {Larry~D.}\ \bibnamefont {McLerran}},\ }\bibfield  {title} {\enquote {\bibinfo {title} {{Quarkyonic effective field theory, quark-nucleon duality, and ghosts}},}\ }\href {\doibase 10.1103/PhysRevD.104.L091901} {\bibfield  {journal} {\bibinfo  {journal} {Phys. Rev. D}\ }\textbf {\bibinfo {volume} {104}},\ \bibinfo {pages} {L091901} (\bibinfo {year} {2021})},\ \Eprint {http://arxiv.org/abs/2103.05679} {arXiv:2103.05679 [nucl-th]} \BibitemShut {NoStop}%
\bibitem [{\citenamefont {Blaschke}\ \emph {et~al.}(2007)\citenamefont {Blaschke}, \citenamefont {Gomez~Dumm}, \citenamefont {Grunfeld}, \citenamefont {Klahn},\ and\ \citenamefont {Scoccola}}]{Blaschke:2007ri}%
  \BibitemOpen
  \bibfield  {author} {\bibinfo {author} {\bibfnamefont {D.~B.}\ \bibnamefont {Blaschke}}, \bibinfo {author} {\bibfnamefont {D.}~\bibnamefont {Gomez~Dumm}}, \bibinfo {author} {\bibfnamefont {A.~G.}\ \bibnamefont {Grunfeld}}, \bibinfo {author} {\bibfnamefont {T.}~\bibnamefont {Klahn}}, \ and\ \bibinfo {author} {\bibfnamefont {N.~N.}\ \bibnamefont {Scoccola}},\ }\bibfield  {title} {\enquote {\bibinfo {title} {{Hybrid stars within a covariant, nonlocal chiral quark model}},}\ }\href {\doibase 10.1103/PhysRevC.75.065804} {\bibfield  {journal} {\bibinfo  {journal} {Phys. Rev. C}\ }\textbf {\bibinfo {volume} {75}},\ \bibinfo {pages} {065804} (\bibinfo {year} {2007})},\ \Eprint {http://arxiv.org/abs/nucl-th/0703088} {arXiv:nucl-th/0703088} \BibitemShut {NoStop}%
\bibitem [{\citenamefont {Kojo}\ \emph {et~al.}(2015)\citenamefont {Kojo}, \citenamefont {Powell}, \citenamefont {Song},\ and\ \citenamefont {Baym}}]{Kojo:2014rca}%
  \BibitemOpen
  \bibfield  {author} {\bibinfo {author} {\bibfnamefont {Toru}\ \bibnamefont {Kojo}}, \bibinfo {author} {\bibfnamefont {Philip~D.}\ \bibnamefont {Powell}}, \bibinfo {author} {\bibfnamefont {Yifan}\ \bibnamefont {Song}}, \ and\ \bibinfo {author} {\bibfnamefont {Gordon}\ \bibnamefont {Baym}},\ }\bibfield  {title} {\enquote {\bibinfo {title} {{Phenomenological QCD equation of state for massive neutron stars}},}\ }\href {\doibase 10.1103/PhysRevD.91.045003} {\bibfield  {journal} {\bibinfo  {journal} {Phys. Rev. D}\ }\textbf {\bibinfo {volume} {91}},\ \bibinfo {pages} {045003} (\bibinfo {year} {2015})},\ \Eprint {http://arxiv.org/abs/1412.1108} {arXiv:1412.1108 [hep-ph]} \BibitemShut {NoStop}%
\bibitem [{\citenamefont {Leonhardt}\ \emph {et~al.}(2020)\citenamefont {Leonhardt}, \citenamefont {Pospiech}, \citenamefont {Schallmo}, \citenamefont {Braun}, \citenamefont {Drischler}, \citenamefont {Hebeler},\ and\ \citenamefont {Schwenk}}]{Leonhardt:2019fua}%
  \BibitemOpen
  \bibfield  {author} {\bibinfo {author} {\bibfnamefont {M.}~\bibnamefont {Leonhardt}}, \bibinfo {author} {\bibfnamefont {M.}~\bibnamefont {Pospiech}}, \bibinfo {author} {\bibfnamefont {B.}~\bibnamefont {Schallmo}}, \bibinfo {author} {\bibfnamefont {J.}~\bibnamefont {Braun}}, \bibinfo {author} {\bibfnamefont {C.}~\bibnamefont {Drischler}}, \bibinfo {author} {\bibfnamefont {K.}~\bibnamefont {Hebeler}}, \ and\ \bibinfo {author} {\bibfnamefont {A.}~\bibnamefont {Schwenk}},\ }\bibfield  {title} {\enquote {\bibinfo {title} {{Symmetric nuclear matter from the strong interaction}},}\ }\href {\doibase 10.1103/PhysRevLett.125.142502} {\bibfield  {journal} {\bibinfo  {journal} {Phys. Rev. Lett.}\ }\textbf {\bibinfo {volume} {125}},\ \bibinfo {pages} {142502} (\bibinfo {year} {2020})},\ \Eprint {http://arxiv.org/abs/1907.05814} {arXiv:1907.05814 [nucl-th]} \BibitemShut {NoStop}%
\bibitem [{\citenamefont {Baym}\ \emph {et~al.}(2019)\citenamefont {Baym}, \citenamefont {Furusawa}, \citenamefont {Hatsuda}, \citenamefont {Kojo},\ and\ \citenamefont {Togashi}}]{Baym:2019iky}%
  \BibitemOpen
  \bibfield  {author} {\bibinfo {author} {\bibfnamefont {Gordon}\ \bibnamefont {Baym}}, \bibinfo {author} {\bibfnamefont {Shun}\ \bibnamefont {Furusawa}}, \bibinfo {author} {\bibfnamefont {Tetsuo}\ \bibnamefont {Hatsuda}}, \bibinfo {author} {\bibfnamefont {Toru}\ \bibnamefont {Kojo}}, \ and\ \bibinfo {author} {\bibfnamefont {Hajime}\ \bibnamefont {Togashi}},\ }\bibfield  {title} {\enquote {\bibinfo {title} {{New Neutron Star Equation of State with Quark-Hadron Crossover}},}\ }\href {\doibase 10.3847/1538-4357/ab441e} {\bibfield  {journal} {\bibinfo  {journal} {Astrophys. J.}\ }\textbf {\bibinfo {volume} {885}},\ \bibinfo {pages} {42} (\bibinfo {year} {2019})},\ \Eprint {http://arxiv.org/abs/1903.08963} {arXiv:1903.08963 [astro-ph.HE]} \BibitemShut {NoStop}%
\bibitem [{\citenamefont {Roupas}\ \emph {et~al.}(2021)\citenamefont {Roupas}, \citenamefont {Panotopoulos},\ and\ \citenamefont {Lopes}}]{Roupas:2020nua}%
  \BibitemOpen
  \bibfield  {author} {\bibinfo {author} {\bibfnamefont {Zacharias}\ \bibnamefont {Roupas}}, \bibinfo {author} {\bibfnamefont {Grigoris}\ \bibnamefont {Panotopoulos}}, \ and\ \bibinfo {author} {\bibfnamefont {Il\'\i{}dio}\ \bibnamefont {Lopes}},\ }\bibfield  {title} {\enquote {\bibinfo {title} {{QCD color superconductivity in compact stars: color-flavor locked quark star candidate for the gravitational-wave signal GW190814}},}\ }\href {\doibase 10.1103/PhysRevD.103.083015} {\bibfield  {journal} {\bibinfo  {journal} {Phys. Rev. D}\ }\textbf {\bibinfo {volume} {103}},\ \bibinfo {pages} {083015} (\bibinfo {year} {2021})},\ \Eprint {http://arxiv.org/abs/2010.11020} {arXiv:2010.11020 [astro-ph.HE]} \BibitemShut {NoStop}%
\bibitem [{\citenamefont {Malfatti}\ \emph {et~al.}(2020)\citenamefont {Malfatti}, \citenamefont {Orsaria}, \citenamefont {Ranea-Sandoval}, \citenamefont {Contrera},\ and\ \citenamefont {Weber}}]{Malfatti:2020onm}%
  \BibitemOpen
  \bibfield  {author} {\bibinfo {author} {\bibfnamefont {German}\ \bibnamefont {Malfatti}}, \bibinfo {author} {\bibfnamefont {Milva~G.}\ \bibnamefont {Orsaria}}, \bibinfo {author} {\bibfnamefont {Ignacio~F.}\ \bibnamefont {Ranea-Sandoval}}, \bibinfo {author} {\bibfnamefont {Gustavo~A.}\ \bibnamefont {Contrera}}, \ and\ \bibinfo {author} {\bibfnamefont {Fridolin}\ \bibnamefont {Weber}},\ }\bibfield  {title} {\enquote {\bibinfo {title} {{Delta baryons and diquark formation in the cores of neutron stars}},}\ }\href {\doibase 10.1103/PhysRevD.102.063008} {\bibfield  {journal} {\bibinfo  {journal} {Phys. Rev. D}\ }\textbf {\bibinfo {volume} {102}},\ \bibinfo {pages} {063008} (\bibinfo {year} {2020})},\ \Eprint {http://arxiv.org/abs/2008.06459} {arXiv:2008.06459 [astro-ph.HE]} \BibitemShut {NoStop}%
\bibitem [{\citenamefont {Ayriyan}\ \emph {et~al.}(2021)\citenamefont {Ayriyan}, \citenamefont {Blaschke}, \citenamefont {Grunfeld}, \citenamefont {Alvarez-Castillo}, \citenamefont {Grigorian},\ and\ \citenamefont {Abgaryan}}]{Ayriyan:2021prr}%
  \BibitemOpen
  \bibfield  {author} {\bibinfo {author} {\bibfnamefont {A.}~\bibnamefont {Ayriyan}}, \bibinfo {author} {\bibfnamefont {D.}~\bibnamefont {Blaschke}}, \bibinfo {author} {\bibfnamefont {A.~G.}\ \bibnamefont {Grunfeld}}, \bibinfo {author} {\bibfnamefont {D.}~\bibnamefont {Alvarez-Castillo}}, \bibinfo {author} {\bibfnamefont {H.}~\bibnamefont {Grigorian}}, \ and\ \bibinfo {author} {\bibfnamefont {V.}~\bibnamefont {Abgaryan}},\ }\bibfield  {title} {\enquote {\bibinfo {title} {{Bayesian analysis of multimessenger M-R data with interpolated hybrid EoS}},}\ }\href {\doibase 10.1140/epja/s10050-021-00619-0} {\bibfield  {journal} {\bibinfo  {journal} {Eur. Phys. J. A}\ }\textbf {\bibinfo {volume} {57}},\ \bibinfo {pages} {318} (\bibinfo {year} {2021})},\ \Eprint {http://arxiv.org/abs/2102.13485} {arXiv:2102.13485 [astro-ph.HE]} \BibitemShut {NoStop}%
\bibitem [{\citenamefont {Bitaghsir~Fadafan}\ \emph {et~al.}(2020)\citenamefont {Bitaghsir~Fadafan}, \citenamefont {Cruz~Rojas},\ and\ \citenamefont {Evans}}]{Fadafa:2019euu}%
  \BibitemOpen
  \bibfield  {author} {\bibinfo {author} {\bibfnamefont {Kazem}\ \bibnamefont {Bitaghsir~Fadafan}}, \bibinfo {author} {\bibfnamefont {Jes\'us}\ \bibnamefont {Cruz~Rojas}}, \ and\ \bibinfo {author} {\bibfnamefont {Nick}\ \bibnamefont {Evans}},\ }\bibfield  {title} {\enquote {\bibinfo {title} {{Deconfined, Massive Quark Phase at High Density and Compact Stars: A Holographic Study}},}\ }\href {\doibase 10.1103/PhysRevD.101.126005} {\bibfield  {journal} {\bibinfo  {journal} {Phys. Rev. D}\ }\textbf {\bibinfo {volume} {101}},\ \bibinfo {pages} {126005} (\bibinfo {year} {2020})},\ \Eprint {http://arxiv.org/abs/1911.12705} {arXiv:1911.12705 [hep-ph]} \BibitemShut {NoStop}%
\bibitem [{\citenamefont {Xia}\ \emph {et~al.}(2021)\citenamefont {Xia}, \citenamefont {Zhu}, \citenamefont {Zhou},\ and\ \citenamefont {Li}}]{Xia:2019xax}%
  \BibitemOpen
  \bibfield  {author} {\bibinfo {author} {\bibfnamefont {Chengjun}\ \bibnamefont {Xia}}, \bibinfo {author} {\bibfnamefont {Zhenyu}\ \bibnamefont {Zhu}}, \bibinfo {author} {\bibfnamefont {Xia}\ \bibnamefont {Zhou}}, \ and\ \bibinfo {author} {\bibfnamefont {Ang}\ \bibnamefont {Li}},\ }\bibfield  {title} {\enquote {\bibinfo {title} {{Sound velocity in dense stellar matter with strangeness and compact stars}},}\ }\href {\doibase 10.1088/1674-1137/abea0d} {\bibfield  {journal} {\bibinfo  {journal} {Chin. Phys. C}\ }\textbf {\bibinfo {volume} {45}},\ \bibinfo {pages} {055104} (\bibinfo {year} {2021})},\ \Eprint {http://arxiv.org/abs/1906.00826} {arXiv:1906.00826 [nucl-th]} \BibitemShut {NoStop}%
\bibitem [{\citenamefont {Stone}\ \emph {et~al.}(2021)\citenamefont {Stone}, \citenamefont {Dexheimer}, \citenamefont {Guichon}, \citenamefont {Thomas},\ and\ \citenamefont {Typel}}]{Stone:2019abq}%
  \BibitemOpen
  \bibfield  {author} {\bibinfo {author} {\bibfnamefont {J.~R.}\ \bibnamefont {Stone}}, \bibinfo {author} {\bibfnamefont {V.}~\bibnamefont {Dexheimer}}, \bibinfo {author} {\bibfnamefont {P.~A~M.}\ \bibnamefont {Guichon}}, \bibinfo {author} {\bibfnamefont {A.~W.}\ \bibnamefont {Thomas}}, \ and\ \bibinfo {author} {\bibfnamefont {S.}~\bibnamefont {Typel}},\ }\bibfield  {title} {\enquote {\bibinfo {title} {{Equation of state of hot dense hyperonic matter in the Quark\textendash{}Meson-Coupling (QMC-A) model}},}\ }\href {\doibase 10.1093/mnras/staa4006} {\bibfield  {journal} {\bibinfo  {journal} {Mon. Not. Roy. Astron. Soc.}\ }\textbf {\bibinfo {volume} {502}},\ \bibinfo {pages} {3476--3490} (\bibinfo {year} {2021})},\ \Eprint {http://arxiv.org/abs/1906.11100} {arXiv:1906.11100 [nucl-th]} \BibitemShut {NoStop}%
\bibitem [{\citenamefont {Li}\ \emph {et~al.}(2020)\citenamefont {Li}, \citenamefont {Zhu}, \citenamefont {Zhou}, \citenamefont {Dong}, \citenamefont {Hu},\ and\ \citenamefont {Xia}}]{Li:2020dst}%
  \BibitemOpen
  \bibfield  {author} {\bibinfo {author} {\bibfnamefont {A.}~\bibnamefont {Li}}, \bibinfo {author} {\bibfnamefont {Z.~Y.}\ \bibnamefont {Zhu}}, \bibinfo {author} {\bibfnamefont {E.~P.}\ \bibnamefont {Zhou}}, \bibinfo {author} {\bibfnamefont {J.~M.}\ \bibnamefont {Dong}}, \bibinfo {author} {\bibfnamefont {J.~N.}\ \bibnamefont {Hu}}, \ and\ \bibinfo {author} {\bibfnamefont {C.~J.}\ \bibnamefont {Xia}},\ }\bibfield  {title} {\enquote {\bibinfo {title} {{Neutron star equation of state: Quark mean-field (QMF) modeling and applications}},}\ }\href {\doibase 10.1016/j.jheap.2020.07.001} {\bibfield  {journal} {\bibinfo  {journal} {JHEAp}\ }\textbf {\bibinfo {volume} {28}},\ \bibinfo {pages} {19--46} (\bibinfo {year} {2020})},\ \Eprint {http://arxiv.org/abs/2007.05116} {arXiv:2007.05116 [nucl-th]} \BibitemShut {NoStop}%
\bibitem [{\citenamefont {Pisarski}(2021)}]{Pisarski:2021aoz}%
  \BibitemOpen
  \bibfield  {author} {\bibinfo {author} {\bibfnamefont {Robert~D.}\ \bibnamefont {Pisarski}},\ }\bibfield  {title} {\enquote {\bibinfo {title} {{Remarks on nuclear matter: How an $\omega_0$ condensate can spike the speed of sound, and a model of $Z(3)$ baryons}},}\ }\href {\doibase 10.1103/PhysRevD.103.L071504} {\bibfield  {journal} {\bibinfo  {journal} {Phys. Rev. D}\ }\textbf {\bibinfo {volume} {103}},\ \bibinfo {pages} {L071504} (\bibinfo {year} {2021})},\ \Eprint {http://arxiv.org/abs/2101.05813} {arXiv:2101.05813 [nucl-th]} \BibitemShut {NoStop}%
\bibitem [{\citenamefont {Pal}\ \emph {et~al.}(2022)\citenamefont {Pal}, \citenamefont {Kadam},\ and\ \citenamefont {Bhattacharyya}}]{Pal:2021qav}%
  \BibitemOpen
  \bibfield  {author} {\bibinfo {author} {\bibfnamefont {Somenath}\ \bibnamefont {Pal}}, \bibinfo {author} {\bibfnamefont {Guruprasad}\ \bibnamefont {Kadam}}, \ and\ \bibinfo {author} {\bibfnamefont {Abhijit}\ \bibnamefont {Bhattacharyya}},\ }\bibfield  {title} {\enquote {\bibinfo {title} {{Hadron resonance gas model with repulsive mean-field interactions: Specific heat, isothermal compressibility and speed of sound}},}\ }\href {\doibase 10.1016/j.nuclphysa.2022.122464} {\bibfield  {journal} {\bibinfo  {journal} {Nucl. Phys. A}\ }\textbf {\bibinfo {volume} {1023}},\ \bibinfo {pages} {122464} (\bibinfo {year} {2022})},\ \Eprint {http://arxiv.org/abs/2104.08531} {arXiv:2104.08531 [hep-ph]} \BibitemShut {NoStop}%
\bibitem [{\citenamefont {Motta}\ \emph {et~al.}(2021)\citenamefont {Motta}, \citenamefont {Guichon},\ and\ \citenamefont {Thomas}}]{Motta:2021xwo}%
  \BibitemOpen
  \bibfield  {author} {\bibinfo {author} {\bibfnamefont {T.~F.}\ \bibnamefont {Motta}}, \bibinfo {author} {\bibfnamefont {P.~A.~M.}\ \bibnamefont {Guichon}}, \ and\ \bibinfo {author} {\bibfnamefont {A.~W.}\ \bibnamefont {Thomas}},\ }\bibfield  {title} {\enquote {\bibinfo {title} {{On the sound speed in hyperonic stars}},}\ }\href {\doibase 10.1016/j.nuclphysa.2021.122157} {\bibfield  {journal} {\bibinfo  {journal} {Nucl. Phys. A}\ }\textbf {\bibinfo {volume} {1009}},\ \bibinfo {pages} {122157} (\bibinfo {year} {2021})},\ \Eprint {http://arxiv.org/abs/2009.10908} {arXiv:2009.10908 [nucl-th]} \BibitemShut {NoStop}%
\bibitem [{\citenamefont {Ma}\ and\ \citenamefont {Rho}(2021)}]{Ma:2021zev}%
  \BibitemOpen
  \bibfield  {author} {\bibinfo {author} {\bibfnamefont {Yong-Liang}\ \bibnamefont {Ma}}\ and\ \bibinfo {author} {\bibfnamefont {Mannque}\ \bibnamefont {Rho}},\ }\bibfield  {title} {\enquote {\bibinfo {title} {{The sound speed and core of massive compact stars: A manifestation of hadron-quark duality}},}\ }\href@noop {} {\  (\bibinfo {year} {2021})},\ \Eprint {http://arxiv.org/abs/2104.13822} {arXiv:2104.13822 [nucl-th]} \BibitemShut {NoStop}%
\bibitem [{\citenamefont {Hippert}\ \emph {et~al.}(2021)\citenamefont {Hippert}, \citenamefont {Fraga},\ and\ \citenamefont {Noronha}}]{Hippert:2021gfs}%
  \BibitemOpen
  \bibfield  {author} {\bibinfo {author} {\bibfnamefont {Maur\'\i{}cio}\ \bibnamefont {Hippert}}, \bibinfo {author} {\bibfnamefont {Eduardo~S.}\ \bibnamefont {Fraga}}, \ and\ \bibinfo {author} {\bibfnamefont {Jorge}\ \bibnamefont {Noronha}},\ }\bibfield  {title} {\enquote {\bibinfo {title} {{Insights on the peak in the speed of sound of ultradense matter}},}\ }\href {\doibase 10.1103/PhysRevD.104.034011} {\bibfield  {journal} {\bibinfo  {journal} {Phys. Rev. D}\ }\textbf {\bibinfo {volume} {104}},\ \bibinfo {pages} {034011} (\bibinfo {year} {2021})},\ \Eprint {http://arxiv.org/abs/2105.04535} {arXiv:2105.04535 [nucl-th]} \BibitemShut {NoStop}%
\bibitem [{\citenamefont {Hoyos}\ \emph {et~al.}(2020)\citenamefont {Hoyos}, \citenamefont {Jokela}, \citenamefont {Jarvinen}, \citenamefont {Subils}, \citenamefont {Tarrio},\ and\ \citenamefont {Vuorinen}}]{Hoyos:2020hmq}%
  \BibitemOpen
  \bibfield  {author} {\bibinfo {author} {\bibfnamefont {Carlos}\ \bibnamefont {Hoyos}}, \bibinfo {author} {\bibfnamefont {Niko}\ \bibnamefont {Jokela}}, \bibinfo {author} {\bibfnamefont {Matti}\ \bibnamefont {Jarvinen}}, \bibinfo {author} {\bibfnamefont {Javier~G.}\ \bibnamefont {Subils}}, \bibinfo {author} {\bibfnamefont {Javier}\ \bibnamefont {Tarrio}}, \ and\ \bibinfo {author} {\bibfnamefont {Aleksi}\ \bibnamefont {Vuorinen}},\ }\bibfield  {title} {\enquote {\bibinfo {title} {{Transport in strongly coupled quark matter}},}\ }\href {\doibase 10.1103/PhysRevLett.125.241601} {\bibfield  {journal} {\bibinfo  {journal} {Phys. Rev. Lett.}\ }\textbf {\bibinfo {volume} {125}},\ \bibinfo {pages} {241601} (\bibinfo {year} {2020})},\ \Eprint {http://arxiv.org/abs/2005.14205} {arXiv:2005.14205 [hep-th]} \BibitemShut {NoStop}%
\bibitem [{\citenamefont {Hoyos}\ \emph {et~al.}(2022)\citenamefont {Hoyos}, \citenamefont {Jokela}, \citenamefont {J\"arvinen}, \citenamefont {Subils}, \citenamefont {Tarrio},\ and\ \citenamefont {Vuorinen}}]{Hoyos:2021njg}%
  \BibitemOpen
  \bibfield  {author} {\bibinfo {author} {\bibfnamefont {Carlos}\ \bibnamefont {Hoyos}}, \bibinfo {author} {\bibfnamefont {Niko}\ \bibnamefont {Jokela}}, \bibinfo {author} {\bibfnamefont {Matti}\ \bibnamefont {J\"arvinen}}, \bibinfo {author} {\bibfnamefont {Javier~G.}\ \bibnamefont {Subils}}, \bibinfo {author} {\bibfnamefont {Javier}\ \bibnamefont {Tarrio}}, \ and\ \bibinfo {author} {\bibfnamefont {Aleksi}\ \bibnamefont {Vuorinen}},\ }\bibfield  {title} {\enquote {\bibinfo {title} {{Holographic approach to transport in dense QCD matter}},}\ }\href {\doibase 10.1103/PhysRevD.105.066014} {\bibfield  {journal} {\bibinfo  {journal} {Phys. Rev. D}\ }\textbf {\bibinfo {volume} {105}},\ \bibinfo {pages} {066014} (\bibinfo {year} {2022})},\ \Eprint {http://arxiv.org/abs/2109.12122} {arXiv:2109.12122 [hep-th]} \BibitemShut {NoStop}%
\bibitem [{\citenamefont {Israel}\ and\ \citenamefont {Stewart}(1979)}]{Israel:1979wp}%
  \BibitemOpen
  \bibfield  {author} {\bibinfo {author} {\bibfnamefont {W.}~\bibnamefont {Israel}}\ and\ \bibinfo {author} {\bibfnamefont {J.~M.}\ \bibnamefont {Stewart}},\ }\bibfield  {title} {\enquote {\bibinfo {title} {{Transient relativistic thermodynamics and kinetic theory}},}\ }\href {\doibase 10.1016/0003-4916(79)90130-1} {\bibfield  {journal} {\bibinfo  {journal} {Annals Phys.}\ }\textbf {\bibinfo {volume} {118}},\ \bibinfo {pages} {341--372} (\bibinfo {year} {1979})}\BibitemShut {NoStop}%
\bibitem [{\citenamefont {Kovtun}\ \emph {et~al.}(2011)\citenamefont {Kovtun}, \citenamefont {Moore},\ and\ \citenamefont {Romatschke}}]{Kovtun:2011np}%
  \BibitemOpen
  \bibfield  {author} {\bibinfo {author} {\bibfnamefont {Pavel}\ \bibnamefont {Kovtun}}, \bibinfo {author} {\bibfnamefont {Guy~D.}\ \bibnamefont {Moore}}, \ and\ \bibinfo {author} {\bibfnamefont {Paul}\ \bibnamefont {Romatschke}},\ }\bibfield  {title} {\enquote {\bibinfo {title} {{The stickiness of sound: An absolute lower limit on viscosity and the breakdown of second order relativistic hydrodynamics}},}\ }\href {\doibase 10.1103/PhysRevD.84.025006} {\bibfield  {journal} {\bibinfo  {journal} {Phys. Rev. D}\ }\textbf {\bibinfo {volume} {84}},\ \bibinfo {pages} {025006} (\bibinfo {year} {2011})},\ \Eprint {http://arxiv.org/abs/1104.1586} {arXiv:1104.1586 [hep-ph]} \BibitemShut {NoStop}%
\bibitem [{\citenamefont {Schmitt}(2015)}]{Schmitt:2014eka}%
  \BibitemOpen
  \bibfield  {author} {\bibinfo {author} {\bibfnamefont {Andreas}\ \bibnamefont {Schmitt}},\ }\href {\doibase 10.1007/978-3-319-07947-9} {\emph {\bibinfo {title} {{Introduction to Superfluidity}: {Field-theoretical approach and applications}}}},\ Vol.\ \bibinfo {volume} {888}\ (\bibinfo {year} {2015})\ \Eprint {http://arxiv.org/abs/1404.1284} {arXiv:1404.1284 [hep-ph]} \BibitemShut {NoStop}%
\bibitem [{\citenamefont {Gavassino}(2022)}]{Gavassino:2022zzz}%
  \BibitemOpen
  \bibfield  {author} {\bibinfo {author} {\bibfnamefont {Lorenzo}\ \bibnamefont {Gavassino}},\ }\bibfield  {title} {\enquote {\bibinfo {title} {{Stability and causality of Carter\textquoteright{}s multifluid theory}},}\ }\href {\doibase 10.1088/1361-6382/ac79f4} {\bibfield  {journal} {\bibinfo  {journal} {Class. Quant. Grav.}\ }\textbf {\bibinfo {volume} {39}},\ \bibinfo {pages} {185008} (\bibinfo {year} {2022})},\ \Eprint {http://arxiv.org/abs/2202.06760} {arXiv:2202.06760 [gr-qc]} \BibitemShut {NoStop}%
\bibitem [{\citenamefont {Son}\ and\ \citenamefont {Stephanov}(2001)}]{Son:2000xc}%
  \BibitemOpen
  \bibfield  {author} {\bibinfo {author} {\bibfnamefont {D.~T.}\ \bibnamefont {Son}}\ and\ \bibinfo {author} {\bibfnamefont {Misha~A.}\ \bibnamefont {Stephanov}},\ }\bibfield  {title} {\enquote {\bibinfo {title} {{QCD at finite isospin density}},}\ }\href {\doibase 10.1103/PhysRevLett.86.592} {\bibfield  {journal} {\bibinfo  {journal} {Phys. Rev. Lett.}\ }\textbf {\bibinfo {volume} {86}},\ \bibinfo {pages} {592--595} (\bibinfo {year} {2001})},\ \Eprint {http://arxiv.org/abs/hep-ph/0005225} {arXiv:hep-ph/0005225} \BibitemShut {NoStop}%
\bibitem [{\citenamefont {Kogut}\ and\ \citenamefont {Sinclair}(2002)}]{Kogut:2002zg}%
  \BibitemOpen
  \bibfield  {author} {\bibinfo {author} {\bibfnamefont {J.~B.}\ \bibnamefont {Kogut}}\ and\ \bibinfo {author} {\bibfnamefont {D.~K.}\ \bibnamefont {Sinclair}},\ }\bibfield  {title} {\enquote {\bibinfo {title} {{Lattice QCD at finite isospin density at zero and finite temperature}},}\ }\href {\doibase 10.1103/PhysRevD.66.034505} {\bibfield  {journal} {\bibinfo  {journal} {Phys. Rev. D}\ }\textbf {\bibinfo {volume} {66}},\ \bibinfo {pages} {034505} (\bibinfo {year} {2002})},\ \Eprint {http://arxiv.org/abs/hep-lat/0202028} {arXiv:hep-lat/0202028} \BibitemShut {NoStop}%
\bibitem [{\citenamefont {Landau}\ and\ \citenamefont {Lifshitz}(1987)}]{LandauLifshitzFluids}%
  \BibitemOpen
  \bibfield  {author} {\bibinfo {author} {\bibfnamefont {L.~D.}\ \bibnamefont {Landau}}\ and\ \bibinfo {author} {\bibfnamefont {E.~M.}\ \bibnamefont {Lifshitz}},\ }\href@noop {} {\emph {\bibinfo {title} {Fluid Mechanics - Volume 6 (Corse of Theoretical Physics)}}},\ \bibinfo {edition} {2nd}\ ed.\ (\bibinfo  {publisher} {Butterworth-Heinemann},\ \bibinfo {year} {1987})\ p.\ \bibinfo {pages} {552}\BibitemShut {NoStop}%
\bibitem [{\citenamefont {Bemfica}\ \emph {et~al.}(2018)\citenamefont {Bemfica}, \citenamefont {Disconzi},\ and\ \citenamefont {Noronha}}]{Bemfica:2017wps}%
  \BibitemOpen
  \bibfield  {author} {\bibinfo {author} {\bibfnamefont {F\'abio~S.}\ \bibnamefont {Bemfica}}, \bibinfo {author} {\bibfnamefont {Marcelo~M.}\ \bibnamefont {Disconzi}}, \ and\ \bibinfo {author} {\bibfnamefont {Jorge}\ \bibnamefont {Noronha}},\ }\bibfield  {title} {\enquote {\bibinfo {title} {{Causality and existence of solutions of relativistic viscous fluid dynamics with gravity}},}\ }\href {\doibase 10.1103/PhysRevD.98.104064} {\bibfield  {journal} {\bibinfo  {journal} {Phys. Rev. D}\ }\textbf {\bibinfo {volume} {98}},\ \bibinfo {pages} {104064} (\bibinfo {year} {2018})},\ \Eprint {http://arxiv.org/abs/1708.06255} {arXiv:1708.06255 [gr-qc]} \BibitemShut {NoStop}%
\bibitem [{\citenamefont {Kovtun}(2019)}]{Kovtun:2019hdm}%
  \BibitemOpen
  \bibfield  {author} {\bibinfo {author} {\bibfnamefont {Pavel}\ \bibnamefont {Kovtun}},\ }\bibfield  {title} {\enquote {\bibinfo {title} {{First-order relativistic hydrodynamics is stable}},}\ }\href {\doibase 10.1007/JHEP10(2019)034} {\bibfield  {journal} {\bibinfo  {journal} {JHEP}\ }\textbf {\bibinfo {volume} {10}},\ \bibinfo {pages} {034} (\bibinfo {year} {2019})},\ \Eprint {http://arxiv.org/abs/1907.08191} {arXiv:1907.08191 [hep-th]} \BibitemShut {NoStop}%
\bibitem [{\citenamefont {Bemfica}\ \emph {et~al.}(2019)\citenamefont {Bemfica}, \citenamefont {Bemfica}, \citenamefont {Disconzi}, \citenamefont {Disconzi}, \citenamefont {Noronha},\ and\ \citenamefont {Noronha}}]{Bemfica:2019knx}%
  \BibitemOpen
  \bibfield  {author} {\bibinfo {author} {\bibfnamefont {F\'abio~S.}\ \bibnamefont {Bemfica}}, \bibinfo {author} {\bibfnamefont {F\'abio~S.}\ \bibnamefont {Bemfica}}, \bibinfo {author} {\bibfnamefont {Marcelo~M.}\ \bibnamefont {Disconzi}}, \bibinfo {author} {\bibfnamefont {Marcelo~M.}\ \bibnamefont {Disconzi}}, \bibinfo {author} {\bibfnamefont {Jorge}\ \bibnamefont {Noronha}}, \ and\ \bibinfo {author} {\bibfnamefont {Jorge}\ \bibnamefont {Noronha}},\ }\bibfield  {title} {\enquote {\bibinfo {title} {{Nonlinear Causality of General First-Order Relativistic Viscous Hydrodynamics}},}\ }\href {\doibase 10.1103/PhysRevD.100.104020} {\bibfield  {journal} {\bibinfo  {journal} {Phys. Rev. D}\ }\textbf {\bibinfo {volume} {100}},\ \bibinfo {pages} {104020} (\bibinfo {year} {2019})},\ \bibinfo {note} {[Erratum: Phys.Rev.D 105, 069902 (2022)]},\ \Eprint {http://arxiv.org/abs/1907.12695} {arXiv:1907.12695 [gr-qc]} \BibitemShut {NoStop}%
\bibitem [{\citenamefont {Hoult}\ and\ \citenamefont {Kovtun}(2020)}]{Hoult:2020eho}%
  \BibitemOpen
  \bibfield  {author} {\bibinfo {author} {\bibfnamefont {Raphael~E.}\ \bibnamefont {Hoult}}\ and\ \bibinfo {author} {\bibfnamefont {Pavel}\ \bibnamefont {Kovtun}},\ }\bibfield  {title} {\enquote {\bibinfo {title} {{Stable and causal relativistic Navier-Stokes equations}},}\ }\href {\doibase 10.1007/JHEP06(2020)067} {\bibfield  {journal} {\bibinfo  {journal} {JHEP}\ }\textbf {\bibinfo {volume} {06}},\ \bibinfo {pages} {067} (\bibinfo {year} {2020})},\ \Eprint {http://arxiv.org/abs/2004.04102} {arXiv:2004.04102 [hep-th]} \BibitemShut {NoStop}%
\bibitem [{\citenamefont {Bemfica}\ \emph {et~al.}(2022)\citenamefont {Bemfica}, \citenamefont {Disconzi},\ and\ \citenamefont {Noronha}}]{Bemfica:2020zjp}%
  \BibitemOpen
  \bibfield  {author} {\bibinfo {author} {\bibfnamefont {Fabio~S.}\ \bibnamefont {Bemfica}}, \bibinfo {author} {\bibfnamefont {Marcelo~M.}\ \bibnamefont {Disconzi}}, \ and\ \bibinfo {author} {\bibfnamefont {Jorge}\ \bibnamefont {Noronha}},\ }\bibfield  {title} {\enquote {\bibinfo {title} {{First-Order General-Relativistic Viscous Fluid Dynamics}},}\ }\href {\doibase 10.1103/PhysRevX.12.021044} {\bibfield  {journal} {\bibinfo  {journal} {Phys. Rev. X}\ }\textbf {\bibinfo {volume} {12}},\ \bibinfo {pages} {021044} (\bibinfo {year} {2022})},\ \Eprint {http://arxiv.org/abs/2009.11388} {arXiv:2009.11388 [gr-qc]} \BibitemShut {NoStop}%
\bibitem [{\citenamefont {Heller}\ \emph {et~al.}(2023{\natexlab{a}})\citenamefont {Heller}, \citenamefont {Serantes}, \citenamefont {Spali\'nski},\ and\ \citenamefont {Withers}}]{Heller:2022ejw}%
  \BibitemOpen
  \bibfield  {author} {\bibinfo {author} {\bibfnamefont {Michal~P.}\ \bibnamefont {Heller}}, \bibinfo {author} {\bibfnamefont {Alexandre}\ \bibnamefont {Serantes}}, \bibinfo {author} {\bibfnamefont {Micha\l{}}\ \bibnamefont {Spali\'nski}}, \ and\ \bibinfo {author} {\bibfnamefont {Benjamin}\ \bibnamefont {Withers}},\ }\bibfield  {title} {\enquote {\bibinfo {title} {{Rigorous Bounds on Transport from Causality}},}\ }\href {\doibase 10.1103/PhysRevLett.130.261601} {\bibfield  {journal} {\bibinfo  {journal} {Phys. Rev. Lett.}\ }\textbf {\bibinfo {volume} {130}},\ \bibinfo {pages} {261601} (\bibinfo {year} {2023}{\natexlab{a}})},\ \Eprint {http://arxiv.org/abs/2212.07434} {arXiv:2212.07434 [hep-th]} \BibitemShut {NoStop}%
\bibitem [{\citenamefont {Heller}\ \emph {et~al.}(2023{\natexlab{b}})\citenamefont {Heller}, \citenamefont {Serantes}, \citenamefont {Spali\'nski},\ and\ \citenamefont {Withers}}]{Heller:2023jtd}%
  \BibitemOpen
  \bibfield  {author} {\bibinfo {author} {\bibfnamefont {Michal~P.}\ \bibnamefont {Heller}}, \bibinfo {author} {\bibfnamefont {Alexandre}\ \bibnamefont {Serantes}}, \bibinfo {author} {\bibfnamefont {Micha\l{}}\ \bibnamefont {Spali\'nski}}, \ and\ \bibinfo {author} {\bibfnamefont {Benjamin}\ \bibnamefont {Withers}},\ }\bibfield  {title} {\enquote {\bibinfo {title} {{The Hydrohedron: Bootstrapping Relativistic Hydrodynamics}},}\ }\href@noop {} {\  (\bibinfo {year} {2023}{\natexlab{b}})},\ \Eprint {http://arxiv.org/abs/2305.07703} {arXiv:2305.07703 [hep-th]} \BibitemShut {NoStop}%
\bibitem [{\citenamefont {Romatschke}\ and\ \citenamefont {Romatschke}(2019)}]{Romatschke:2017ejr}%
  \BibitemOpen
  \bibfield  {author} {\bibinfo {author} {\bibfnamefont {Paul}\ \bibnamefont {Romatschke}}\ and\ \bibinfo {author} {\bibfnamefont {Ulrike}\ \bibnamefont {Romatschke}},\ }\href {\doibase 10.1017/9781108651998} {\emph {\bibinfo {title} {{Relativistic Fluid Dynamics In and Out of Equilibrium}}}},\ Cambridge Monographs on Mathematical Physics\ (\bibinfo  {publisher} {Cambridge University Press},\ \bibinfo {year} {2019})\ \Eprint {http://arxiv.org/abs/1712.05815} {arXiv:1712.05815 [nucl-th]} \BibitemShut {NoStop}%
\bibitem [{\citenamefont {York}\ and\ \citenamefont {Moore}(2009)}]{York:2008rr}%
  \BibitemOpen
  \bibfield  {author} {\bibinfo {author} {\bibfnamefont {Mark~Abraao}\ \bibnamefont {York}}\ and\ \bibinfo {author} {\bibfnamefont {Guy~D.}\ \bibnamefont {Moore}},\ }\bibfield  {title} {\enquote {\bibinfo {title} {{Second order hydrodynamic coefficients from kinetic theory}},}\ }\href {\doibase 10.1103/PhysRevD.79.054011} {\bibfield  {journal} {\bibinfo  {journal} {Phys. Rev. D}\ }\textbf {\bibinfo {volume} {79}},\ \bibinfo {pages} {054011} (\bibinfo {year} {2009})},\ \Eprint {http://arxiv.org/abs/0811.0729} {arXiv:0811.0729 [hep-ph]} \BibitemShut {NoStop}%
\bibitem [{\citenamefont {Jaiswal}\ \emph {et~al.}(2014)\citenamefont {Jaiswal}, \citenamefont {Ryblewski},\ and\ \citenamefont {Strickland}}]{Jaiswal:2014isa}%
  \BibitemOpen
  \bibfield  {author} {\bibinfo {author} {\bibfnamefont {Amaresh}\ \bibnamefont {Jaiswal}}, \bibinfo {author} {\bibfnamefont {Radoslaw}\ \bibnamefont {Ryblewski}}, \ and\ \bibinfo {author} {\bibfnamefont {Michael}\ \bibnamefont {Strickland}},\ }\bibfield  {title} {\enquote {\bibinfo {title} {{Transport coefficients for bulk viscous evolution in the relaxation time approximation}},}\ }\href {\doibase 10.1103/PhysRevC.90.044908} {\bibfield  {journal} {\bibinfo  {journal} {Phys. Rev. C}\ }\textbf {\bibinfo {volume} {90}},\ \bibinfo {pages} {044908} (\bibinfo {year} {2014})},\ \Eprint {http://arxiv.org/abs/1407.7231} {arXiv:1407.7231 [hep-ph]} \BibitemShut {NoStop}%
\bibitem [{\citenamefont {Gavassino}(2021)}]{Gavassino:2021cli}%
  \BibitemOpen
  \bibfield  {author} {\bibinfo {author} {\bibfnamefont {Lorenzo}\ \bibnamefont {Gavassino}},\ }\bibfield  {title} {\enquote {\bibinfo {title} {{Applying the Gibbs stability criterion to relativistic hydrodynamics}},}\ }\href {\doibase 10.1088/1361-6382/ac2b0e} {\bibfield  {journal} {\bibinfo  {journal} {Class. Quant. Grav.}\ }\textbf {\bibinfo {volume} {38}},\ \bibinfo {pages} {21LT02} (\bibinfo {year} {2021})},\ \Eprint {http://arxiv.org/abs/2104.09142} {arXiv:2104.09142 [gr-qc]} \BibitemShut {NoStop}%
\bibitem [{\citenamefont {Gavassino}\ \emph {et~al.}(2022)\citenamefont {Gavassino}, \citenamefont {Antonelli},\ and\ \citenamefont {Haskell}}]{Gavassino:2021kjm}%
  \BibitemOpen
  \bibfield  {author} {\bibinfo {author} {\bibfnamefont {Lorenzo}\ \bibnamefont {Gavassino}}, \bibinfo {author} {\bibfnamefont {Marco}\ \bibnamefont {Antonelli}}, \ and\ \bibinfo {author} {\bibfnamefont {Brynmor}\ \bibnamefont {Haskell}},\ }\bibfield  {title} {\enquote {\bibinfo {title} {{Thermodynamic Stability Implies Causality}},}\ }\href {\doibase 10.1103/PhysRevLett.128.010606} {\bibfield  {journal} {\bibinfo  {journal} {Phys. Rev. Lett.}\ }\textbf {\bibinfo {volume} {128}},\ \bibinfo {pages} {010606} (\bibinfo {year} {2022})},\ \Eprint {http://arxiv.org/abs/2105.14621} {arXiv:2105.14621 [gr-qc]} \BibitemShut {NoStop}%
\bibitem [{\citenamefont {Hiscock}\ and\ \citenamefont {Lindblom}(1983)}]{Hiscock:1983zz}%
  \BibitemOpen
  \bibfield  {author} {\bibinfo {author} {\bibfnamefont {W.~A.}\ \bibnamefont {Hiscock}}\ and\ \bibinfo {author} {\bibfnamefont {L.}~\bibnamefont {Lindblom}},\ }\bibfield  {title} {\enquote {\bibinfo {title} {{Stability and causality in dissipative relativistic fluids}},}\ }\href {\doibase 10.1016/0003-4916(83)90288-9} {\bibfield  {journal} {\bibinfo  {journal} {Annals Phys.}\ }\textbf {\bibinfo {volume} {151}},\ \bibinfo {pages} {466--496} (\bibinfo {year} {1983})}\BibitemShut {NoStop}%
\bibitem [{\citenamefont {Olson}(1990)}]{Olson:1990rzl}%
  \BibitemOpen
  \bibfield  {author} {\bibinfo {author} {\bibfnamefont {Timothy~S.}\ \bibnamefont {Olson}},\ }\bibfield  {title} {\enquote {\bibinfo {title} {{STABILITY AND CAUSALITY IN THE ISRAEL-STEWART ENERGY FRAME THEORY}},}\ }\href {\doibase 10.1016/0003-4916(90)90366-V} {\bibfield  {journal} {\bibinfo  {journal} {Annals Phys.}\ }\textbf {\bibinfo {volume} {199}},\ \bibinfo {pages} {18} (\bibinfo {year} {1990})}\BibitemShut {NoStop}%
\bibitem [{\citenamefont {Romatschke}(2010)}]{Romatschke:2009im}%
  \BibitemOpen
  \bibfield  {author} {\bibinfo {author} {\bibfnamefont {Paul}\ \bibnamefont {Romatschke}},\ }\bibfield  {title} {\enquote {\bibinfo {title} {{New Developments in Relativistic Viscous Hydrodynamics}},}\ }\href {\doibase 10.1142/S0218301310014613} {\bibfield  {journal} {\bibinfo  {journal} {Int. J. Mod. Phys. E}\ }\textbf {\bibinfo {volume} {19}},\ \bibinfo {pages} {1--53} (\bibinfo {year} {2010})},\ \Eprint {http://arxiv.org/abs/0902.3663} {arXiv:0902.3663 [hep-ph]} \BibitemShut {NoStop}%
\bibitem [{\citenamefont {Landau}\ and\ \citenamefont {Lifshitz}(2013)}]{landau5}%
  \BibitemOpen
  \bibfield  {author} {\bibinfo {author} {\bibfnamefont {L.D.}\ \bibnamefont {Landau}}\ and\ \bibinfo {author} {\bibfnamefont {E.M.}\ \bibnamefont {Lifshitz}},\ }\href {https://books.google.pl/books?id=VzgJN-XPTRsC} {\emph {\bibinfo {title} {Statistical Physics}}},\ \bibinfo {number} {v. 5}\ (\bibinfo  {publisher} {Elsevier Science},\ \bibinfo {year} {2013})\BibitemShut {NoStop}%
\bibitem [{\citenamefont {Frenkel}(1955)}]{Frenkel:106808}%
  \BibitemOpen
  \bibfield  {author} {\bibinfo {author} {\bibfnamefont {Yakovich~Ilich}\ \bibnamefont {Frenkel}},\ }\href {https://cds.cern.ch/record/106808} {\emph {\bibinfo {title} {{Kinetic theory of liquids}}}}\ (\bibinfo  {publisher} {Dover},\ \bibinfo {address} {New York, NY},\ \bibinfo {year} {1955})\BibitemShut {NoStop}%
\bibitem [{\citenamefont {Trachenko}\ and\ \citenamefont {Brazhkin}(2015)}]{trachenko2015collective}%
  \BibitemOpen
  \bibfield  {author} {\bibinfo {author} {\bibfnamefont {K}~\bibnamefont {Trachenko}}\ and\ \bibinfo {author} {\bibfnamefont {VV}~\bibnamefont {Brazhkin}},\ }\bibfield  {title} {\enquote {\bibinfo {title} {Collective modes and thermodynamics of the liquid state},}\ }\href@noop {} {\bibfield  {journal} {\bibinfo  {journal} {Reports on Progress in Physics}\ }\textbf {\bibinfo {volume} {79}},\ \bibinfo {pages} {016502} (\bibinfo {year} {2015})}\BibitemShut {NoStop}%
\bibitem [{\citenamefont {Baggioli}\ \emph {et~al.}(2020)\citenamefont {Baggioli}, \citenamefont {Vasin}, \citenamefont {Brazhkin},\ and\ \citenamefont {Trachenko}}]{Baggioli:2019jcm}%
  \BibitemOpen
  \bibfield  {author} {\bibinfo {author} {\bibfnamefont {Matteo}\ \bibnamefont {Baggioli}}, \bibinfo {author} {\bibfnamefont {Mikhail}\ \bibnamefont {Vasin}}, \bibinfo {author} {\bibfnamefont {Vadim~V.}\ \bibnamefont {Brazhkin}}, \ and\ \bibinfo {author} {\bibfnamefont {Kostya}\ \bibnamefont {Trachenko}},\ }\bibfield  {title} {\enquote {\bibinfo {title} {{Gapped momentum states}},}\ }\href {\doibase 10.1016/j.physrep.2020.04.002} {\bibfield  {journal} {\bibinfo  {journal} {Phys. Rept.}\ }\textbf {\bibinfo {volume} {865}},\ \bibinfo {pages} {1--44} (\bibinfo {year} {2020})},\ \Eprint {http://arxiv.org/abs/1904.01419} {arXiv:1904.01419 [cond-mat.stat-mech]} \BibitemShut {NoStop}%
\bibitem [{\citenamefont {Rocha}\ \emph {et~al.}(2023)\citenamefont {Rocha}, \citenamefont {Wagner}, \citenamefont {Denicol}, \citenamefont {Noronha},\ and\ \citenamefont {Rischke}}]{Rocha:2023ilf}%
  \BibitemOpen
  \bibfield  {author} {\bibinfo {author} {\bibfnamefont {Gabriel~S.}\ \bibnamefont {Rocha}}, \bibinfo {author} {\bibfnamefont {David}\ \bibnamefont {Wagner}}, \bibinfo {author} {\bibfnamefont {Gabriel~S.}\ \bibnamefont {Denicol}}, \bibinfo {author} {\bibfnamefont {Jorge}\ \bibnamefont {Noronha}}, \ and\ \bibinfo {author} {\bibfnamefont {Dirk~H.}\ \bibnamefont {Rischke}},\ }\bibfield  {title} {\enquote {\bibinfo {title} {{Theories of Relativistic Dissipative Fluid Dynamics}},}\ }\href@noop {} {\  (\bibinfo {year} {2023})},\ \Eprint {http://arxiv.org/abs/2311.15063} {arXiv:2311.15063 [nucl-th]} \BibitemShut {NoStop}%
\bibitem [{\citenamefont {Detmold}\ \emph {et~al.}(2012)\citenamefont {Detmold}, \citenamefont {Orginos},\ and\ \citenamefont {Shi}}]{Detmold:2012wc}%
  \BibitemOpen
  \bibfield  {author} {\bibinfo {author} {\bibfnamefont {William}\ \bibnamefont {Detmold}}, \bibinfo {author} {\bibfnamefont {Kostas}\ \bibnamefont {Orginos}}, \ and\ \bibinfo {author} {\bibfnamefont {Zhifeng}\ \bibnamefont {Shi}},\ }\bibfield  {title} {\enquote {\bibinfo {title} {{Lattice QCD at non-zero isospin chemical potential}},}\ }\href {\doibase 10.1103/PhysRevD.86.054507} {\bibfield  {journal} {\bibinfo  {journal} {Phys. Rev. D}\ }\textbf {\bibinfo {volume} {86}},\ \bibinfo {pages} {054507} (\bibinfo {year} {2012})},\ \Eprint {http://arxiv.org/abs/1205.4224} {arXiv:1205.4224 [hep-lat]} \BibitemShut {NoStop}%
\bibitem [{\citenamefont {Abbott}\ \emph {et~al.}(2023)\citenamefont {Abbott}, \citenamefont {Detmold}, \citenamefont {Romero-L\'opez}, \citenamefont {Davoudi}, \citenamefont {Illa}, \citenamefont {Parre\~no}, \citenamefont {Perry}, \citenamefont {Shanahan},\ and\ \citenamefont {Wagman}}]{Abbott:2023coj}%
  \BibitemOpen
  \bibfield  {author} {\bibinfo {author} {\bibfnamefont {Ryan}\ \bibnamefont {Abbott}}, \bibinfo {author} {\bibfnamefont {William}\ \bibnamefont {Detmold}}, \bibinfo {author} {\bibfnamefont {Fernando}\ \bibnamefont {Romero-L\'opez}}, \bibinfo {author} {\bibfnamefont {Zohreh}\ \bibnamefont {Davoudi}}, \bibinfo {author} {\bibfnamefont {Marc}\ \bibnamefont {Illa}}, \bibinfo {author} {\bibfnamefont {Assumpta}\ \bibnamefont {Parre\~no}}, \bibinfo {author} {\bibfnamefont {Robert~J.}\ \bibnamefont {Perry}}, \bibinfo {author} {\bibfnamefont {Phiala~E.}\ \bibnamefont {Shanahan}}, \ and\ \bibinfo {author} {\bibfnamefont {Michael~L.}\ \bibnamefont {Wagman}} (\bibinfo {collaboration} {NPLQCD}),\ }\bibfield  {title} {\enquote {\bibinfo {title} {{Lattice quantum chromodynamics at large isospin density}},}\ }\href {\doibase 10.1103/PhysRevD.108.114506} {\bibfield  {journal} {\bibinfo  {journal} {Phys. Rev. D}\ }\textbf {\bibinfo {volume} {108}},\ \bibinfo {pages} {114506} (\bibinfo {year} {2023})},\ \Eprint
  {http://arxiv.org/abs/2307.15014} {arXiv:2307.15014 [hep-lat]} \BibitemShut {NoStop}%
\bibitem [{\citenamefont {Brandt}\ \emph {et~al.}(2018)\citenamefont {Brandt}, \citenamefont {Endrodi}, \citenamefont {Fraga}, \citenamefont {Hippert}, \citenamefont {Schaffner-Bielich},\ and\ \citenamefont {Schmalzbauer}}]{Brandt:2018bwq}%
  \BibitemOpen
  \bibfield  {author} {\bibinfo {author} {\bibfnamefont {Bastian~B.}\ \bibnamefont {Brandt}}, \bibinfo {author} {\bibfnamefont {Gergely}\ \bibnamefont {Endrodi}}, \bibinfo {author} {\bibfnamefont {Eduardo~S.}\ \bibnamefont {Fraga}}, \bibinfo {author} {\bibfnamefont {Mauricio}\ \bibnamefont {Hippert}}, \bibinfo {author} {\bibfnamefont {Jurgen}\ \bibnamefont {Schaffner-Bielich}}, \ and\ \bibinfo {author} {\bibfnamefont {Sebastian}\ \bibnamefont {Schmalzbauer}},\ }\bibfield  {title} {\enquote {\bibinfo {title} {{New class of compact stars: Pion stars}},}\ }\href {\doibase 10.1103/PhysRevD.98.094510} {\bibfield  {journal} {\bibinfo  {journal} {Phys. Rev. D}\ }\textbf {\bibinfo {volume} {98}},\ \bibinfo {pages} {094510} (\bibinfo {year} {2018})},\ \Eprint {http://arxiv.org/abs/1802.06685} {arXiv:1802.06685 [hep-ph]} \BibitemShut {NoStop}%
\bibitem [{\citenamefont {Brandt}\ \emph {et~al.}(2023)\citenamefont {Brandt}, \citenamefont {Cuteri},\ and\ \citenamefont {Endrodi}}]{Brandt:2022hwy}%
  \BibitemOpen
  \bibfield  {author} {\bibinfo {author} {\bibfnamefont {Bastian~B.}\ \bibnamefont {Brandt}}, \bibinfo {author} {\bibfnamefont {Francesca}\ \bibnamefont {Cuteri}}, \ and\ \bibinfo {author} {\bibfnamefont {Gergely}\ \bibnamefont {Endrodi}},\ }\bibfield  {title} {\enquote {\bibinfo {title} {{Equation of state and speed of sound of isospin-asymmetric QCD on the lattice}},}\ }\href {\doibase 10.1007/JHEP07(2023)055} {\bibfield  {journal} {\bibinfo  {journal} {JHEP}\ }\textbf {\bibinfo {volume} {07}},\ \bibinfo {pages} {055} (\bibinfo {year} {2023})},\ \Eprint {http://arxiv.org/abs/2212.14016} {arXiv:2212.14016 [hep-lat]} \BibitemShut {NoStop}%
\bibitem [{\citenamefont {Andersen}\ \emph {et~al.}(2023)\citenamefont {Andersen}, \citenamefont {Yu},\ and\ \citenamefont {Zhou}}]{Andersen:2023ivj}%
  \BibitemOpen
  \bibfield  {author} {\bibinfo {author} {\bibfnamefont {Jens~O.}\ \bibnamefont {Andersen}}, \bibinfo {author} {\bibfnamefont {Qing}\ \bibnamefont {Yu}}, \ and\ \bibinfo {author} {\bibfnamefont {Hua}\ \bibnamefont {Zhou}},\ }\bibfield  {title} {\enquote {\bibinfo {title} {{Pion condensation in dense QCD, the dilute Bose gas, and speedy Goldstone bosons}},}\ }\href@noop {} {\  (\bibinfo {year} {2023})},\ \Eprint {http://arxiv.org/abs/2306.14472} {arXiv:2306.14472 [hep-ph]} \BibitemShut {NoStop}%
\bibitem [{\citenamefont {Kim}\ \emph {et~al.}(2023)\citenamefont {Kim}, \citenamefont {Pattanaik},\ and\ \citenamefont {Unger}}]{Kim:2023dnq}%
  \BibitemOpen
  \bibfield  {author} {\bibinfo {author} {\bibfnamefont {Jangho}\ \bibnamefont {Kim}}, \bibinfo {author} {\bibfnamefont {Pratitee}\ \bibnamefont {Pattanaik}}, \ and\ \bibinfo {author} {\bibfnamefont {Wolfgang}\ \bibnamefont {Unger}},\ }\bibfield  {title} {\enquote {\bibinfo {title} {{Nuclear liquid-gas transition in the strong coupling regime of lattice QCD}},}\ }\href {\doibase 10.1103/PhysRevD.107.094514} {\bibfield  {journal} {\bibinfo  {journal} {Phys. Rev. D}\ }\textbf {\bibinfo {volume} {107}},\ \bibinfo {pages} {094514} (\bibinfo {year} {2023})},\ \Eprint {http://arxiv.org/abs/2303.01467} {arXiv:2303.01467 [hep-lat]} \BibitemShut {NoStop}%
\bibitem [{\citenamefont {Chiba}\ and\ \citenamefont {Kojo}(2023)}]{Chiba:2023ftg}%
  \BibitemOpen
  \bibfield  {author} {\bibinfo {author} {\bibfnamefont {Ryuji}\ \bibnamefont {Chiba}}\ and\ \bibinfo {author} {\bibfnamefont {Toru}\ \bibnamefont {Kojo}},\ }\bibfield  {title} {\enquote {\bibinfo {title} {{Sound velocity peak and conformality in isospin QCD}},}\ }\href@noop {} {\  (\bibinfo {year} {2023})},\ \Eprint {http://arxiv.org/abs/2304.13920} {arXiv:2304.13920 [hep-ph]} \BibitemShut {NoStop}%
\bibitem [{\citenamefont {Ayala}\ \emph {et~al.}(2023)\citenamefont {Ayala}, \citenamefont {Lopes}, \citenamefont {Farias},\ and\ \citenamefont {Parra}}]{Ayala:2023mms}%
  \BibitemOpen
  \bibfield  {author} {\bibinfo {author} {\bibfnamefont {Alejandro}\ \bibnamefont {Ayala}}, \bibinfo {author} {\bibfnamefont {Bruno~S.}\ \bibnamefont {Lopes}}, \bibinfo {author} {\bibfnamefont {Ricardo L.~S.}\ \bibnamefont {Farias}}, \ and\ \bibinfo {author} {\bibfnamefont {Luis~C.}\ \bibnamefont {Parra}},\ }\bibfield  {title} {\enquote {\bibinfo {title} {{Describing the speed of sound peak of isospin-asymmetric cold strongly interacting matter using effective models}},}\ }\href@noop {} {\  (\bibinfo {year} {2023})},\ \Eprint {http://arxiv.org/abs/2310.13130} {arXiv:2310.13130 [hep-ph]} \BibitemShut {NoStop}%
\bibitem [{\citenamefont {Lawrence}\ and\ \citenamefont {Romatschke}(2023)}]{Lawrence:2022vwa}%
  \BibitemOpen
  \bibfield  {author} {\bibinfo {author} {\bibfnamefont {Scott}\ \bibnamefont {Lawrence}}\ and\ \bibinfo {author} {\bibfnamefont {Paul}\ \bibnamefont {Romatschke}},\ }\bibfield  {title} {\enquote {\bibinfo {title} {{Gravitational-wave-to-matter coupling of superfluid Fermi gases near unitarity}},}\ }\href {\doibase 10.1103/PhysRevA.107.033327} {\bibfield  {journal} {\bibinfo  {journal} {Phys. Rev. A}\ }\textbf {\bibinfo {volume} {107}},\ \bibinfo {pages} {033327} (\bibinfo {year} {2023})},\ \Eprint {http://arxiv.org/abs/2206.04765} {arXiv:2206.04765 [cond-mat.str-el]} \BibitemShut {NoStop}%
\bibitem [{\citenamefont {Gorda}\ \emph {et~al.}(2021{\natexlab{a}})\citenamefont {Gorda}, \citenamefont {Kurkela}, \citenamefont {Paatelainen}, \citenamefont {S\"appi},\ and\ \citenamefont {Vuorinen}}]{Gorda:2021kme}%
  \BibitemOpen
  \bibfield  {author} {\bibinfo {author} {\bibfnamefont {Tyler}\ \bibnamefont {Gorda}}, \bibinfo {author} {\bibfnamefont {Aleksi}\ \bibnamefont {Kurkela}}, \bibinfo {author} {\bibfnamefont {Risto}\ \bibnamefont {Paatelainen}}, \bibinfo {author} {\bibfnamefont {Saga}\ \bibnamefont {S\"appi}}, \ and\ \bibinfo {author} {\bibfnamefont {Aleksi}\ \bibnamefont {Vuorinen}},\ }\bibfield  {title} {\enquote {\bibinfo {title} {{Cold quark matter at N3LO: Soft contributions}},}\ }\href {\doibase 10.1103/PhysRevD.104.074015} {\bibfield  {journal} {\bibinfo  {journal} {Phys. Rev. D}\ }\textbf {\bibinfo {volume} {104}},\ \bibinfo {pages} {074015} (\bibinfo {year} {2021}{\natexlab{a}})},\ \Eprint {http://arxiv.org/abs/2103.07427} {arXiv:2103.07427 [hep-ph]} \BibitemShut {NoStop}%
\bibitem [{\citenamefont {Gorda}\ \emph {et~al.}(2021{\natexlab{b}})\citenamefont {Gorda}, \citenamefont {Kurkela}, \citenamefont {Paatelainen}, \citenamefont {S\"appi},\ and\ \citenamefont {Vuorinen}}]{Gorda:2021znl}%
  \BibitemOpen
  \bibfield  {author} {\bibinfo {author} {\bibfnamefont {Tyler}\ \bibnamefont {Gorda}}, \bibinfo {author} {\bibfnamefont {Aleksi}\ \bibnamefont {Kurkela}}, \bibinfo {author} {\bibfnamefont {Risto}\ \bibnamefont {Paatelainen}}, \bibinfo {author} {\bibfnamefont {Saga}\ \bibnamefont {S\"appi}}, \ and\ \bibinfo {author} {\bibfnamefont {Aleksi}\ \bibnamefont {Vuorinen}},\ }\bibfield  {title} {\enquote {\bibinfo {title} {{Soft Interactions in Cold Quark Matter}},}\ }\href {\doibase 10.1103/PhysRevLett.127.162003} {\bibfield  {journal} {\bibinfo  {journal} {Phys. Rev. Lett.}\ }\textbf {\bibinfo {volume} {127}},\ \bibinfo {pages} {162003} (\bibinfo {year} {2021}{\natexlab{b}})},\ \Eprint {http://arxiv.org/abs/2103.05658} {arXiv:2103.05658 [hep-ph]} \BibitemShut {NoStop}%
\bibitem [{\citenamefont {Gorda}\ \emph {et~al.}(2023)\citenamefont {Gorda}, \citenamefont {Komoltsev}, \citenamefont {Kurkela},\ and\ \citenamefont {Mazeliauskas}}]{Gorda:2023usm}%
  \BibitemOpen
  \bibfield  {author} {\bibinfo {author} {\bibfnamefont {Tyler}\ \bibnamefont {Gorda}}, \bibinfo {author} {\bibfnamefont {Oleg}\ \bibnamefont {Komoltsev}}, \bibinfo {author} {\bibfnamefont {Aleksi}\ \bibnamefont {Kurkela}}, \ and\ \bibinfo {author} {\bibfnamefont {Aleksas}\ \bibnamefont {Mazeliauskas}},\ }\bibfield  {title} {\enquote {\bibinfo {title} {{Bayesian uncertainty quantification of perturbative QCD input to the neutron-star equation of state}},}\ }\href {\doibase 10.1007/JHEP06(2023)002} {\bibfield  {journal} {\bibinfo  {journal} {JHEP}\ }\textbf {\bibinfo {volume} {06}},\ \bibinfo {pages} {002} (\bibinfo {year} {2023})},\ \Eprint {http://arxiv.org/abs/2303.02175} {arXiv:2303.02175 [hep-ph]} \BibitemShut {NoStop}%
\bibitem [{\citenamefont {Cotter}\ \emph {et~al.}(2013)\citenamefont {Cotter}, \citenamefont {Giudice}, \citenamefont {Hands},\ and\ \citenamefont {Skullerud}}]{Cotter:2012mb}%
  \BibitemOpen
  \bibfield  {author} {\bibinfo {author} {\bibfnamefont {Seamus}\ \bibnamefont {Cotter}}, \bibinfo {author} {\bibfnamefont {Pietro}\ \bibnamefont {Giudice}}, \bibinfo {author} {\bibfnamefont {Simon}\ \bibnamefont {Hands}}, \ and\ \bibinfo {author} {\bibfnamefont {Jon-Ivar}\ \bibnamefont {Skullerud}},\ }\bibfield  {title} {\enquote {\bibinfo {title} {{Towards the phase diagram of dense two-color matter}},}\ }\href {\doibase 10.1103/PhysRevD.87.034507} {\bibfield  {journal} {\bibinfo  {journal} {Phys. Rev. D}\ }\textbf {\bibinfo {volume} {87}},\ \bibinfo {pages} {034507} (\bibinfo {year} {2013})},\ \Eprint {http://arxiv.org/abs/1210.4496} {arXiv:1210.4496 [hep-lat]} \BibitemShut {NoStop}%
\bibitem [{\citenamefont {Miller}\ \emph {et~al.}(2019)\citenamefont {Miller} \emph {et~al.}}]{Miller:2019cac}%
  \BibitemOpen
  \bibfield  {author} {\bibinfo {author} {\bibfnamefont {M.~C.}\ \bibnamefont {Miller}} \emph {et~al.},\ }\bibfield  {title} {\enquote {\bibinfo {title} {{PSR J0030+0451 Mass and Radius from $NICER$ Data and Implications for the Properties of Neutron Star Matter}},}\ }\href {\doibase 10.3847/2041-8213/ab50c5} {\bibfield  {journal} {\bibinfo  {journal} {Astrophys. J. Lett.}\ }\textbf {\bibinfo {volume} {887}},\ \bibinfo {pages} {L24} (\bibinfo {year} {2019})},\ \Eprint {http://arxiv.org/abs/1912.05705} {arXiv:1912.05705 [astro-ph.HE]} \BibitemShut {NoStop}%
\bibitem [{\citenamefont {Riley}\ \emph {et~al.}(2019)\citenamefont {Riley} \emph {et~al.}}]{Riley:2019yda}%
  \BibitemOpen
  \bibfield  {author} {\bibinfo {author} {\bibfnamefont {Thomas~E.}\ \bibnamefont {Riley}} \emph {et~al.},\ }\bibfield  {title} {\enquote {\bibinfo {title} {{A $NICER$ View of PSR J0030+0451: Millisecond Pulsar Parameter Estimation}},}\ }\href {\doibase 10.3847/2041-8213/ab481c} {\bibfield  {journal} {\bibinfo  {journal} {Astrophys. J. Lett.}\ }\textbf {\bibinfo {volume} {887}},\ \bibinfo {pages} {L21} (\bibinfo {year} {2019})},\ \Eprint {http://arxiv.org/abs/1912.05702} {arXiv:1912.05702 [astro-ph.HE]} \BibitemShut {NoStop}%
\bibitem [{\citenamefont {Miller}\ \emph {et~al.}(2021)\citenamefont {Miller} \emph {et~al.}}]{Miller:2021qha}%
  \BibitemOpen
  \bibfield  {author} {\bibinfo {author} {\bibfnamefont {M.~C.}\ \bibnamefont {Miller}} \emph {et~al.},\ }\bibfield  {title} {\enquote {\bibinfo {title} {{The Radius of PSR J0740+6620 from NICER and XMM-Newton Data}},}\ }\href {\doibase 10.3847/2041-8213/ac089b} {\bibfield  {journal} {\bibinfo  {journal} {Astrophys. J. Lett.}\ }\textbf {\bibinfo {volume} {918}},\ \bibinfo {pages} {L28} (\bibinfo {year} {2021})},\ \Eprint {http://arxiv.org/abs/2105.06979} {arXiv:2105.06979 [astro-ph.HE]} \BibitemShut {NoStop}%
\bibitem [{\citenamefont {Riley}\ \emph {et~al.}(2021)\citenamefont {Riley} \emph {et~al.}}]{Riley:2021pdl}%
  \BibitemOpen
  \bibfield  {author} {\bibinfo {author} {\bibfnamefont {Thomas~E.}\ \bibnamefont {Riley}} \emph {et~al.},\ }\bibfield  {title} {\enquote {\bibinfo {title} {{A NICER View of the Massive Pulsar PSR J0740+6620 Informed by Radio Timing and XMM-Newton Spectroscopy}},}\ }\href {\doibase 10.3847/2041-8213/ac0a81} {\bibfield  {journal} {\bibinfo  {journal} {Astrophys. J. Lett.}\ }\textbf {\bibinfo {volume} {918}},\ \bibinfo {pages} {L27} (\bibinfo {year} {2021})},\ \Eprint {http://arxiv.org/abs/2105.06980} {arXiv:2105.06980 [astro-ph.HE]} \BibitemShut {NoStop}%
\bibitem [{\citenamefont {Alsing}\ \emph {et~al.}(2018)\citenamefont {Alsing}, \citenamefont {Silva},\ and\ \citenamefont {Berti}}]{Alsing:2017bbc}%
  \BibitemOpen
  \bibfield  {author} {\bibinfo {author} {\bibfnamefont {Justin}\ \bibnamefont {Alsing}}, \bibinfo {author} {\bibfnamefont {Hector~O.}\ \bibnamefont {Silva}}, \ and\ \bibinfo {author} {\bibfnamefont {Emanuele}\ \bibnamefont {Berti}},\ }\bibfield  {title} {\enquote {\bibinfo {title} {{Evidence for a maximum mass cut-off in the neutron star mass distribution and constraints on the equation of state}},}\ }\href {\doibase 10.1093/mnras/sty1065} {\bibfield  {journal} {\bibinfo  {journal} {Mon. Not. Roy. Astron. Soc.}\ }\textbf {\bibinfo {volume} {478}},\ \bibinfo {pages} {1377--1391} (\bibinfo {year} {2018})},\ \Eprint {http://arxiv.org/abs/1709.07889} {arXiv:1709.07889 [astro-ph.HE]} \BibitemShut {NoStop}%
\bibitem [{\citenamefont {Ma}\ and\ \citenamefont {Rho}(2019)}]{Ma:2018qkg}%
  \BibitemOpen
  \bibfield  {author} {\bibinfo {author} {\bibfnamefont {Yong-Liang}\ \bibnamefont {Ma}}\ and\ \bibinfo {author} {\bibfnamefont {Mannque}\ \bibnamefont {Rho}},\ }\bibfield  {title} {\enquote {\bibinfo {title} {{Sound velocity and tidal deformability in compact stars}},}\ }\href {\doibase 10.1103/PhysRevD.100.114003} {\bibfield  {journal} {\bibinfo  {journal} {Phys. Rev. D}\ }\textbf {\bibinfo {volume} {100}},\ \bibinfo {pages} {114003} (\bibinfo {year} {2019})},\ \Eprint {http://arxiv.org/abs/1811.07071} {arXiv:1811.07071 [nucl-th]} \BibitemShut {NoStop}%
\bibitem [{\citenamefont {Zhang}\ \emph {et~al.}(2019)\citenamefont {Zhang}, \citenamefont {Wen},\ and\ \citenamefont {Chen}}]{Zhang:2019udy}%
  \BibitemOpen
  \bibfield  {author} {\bibinfo {author} {\bibfnamefont {Na}~\bibnamefont {Zhang}}, \bibinfo {author} {\bibfnamefont {Dehua}\ \bibnamefont {Wen}}, \ and\ \bibinfo {author} {\bibfnamefont {Houyuan}\ \bibnamefont {Chen}},\ }\bibfield  {title} {\enquote {\bibinfo {title} {{Imprint of the speed of sound in nuclear matter on global properties of neutron stars}},}\ }\href {\doibase 10.1103/PhysRevC.99.035803} {\bibfield  {journal} {\bibinfo  {journal} {Phys. Rev. C}\ }\textbf {\bibinfo {volume} {99}},\ \bibinfo {pages} {035803} (\bibinfo {year} {2019})}\BibitemShut {NoStop}%
\bibitem [{\citenamefont {Alford}\ \emph {et~al.}(2015)\citenamefont {Alford}, \citenamefont {Burgio}, \citenamefont {Han}, \citenamefont {Taranto},\ and\ \citenamefont {Zappal\`a}}]{Alford:2015dpa}%
  \BibitemOpen
  \bibfield  {author} {\bibinfo {author} {\bibfnamefont {Mark~G.}\ \bibnamefont {Alford}}, \bibinfo {author} {\bibfnamefont {G.~F.}\ \bibnamefont {Burgio}}, \bibinfo {author} {\bibfnamefont {Sophia}\ \bibnamefont {Han}}, \bibinfo {author} {\bibfnamefont {Gabriele}\ \bibnamefont {Taranto}}, \ and\ \bibinfo {author} {\bibfnamefont {Dario}\ \bibnamefont {Zappal\`a}},\ }\bibfield  {title} {\enquote {\bibinfo {title} {{Constraining and applying a generic high-density equation of state}},}\ }\href {\doibase 10.1103/PhysRevD.92.083002} {\bibfield  {journal} {\bibinfo  {journal} {Phys. Rev. D}\ }\textbf {\bibinfo {volume} {92}},\ \bibinfo {pages} {083002} (\bibinfo {year} {2015})},\ \Eprint {http://arxiv.org/abs/1501.07902} {arXiv:1501.07902 [nucl-th]} \BibitemShut {NoStop}%
\bibitem [{\citenamefont {Grill}\ \emph {et~al.}(2014)\citenamefont {Grill}, \citenamefont {Pais}, \citenamefont {Provid\^encia}, \citenamefont {Vida\~na},\ and\ \citenamefont {Avancini}}]{Grill:2014aea}%
  \BibitemOpen
  \bibfield  {author} {\bibinfo {author} {\bibfnamefont {Fabrizio}\ \bibnamefont {Grill}}, \bibinfo {author} {\bibfnamefont {Helena}\ \bibnamefont {Pais}}, \bibinfo {author} {\bibfnamefont {Constan\c{c}a}\ \bibnamefont {Provid\^encia}}, \bibinfo {author} {\bibfnamefont {Isaac}\ \bibnamefont {Vida\~na}}, \ and\ \bibinfo {author} {\bibfnamefont {Sidney~S.}\ \bibnamefont {Avancini}},\ }\bibfield  {title} {\enquote {\bibinfo {title} {{Equation of state and thickness of the inner crust of neutron stars}},}\ }\href {\doibase 10.1103/PhysRevC.90.045803} {\bibfield  {journal} {\bibinfo  {journal} {Phys. Rev. C}\ }\textbf {\bibinfo {volume} {90}},\ \bibinfo {pages} {045803} (\bibinfo {year} {2014})},\ \Eprint {http://arxiv.org/abs/1404.2753} {arXiv:1404.2753 [nucl-th]} \BibitemShut {NoStop}%
\bibitem [{\citenamefont {Shen}\ \emph {et~al.}(2020)\citenamefont {Shen}, \citenamefont {Ji}, \citenamefont {Hu},\ and\ \citenamefont {Sumiyoshi}}]{Shen:2020sec}%
  \BibitemOpen
  \bibfield  {author} {\bibinfo {author} {\bibfnamefont {Hong}\ \bibnamefont {Shen}}, \bibinfo {author} {\bibfnamefont {Fan}\ \bibnamefont {Ji}}, \bibinfo {author} {\bibfnamefont {Jinniu}\ \bibnamefont {Hu}}, \ and\ \bibinfo {author} {\bibfnamefont {Kohsuke}\ \bibnamefont {Sumiyoshi}},\ }\bibfield  {title} {\enquote {\bibinfo {title} {{Effects of symmetry energy on equation of state for simulations of core-collapse supernovae and neutron-star mergers}},}\ }\href {\doibase 10.3847/1538-4357/ab72fd} {\bibfield  {journal} {\bibinfo  {journal} {Astrophys. J.}\ }\textbf {\bibinfo {volume} {891}},\ \bibinfo {pages} {148} (\bibinfo {year} {2020})},\ \Eprint {http://arxiv.org/abs/2001.10143} {arXiv:2001.10143 [nucl-th]} \BibitemShut {NoStop}%
\bibitem [{\citenamefont {Boukari}\ \emph {et~al.}(2021)\citenamefont {Boukari}, \citenamefont {Pais}, \citenamefont {Anti\'c},\ and\ \citenamefont {Provid\^encia}}]{Boukari:2020iut}%
  \BibitemOpen
  \bibfield  {author} {\bibinfo {author} {\bibfnamefont {Olfa}\ \bibnamefont {Boukari}}, \bibinfo {author} {\bibfnamefont {Helena}\ \bibnamefont {Pais}}, \bibinfo {author} {\bibfnamefont {Sofija}\ \bibnamefont {Anti\'c}}, \ and\ \bibinfo {author} {\bibfnamefont {Constan\c{c}a}\ \bibnamefont {Provid\^encia}},\ }\bibfield  {title} {\enquote {\bibinfo {title} {{Critical properties of calibrated relativistic mean-field models for the transition to warm, nonhomogeneous nuclear and stellar matter}},}\ }\href {\doibase 10.1103/PhysRevC.103.055804} {\bibfield  {journal} {\bibinfo  {journal} {Phys. Rev. C}\ }\textbf {\bibinfo {volume} {103}},\ \bibinfo {pages} {055804} (\bibinfo {year} {2021})},\ \Eprint {http://arxiv.org/abs/2007.08852} {arXiv:2007.08852 [nucl-th]} \BibitemShut {NoStop}%
\bibitem [{\citenamefont {Pearson}\ \emph {et~al.}(2018)\citenamefont {Pearson}, \citenamefont {Chamel}, \citenamefont {Potekhin}, \citenamefont {Fantina}, \citenamefont {Ducoin}, \citenamefont {Dutta},\ and\ \citenamefont {Goriely}}]{Pearson:2018tkr}%
  \BibitemOpen
  \bibfield  {author} {\bibinfo {author} {\bibfnamefont {J.~M.}\ \bibnamefont {Pearson}}, \bibinfo {author} {\bibfnamefont {N.}~\bibnamefont {Chamel}}, \bibinfo {author} {\bibfnamefont {A.~Y.}\ \bibnamefont {Potekhin}}, \bibinfo {author} {\bibfnamefont {A.~F.}\ \bibnamefont {Fantina}}, \bibinfo {author} {\bibfnamefont {C.}~\bibnamefont {Ducoin}}, \bibinfo {author} {\bibfnamefont {A.~K.}\ \bibnamefont {Dutta}}, \ and\ \bibinfo {author} {\bibfnamefont {S.}~\bibnamefont {Goriely}},\ }\bibfield  {title} {\enquote {\bibinfo {title} {{Unified equations of state for cold non-accreting neutron stars with Brussels\textendash{}Montreal functionals \textendash{} I. Role of symmetry energy}},}\ }\href {\doibase 10.1093/mnras/sty2413} {\bibfield  {journal} {\bibinfo  {journal} {Mon. Not. Roy. Astron. Soc.}\ }\textbf {\bibinfo {volume} {481}},\ \bibinfo {pages} {2994--3026} (\bibinfo {year} {2018})},\ \bibinfo {note} {[Erratum: Mon.Not.Roy.Astron.Soc. 486, 768 (2019)]},\ \Eprint {http://arxiv.org/abs/1903.04981}
  {arXiv:1903.04981 [astro-ph.HE]} \BibitemShut {NoStop}%
\bibitem [{\citenamefont {Oertel}\ \emph {et~al.}(2017)\citenamefont {Oertel}, \citenamefont {Hempel}, \citenamefont {Kl\"ahn},\ and\ \citenamefont {Typel}}]{Oertel:2016bki}%
  \BibitemOpen
  \bibfield  {author} {\bibinfo {author} {\bibfnamefont {M.}~\bibnamefont {Oertel}}, \bibinfo {author} {\bibfnamefont {M.}~\bibnamefont {Hempel}}, \bibinfo {author} {\bibfnamefont {T.}~\bibnamefont {Kl\"ahn}}, \ and\ \bibinfo {author} {\bibfnamefont {S.}~\bibnamefont {Typel}},\ }\bibfield  {title} {\enquote {\bibinfo {title} {{Equations of state for supernovae and compact stars}},}\ }\href {\doibase 10.1103/RevModPhys.89.015007} {\bibfield  {journal} {\bibinfo  {journal} {Rev. Mod. Phys.}\ }\textbf {\bibinfo {volume} {89}},\ \bibinfo {pages} {015007} (\bibinfo {year} {2017})},\ \Eprint {http://arxiv.org/abs/1610.03361} {arXiv:1610.03361 [astro-ph.HE]} \BibitemShut {NoStop}%
\bibitem [{\citenamefont {Moustakidis}\ \emph {et~al.}(2017)\citenamefont {Moustakidis}, \citenamefont {Gaitanos}, \citenamefont {Margaritis},\ and\ \citenamefont {Lalazissis}}]{Moustakidis:2016sab}%
  \BibitemOpen
  \bibfield  {author} {\bibinfo {author} {\bibfnamefont {Ch.~C.}\ \bibnamefont {Moustakidis}}, \bibinfo {author} {\bibfnamefont {T.}~\bibnamefont {Gaitanos}}, \bibinfo {author} {\bibfnamefont {Ch.}\ \bibnamefont {Margaritis}}, \ and\ \bibinfo {author} {\bibfnamefont {G.~A.}\ \bibnamefont {Lalazissis}},\ }\bibfield  {title} {\enquote {\bibinfo {title} {{Bounds on the speed of sound in dense matter, and neutron star structure}},}\ }\href {\doibase 10.1103/PhysRevC.95.045801} {\bibfield  {journal} {\bibinfo  {journal} {Phys. Rev. C}\ }\textbf {\bibinfo {volume} {95}},\ \bibinfo {pages} {045801} (\bibinfo {year} {2017})},\ \bibinfo {note} {[Erratum: Phys.Rev.C 95, 059904 (2017)]},\ \Eprint {http://arxiv.org/abs/1608.00344} {arXiv:1608.00344 [nucl-th]} \BibitemShut {NoStop}%
\bibitem [{\citenamefont {Kovtun}\ and\ \citenamefont {Shukla}(2018)}]{Kovtun:2018dvd}%
  \BibitemOpen
  \bibfield  {author} {\bibinfo {author} {\bibfnamefont {Pavel}\ \bibnamefont {Kovtun}}\ and\ \bibinfo {author} {\bibfnamefont {Ashish}\ \bibnamefont {Shukla}},\ }\bibfield  {title} {\enquote {\bibinfo {title} {{Kubo formulas for thermodynamic transport coefficients}},}\ }\href {\doibase 10.1007/JHEP10(2018)007} {\bibfield  {journal} {\bibinfo  {journal} {JHEP}\ }\textbf {\bibinfo {volume} {10}},\ \bibinfo {pages} {007} (\bibinfo {year} {2018})},\ \Eprint {http://arxiv.org/abs/1806.05774} {arXiv:1806.05774 [hep-th]} \BibitemShut {NoStop}%
\bibitem [{\citenamefont {Grieninger}\ and\ \citenamefont {Shukla}(2021)}]{Grieninger:2021rxd}%
  \BibitemOpen
  \bibfield  {author} {\bibinfo {author} {\bibfnamefont {Sebastian}\ \bibnamefont {Grieninger}}\ and\ \bibinfo {author} {\bibfnamefont {Ashish}\ \bibnamefont {Shukla}},\ }\bibfield  {title} {\enquote {\bibinfo {title} {{Second order equilibrium transport in strongly coupled $ \mathcal{N} $ = 4 supersymmetric SU(N$_{c}$) Yang-Mills plasma via holography}},}\ }\href {\doibase 10.1007/JHEP08(2021)108} {\bibfield  {journal} {\bibinfo  {journal} {JHEP}\ }\textbf {\bibinfo {volume} {08}},\ \bibinfo {pages} {108} (\bibinfo {year} {2021})},\ \Eprint {http://arxiv.org/abs/2105.08673} {arXiv:2105.08673 [hep-th]} \BibitemShut {NoStop}%
\bibitem [{\citenamefont {Weiner}(2023)}]{Weiner:2023wew}%
  \BibitemOpen
  \bibfield  {author} {\bibinfo {author} {\bibfnamefont {Max}\ \bibnamefont {Weiner}},\ }\bibfield  {title} {\enquote {\bibinfo {title} {{Thermodynamic Susceptibilities for a Unitary Fermi Gas}},}\ }\href@noop {} {\  (\bibinfo {year} {2023})},\ \Eprint {http://arxiv.org/abs/2311.10906} {arXiv:2311.10906 [nucl-th]} \BibitemShut {NoStop}%
\bibitem [{\citenamefont {Ratti}\ and\ \citenamefont {Weise}(2004)}]{Ratti:2004ra}%
  \BibitemOpen
  \bibfield  {author} {\bibinfo {author} {\bibfnamefont {Claudia}\ \bibnamefont {Ratti}}\ and\ \bibinfo {author} {\bibfnamefont {Wolfram}\ \bibnamefont {Weise}},\ }\bibfield  {title} {\enquote {\bibinfo {title} {{Thermodynamics of two-colour QCD and the Nambu Jona-Lasinio model}},}\ }\href {\doibase 10.1103/PhysRevD.70.054013} {\bibfield  {journal} {\bibinfo  {journal} {Phys. Rev. D}\ }\textbf {\bibinfo {volume} {70}},\ \bibinfo {pages} {054013} (\bibinfo {year} {2004})},\ \Eprint {http://arxiv.org/abs/hep-ph/0406159} {arXiv:hep-ph/0406159} \BibitemShut {NoStop}%
\bibitem [{\citenamefont {Hebeler}\ \emph {et~al.}(2010)\citenamefont {Hebeler}, \citenamefont {Lattimer}, \citenamefont {Pethick},\ and\ \citenamefont {Schwenk}}]{Hebeler:2010jx}%
  \BibitemOpen
  \bibfield  {author} {\bibinfo {author} {\bibfnamefont {K.}~\bibnamefont {Hebeler}}, \bibinfo {author} {\bibfnamefont {J.~M.}\ \bibnamefont {Lattimer}}, \bibinfo {author} {\bibfnamefont {C.~J.}\ \bibnamefont {Pethick}}, \ and\ \bibinfo {author} {\bibfnamefont {A.}~\bibnamefont {Schwenk}},\ }\bibfield  {title} {\enquote {\bibinfo {title} {{Constraints on neutron star radii based on chiral effective field theory interactions}},}\ }\href {\doibase 10.1103/PhysRevLett.105.161102} {\bibfield  {journal} {\bibinfo  {journal} {Phys. Rev. Lett.}\ }\textbf {\bibinfo {volume} {105}},\ \bibinfo {pages} {161102} (\bibinfo {year} {2010})},\ \Eprint {http://arxiv.org/abs/1007.1746} {arXiv:1007.1746 [nucl-th]} \BibitemShut {NoStop}%
\bibitem [{com()}]{compose}%
  \BibitemOpen
  \href@noop {} {\enquote {\bibinfo {title} {{CompOSE -- CompStar Supernovae Equations of State}},}\ }\bibinfo {howpublished} {\url{compose.obspm.fr}}\BibitemShut {NoStop}%
\bibitem [{\citenamefont {Douchin}\ and\ \citenamefont {Haensel}(2001)}]{Douchin:2001sv}%
  \BibitemOpen
  \bibfield  {author} {\bibinfo {author} {\bibfnamefont {F.}~\bibnamefont {Douchin}}\ and\ \bibinfo {author} {\bibfnamefont {P.}~\bibnamefont {Haensel}},\ }\bibfield  {title} {\enquote {\bibinfo {title} {{A unified equation of state of dense matter and neutron star structure}},}\ }\href {\doibase 10.1051/0004-6361:20011402} {\bibfield  {journal} {\bibinfo  {journal} {Astron. Astrophys.}\ }\textbf {\bibinfo {volume} {380}},\ \bibinfo {pages} {151} (\bibinfo {year} {2001})},\ \Eprint {http://arxiv.org/abs/astro-ph/0111092} {arXiv:astro-ph/0111092} \BibitemShut {NoStop}%
\end{thebibliography}%

\end{document}